\documentclass{aa}  
\usepackage{natbib}
\bibpunct{(}{)}{;}{a}{}{,} 
\usepackage{orcidlink}
\usepackage{multirow}
\usepackage{hyperref}
\usepackage{xcolor}
\usepackage[utf8]{inputenc}
\DeclareUnicodeCharacter{2212}{\ensuremath{-}}
\usepackage{ulem}
\usepackage[document]{}
\usepackage{graphicx}
\usepackage{arydshln}
\usepackage{pifont}
\usepackage{txfonts}
\begin{document}



    \title{Magnetic field morphological diagnostics with ALMA in the G327.29 protocluster: VGT versus dust polarization}
    
    
   \author{A. Koley$^{1}$\thanks{Fondecyt Postdoctoral fellow}\orcidlink{0000-0003-2713-0211}, A. M. Stutz$^{1}$\orcidlink{0000-0003-2300-8200}, A. Lazarian$^{2}$\orcidlink{0000-0002-7336-6674}, Y. Hu $^{3}$\orcidlink{0000-0002-8455-0805}, P. Sanhueza$^{4}$\orcidlink{0000-0002-7125-7685}, P. Saha$^{5,6}$\orcidlink{0000-0002-0028-1354}, R. H. $\acute{\text{A}}$lvarez-Guti$\acute{\text{e}}$rrez$^{7}$\orcidlink{0000-0002-9386-8612},  N. Sandoval-Garrido$^{1}$\orcidlink{0000-0001-9600-2796}, N. Castro-Toledo$^{1}$\orcidlink{0009-0005-6363-5104}, G. Bernal Mesina$^{1}$\\
   \vspace{5mm}
    (Affiliations can be found after the references)}
    \authorrunning{Koley et al.}{}
          \institute{}
    \vspace{5mm}

   \date{Received xxxxxxx; accepted xxxxxx}{}

 
  \abstract
    {Magnetic fields and turbulence may play a key role in the evolution of protoclusters, influencing the formation of dense cores and stars. Here, we examine the morphology of the magnetic fields in the G327.29 protocluster using both the velocity gradient technique (VGT) extracted from molecular line emissions and linear polarization in the dust continuum emission. The VGT analysis is performed using four molecular tracers: DCN (3$-$2), C$^{18}$O (2$-$1), HN$^{13}$C (3$-$2), and H$^{13}$CO$^{+}$ (3$-$2) $-$ observed with the ALMA 12 m array. Owing to its sensitivity to gas dynamics, a comparison between VGT and dust polarization provides a powerful probe of the evolutionary processes in massive star-forming regions. Our analysis reveals a complex magnetic-field structure shaped by the combined influence of turbulence and gravity, as evidenced by the distributions of both the alignment measure ($AM$) and the relative orientation angle ($\theta_\text{r}$), between the magnetic fields inferred from the VGT and dust polarization observations. The $\theta_\text{r}$ distributions are clearly double-peaked for DCN and HN$^{13}$C, whereas C$^{18}$O and H$^{13}$CO$^+$ predominantly exhibit a preferred perpendicular orientation.
 In addition, it also appears that there is a large-scale (beyond the core scale) gravitational infall from the surrounding medium on to the filament and the central densest region. Furthermore, we observe that cores are dominated by a mix of turbulence and gravity. Overall, this work presents, for the first time, the application of VGT to a massive protocluster, G327.29, using high-resolution ALMA observations.} 


   \maketitle
%

\section{Introduction}\label{section_1}

Turbulence and magnetic field ($B$-field) are ubiquitous in the universe. These two actors are believed to play a crucial role in many processes, including star formation, propagation of cosmic rays, and the evolution of galaxies \citep{2016A&A...590A...2S,2018MNRAS.473.4890S,2021ApJ...923...53L,2022MNRAS.516..185K,2022MNRAS.516L..48K,2023PASA...40...53K,2025A&A...702A.133K}. In molecular clouds, turbulence and magnetic fields acting together with gravity make the star formation process more complex. Even though numerous theoretical, numerical, and observational studies have been performed over several decades, the importance of magnetic fields and turbulence in star formation is still being hotly debated \citep{2009Sci...324.1408G,2017ApJ...842L...9H,2021MNRAS.501.4825K}. While turbulence measurements are possible using several techniques, including intensity and velocity power spectrum analysis \citep{2000ApJ...537..720L,
2015ApJ...810...33C,2018ApJ...856..136P}, Principal Component Analysis (PCA) \citep{2004ApJ...615L..45H}, spectral line width estimation \citep{2019MNRAS.483..593K,2022MNRAS.516..185K,2023PASA...40...53K,2023PASA...40...46K}, measurements of magnetic fields and their 3D morphology remain challenging \citep{2010ApJ...725..466C,2023ASPC..534..193P}. Thus, techniques that attempt to extract physical parameters related to the nature of magnetic fields are precious to the endeavor of constraining the physics of star formation.\\

Nowadays, dust polarization is the most effective tool for studying the morphology of the magnetic field on the plane of the sky ($B_{\text{POS}}$). The analysis of $B_{\text{POS}}$ sheds light on many aspects of star formation. For example, the observed hourglass $B$-field morphology formed by the pinching effect of gravity is a crucial evidence of the importance of the $B$-field in star formation \citep{2009Sci...324.1408G,2024ApJ...972L...6S}. Similarly, a chaotic orientation of the $B$-field indicates the dominance of turbulence over the $B$-field in cloud and core dynamics \citep{2017ApJ...842L...9H}. On a comparatively small scale, the relative orientation of the outflow and the $B$-field predicts the efficiency of magnetic braking during the disk formation of protostars \citep{2009ApJ...698..922M,2009Sci...324.1408G,2018A&A...616A.139G}. Aside from morphological diagnostics, indirect estimation of the magnetic field strength using dust polarization allows us to assess the dynamically evolving stages of a system by measuring the mass-to-magnetic flux ratio, the virial parameter, and the ratio of $B$-field and turbulence energies \citep{2021ApJ...915L..10S, LYP22, 2025ApJ...980...87S}. For several decades, this type of analysis has been carried out with single-dish and interferometric telescopes in massive star forming regions \citep{2013ApJ...779..182Q,2019ApJ...883...95S,2021ApJ...923..204C,2021ApJ...913...29F,2022ApJ...926L...6S, 2024ApJ...967..157L,2024ApJ...974..257Z,2024ApJ...972..115C,2025arXiv251025078H,2025arXiv251005933S}.

\begin{figure*}[ht!]
	\centering 
	\includegraphics[width=7.2in,height=2.6in,angle=0]{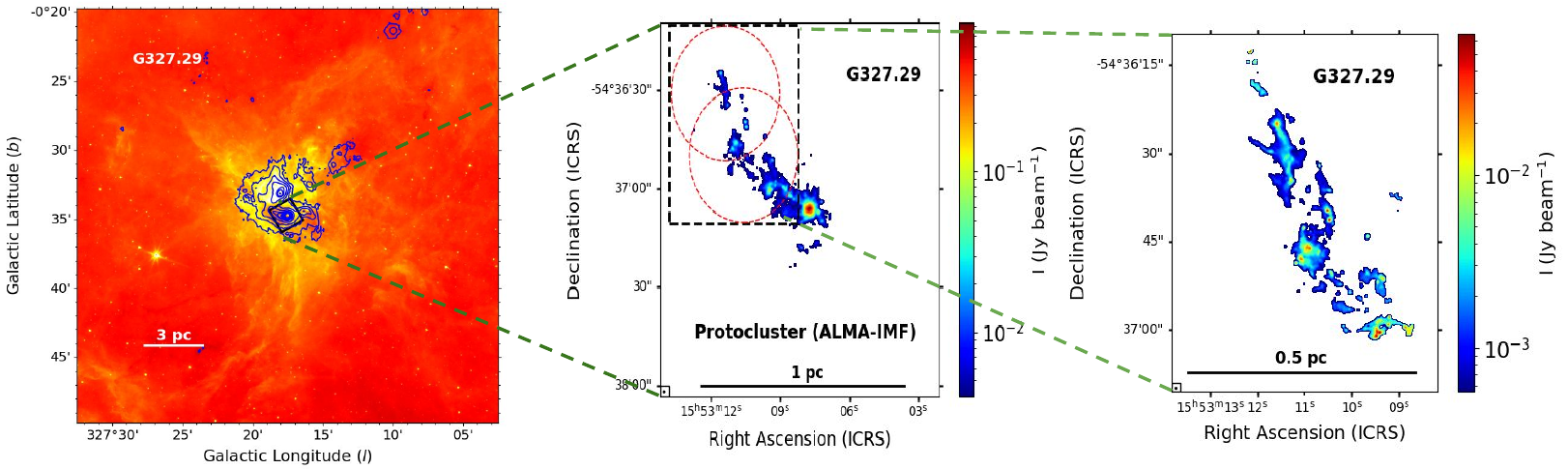}\\
	\caption{Left: Color image shows the  Spitzer (GLIMPSE) 8 $\mu$m emission of G327.29 star-forming region \citep{2003PASP..115..953B}. Contours represent the ATLASGAL 870 $\mu$m emission obtained from the work of \cite{2009A&A...504..415S}. Contour levels are 0.45, 1, 2, 4, 8, 10, 14, 16, 18, 23, 28, and  34 Jy beam$^{-1}$ respectively with a 19.2$''$ beam. Middle: 1.2 mm continuum emission towards the central clump (protocluster) obtained from the ALMA-IMF large program \citep{2022A&A...662A...8M}. Here the two maroon dashed circles represent the area where the polarization measurements are carried out using ALMA 12m configurations in this study. Right: 1.2 mm continuum emission obtained from this work after mosaicking the two pointings denoted in blue dashed circles in the bottom left figure. }
	\label{fig:fig1}
\end{figure*}

\begin{figure}
	\includegraphics[width=3.6in,height=3.6in,angle=0]{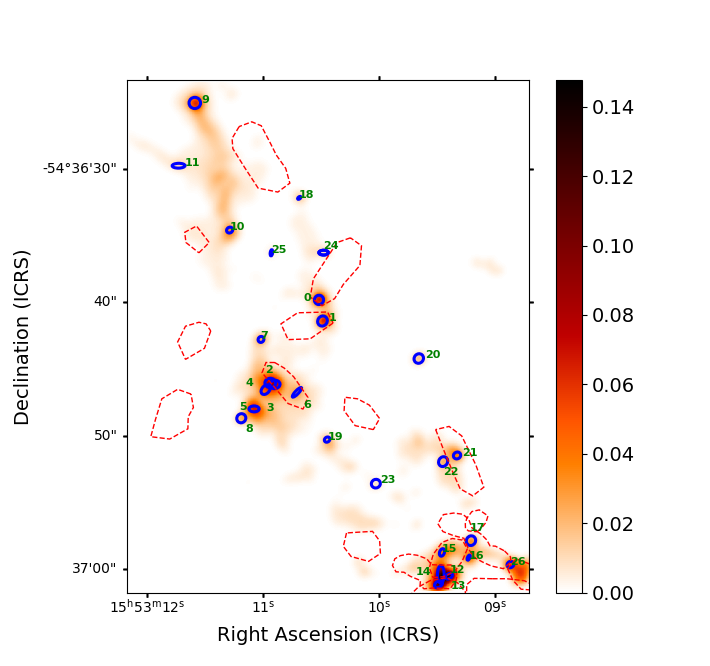}
	\caption{Spatial distribution of 1.2 mm cores in this region obtained from \cite{2026arXiv260323747K}. Here, blue circles/ellipses indicate 1.2 mm cores, and the numerical number associated with each core indicates its ID mentioned in Table \ref{tab:table1}. The red dashed polygon regions indicate the possible outflow regions identified from the SiO (5$-$4) emission \citep{2024ApJ...960...48T} and the background color image represents the integrated intensity of the 1.2 mm continuum.}
       \label{fig:fig2}
\end{figure}

\begin{figure*}
    \centering
	\includegraphics[width=3.3in,height=2.9in,angle=0]{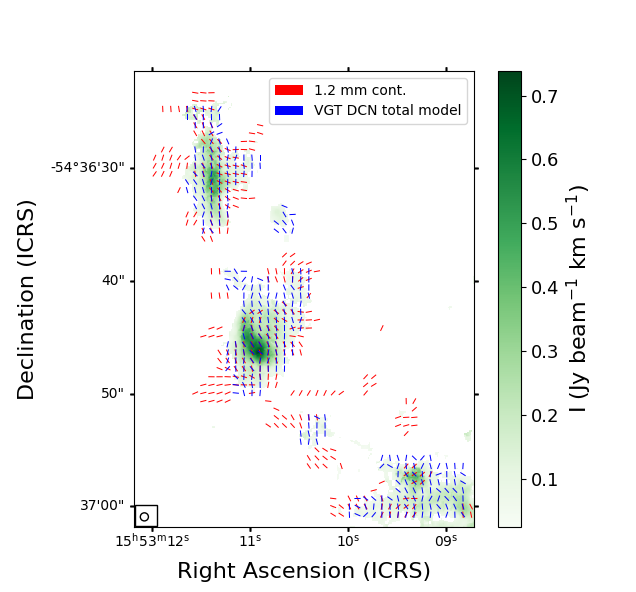} \includegraphics[width=3.3in,height=2.9in,angle=0]{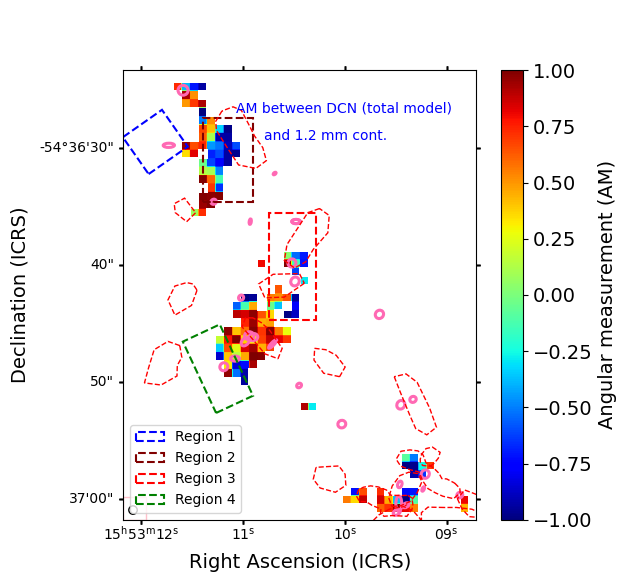}\\
    \includegraphics[width=3.3in,height=2.9in,angle=0]{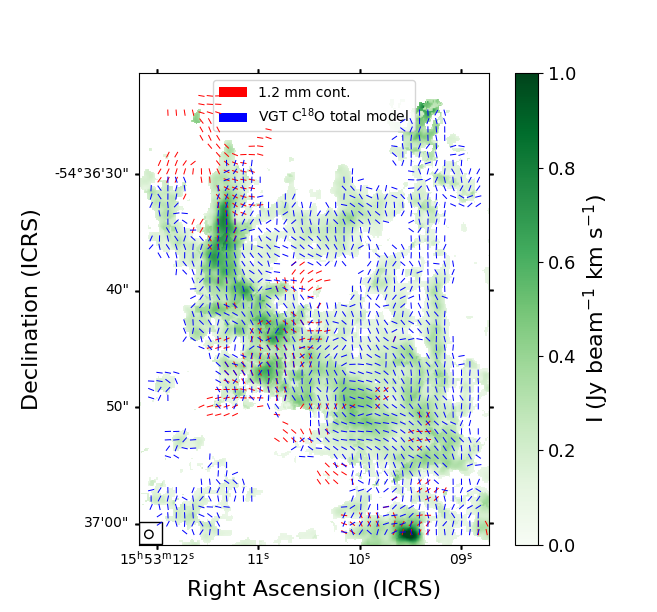} \includegraphics[width=3.3in,height=2.8in,angle=0]{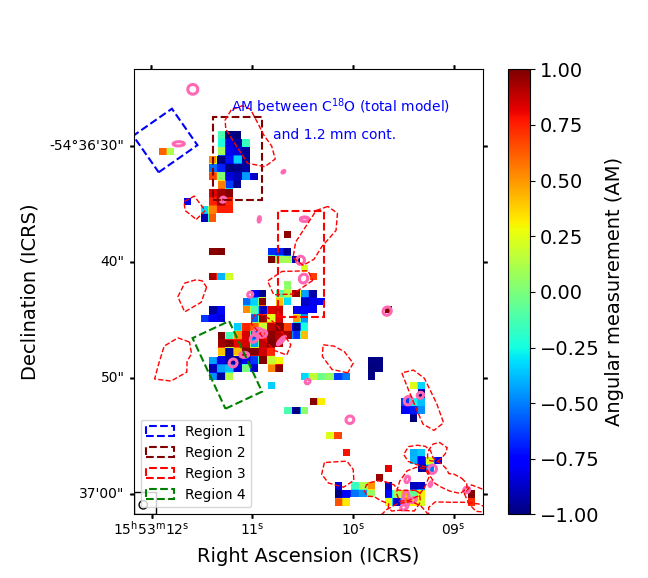}\\ 
    \includegraphics[width=3.3in,height=2.9in,angle=0]{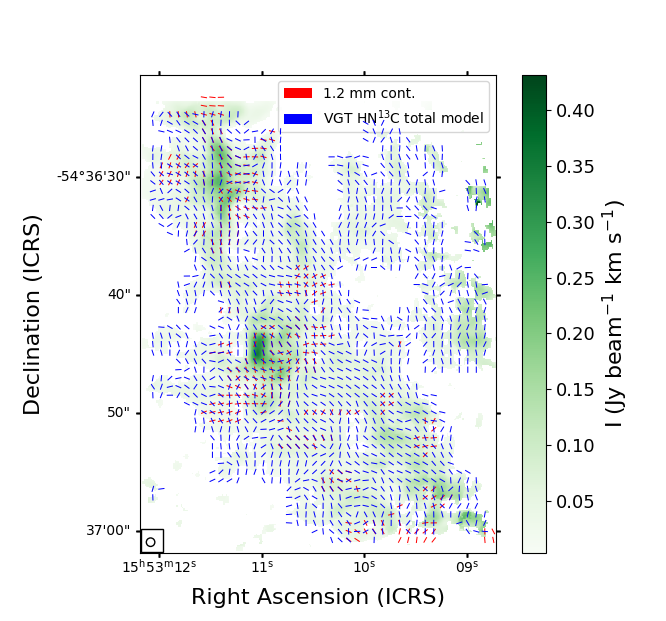} \includegraphics[width=3.3in,height=2.9in,angle=0]{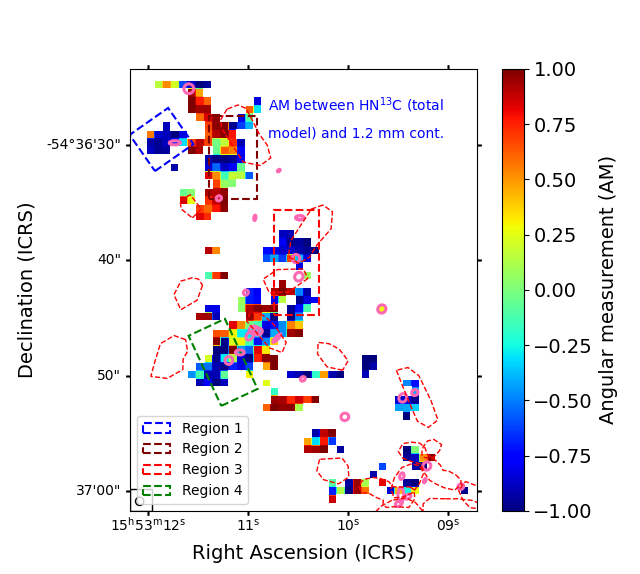}\\ 
	\caption{left column: Overplots of the $B_{\text{pos}}$ obtained from 1.2 mm dust continuum emission (red pseudo-vectors) and VGT of (i) DCN (3$-$2), (ii) C$^{18}$O (2$-$1), and (iii) HN$^{13}$C (3$-$2) line (\texttt{total model}) emissions (blue pseudo-vectors). Here, the color map in each figure indicates the integrated intensity map (moment 0) of the respective line emissions. Right column: Spatial distributions of angular measurements (AM) between dust and corresponding line emissions. Four rectangular regions: \texttt{Region 1}, \texttt{Region 2}, \texttt{Region 3} and \texttt{Region 4} showed in four different colors. A gravitational infall signature is observed towards these regions which are the arc of a filament or arc of a dense central structure. Here, the 1.2 mm continuum cores are overplotted in pink color and SiO outflow polygon regions are in red dashed color. }
    \label{fig:fig3}
\end{figure*}

\begin{figure*}
    \centering
     \includegraphics[width=3.2in,height=2.9in,angle=0]{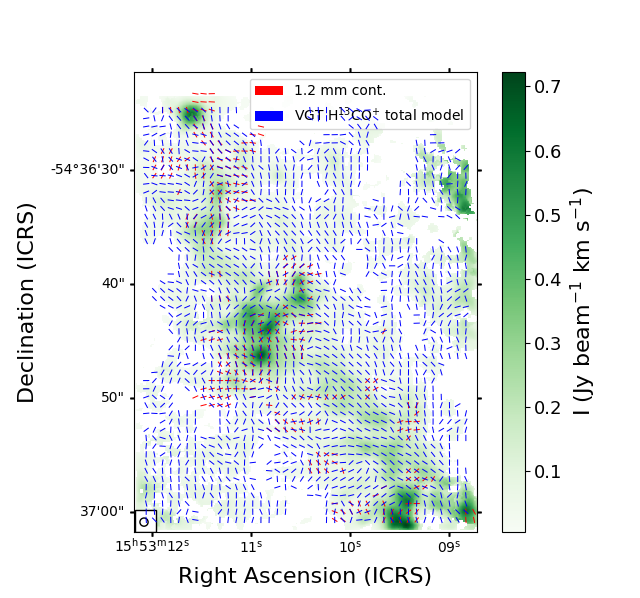} \includegraphics[width=3.2in,height=2.9in,angle=0]{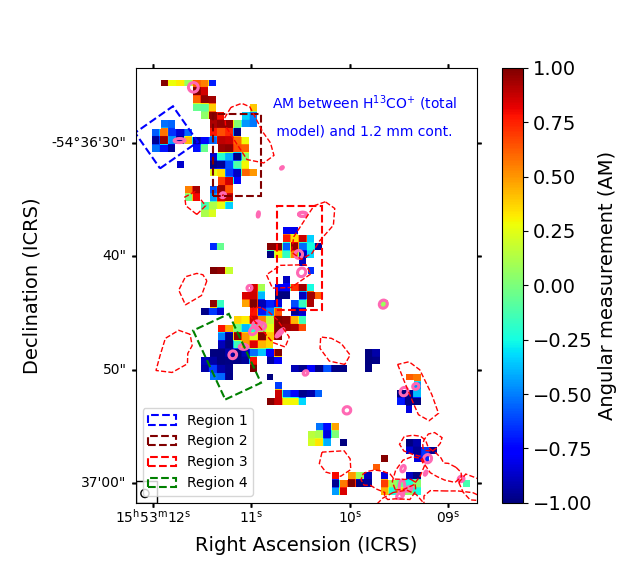}\\ 
	\caption{Continuation of Fig. \ref{fig:fig3} but for  H$^{13}$CO$^{+}$ (3$-$2) line emission. }
    \label{fig:fig4}
\end{figure*}

\begin{figure*}
    \centering
     \includegraphics[width=3.2in,height=2.9in,angle=0]{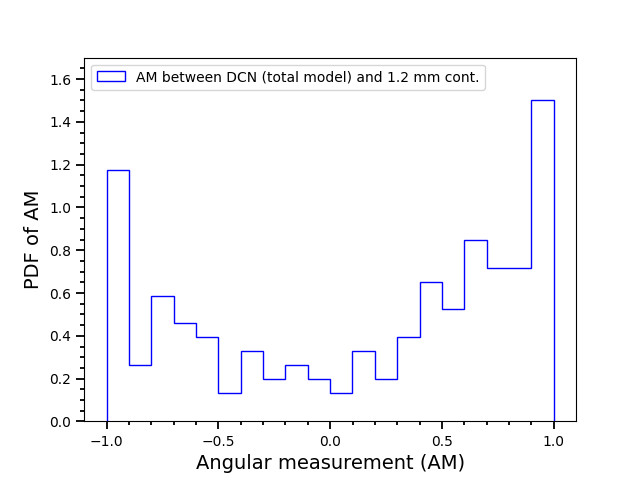} \includegraphics[width=3.2in,height=2.9in,angle=0]{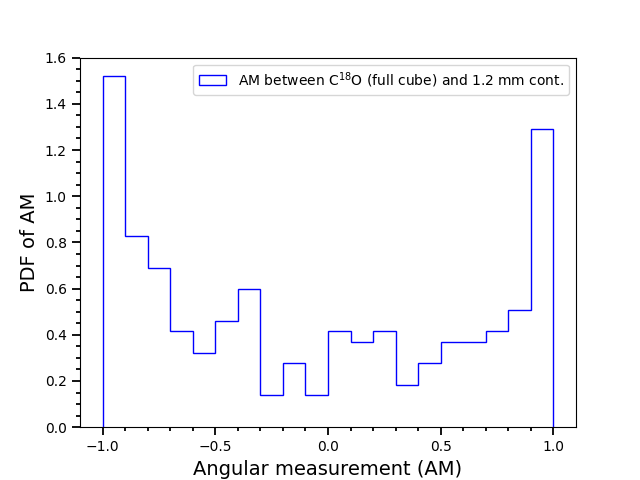}\\ 
          \includegraphics[width=3.2in,height=2.9in,angle=0]{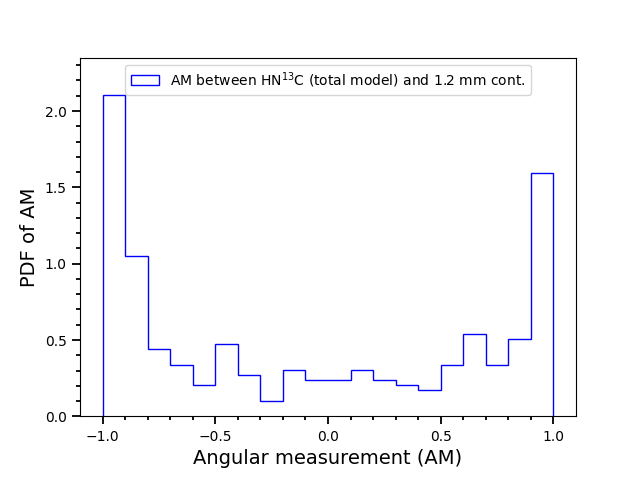} \includegraphics[width=3.2in,height=2.9in,angle=0]{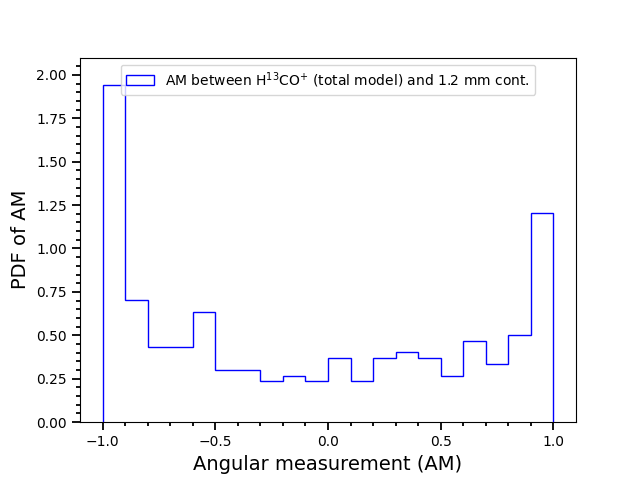}\\ 
	\caption{Histogram plots of angular measurement (AM) between dust and corresponding line emissions. In each figure, peaks are observed at approximately $\sim -1$ and $\sim +1$.}
    \label{fig:fig5}
\end{figure*}

\begin{figure*}
    \centering
     \includegraphics[width=3.2in,height=2.9in,angle=0]{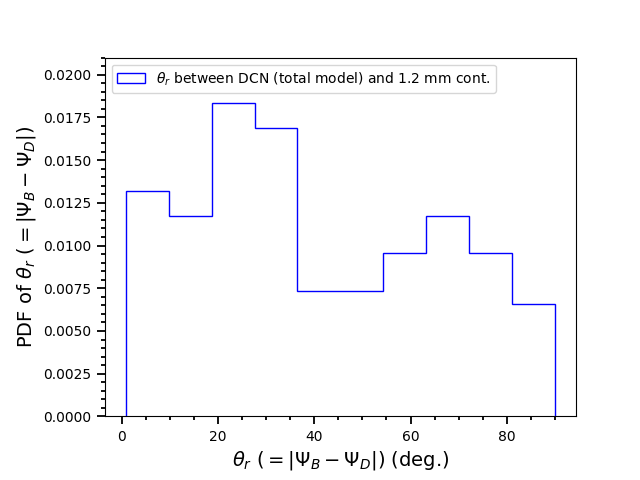} \includegraphics[width=3.2in,height=2.9in,angle=0]{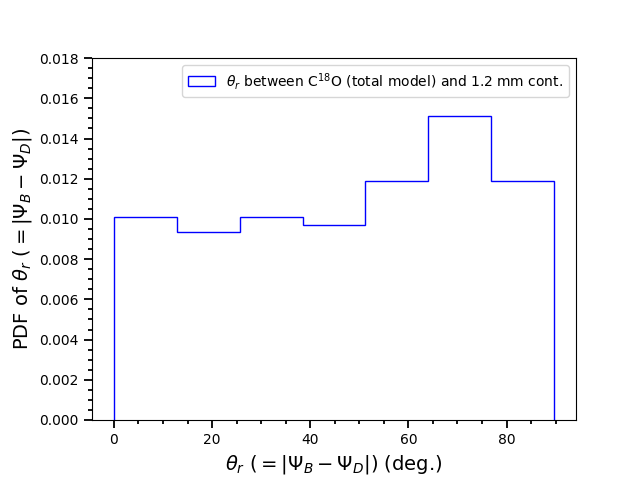}\\ 
          \includegraphics[width=3.2in,height=2.9in,angle=0]{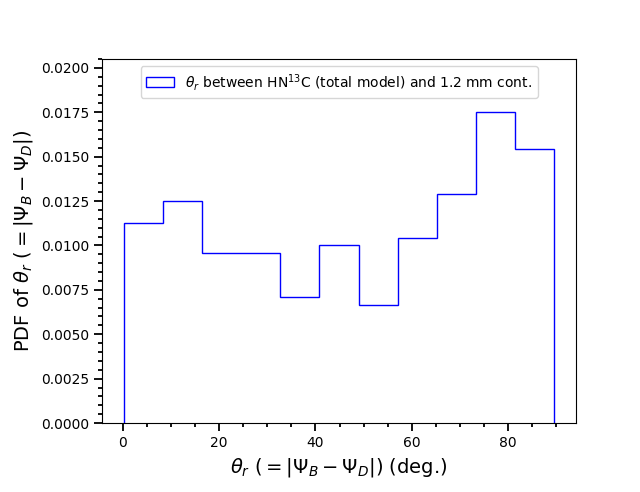} \includegraphics[width=3.2in,height=2.9in,angle=0]{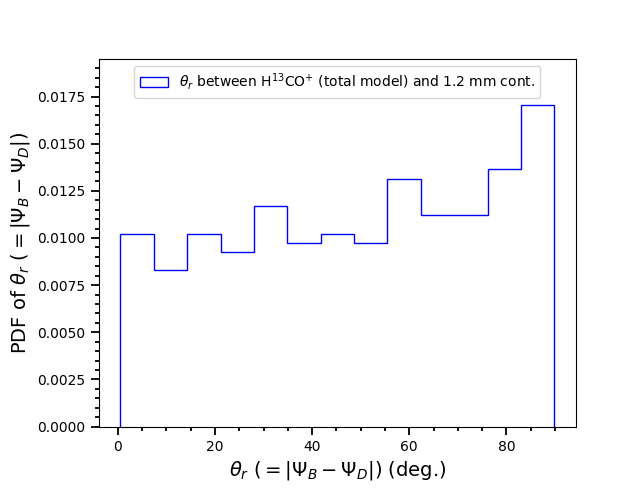}\\ 
	\caption{Histograms of the relative orientation angle, $\theta_{\rm r}$, between the VGT derived and the dust polarization-derived magnetic field orientations for the four spectral lines. A peak near $0^{\circ}$ indicates a turbulence-dominated regime, whereas a peak near $90^{\circ}$ indicates a gravity-dominated regime.}
    \label{fig:fig5_0}
\end{figure*}

\begin{figure*}[!ht]
~~~~~~~~~~~~~~~~~~~~~~~~~~~~~~~~~~~~~~~~~~~~~~\includegraphics[width=3.4in,height=2.8in,angle=0]{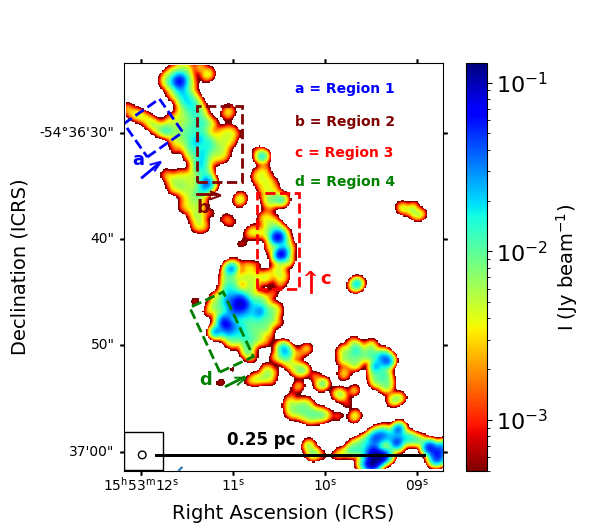}\\
\includegraphics[width=3.4in,height=2.8in,angle=0]{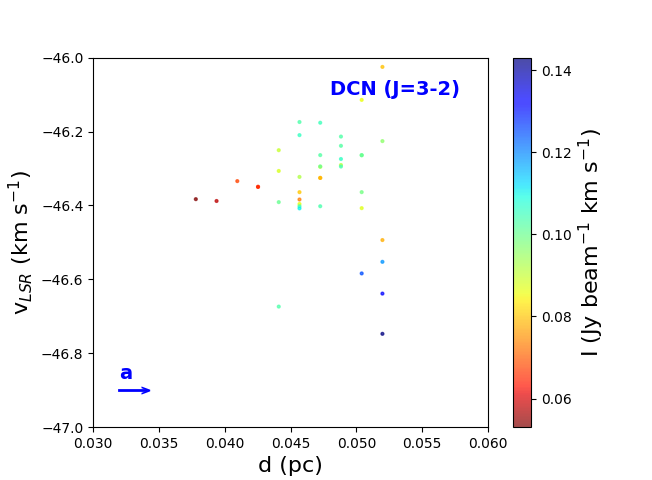}\includegraphics[width=3.4in,height=2.8in,angle=0]{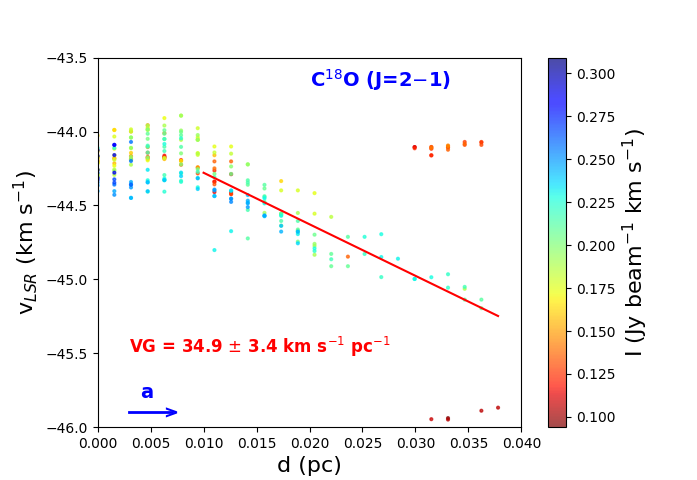}\\
\includegraphics[width=3.4in,height=2.8in,angle=0]{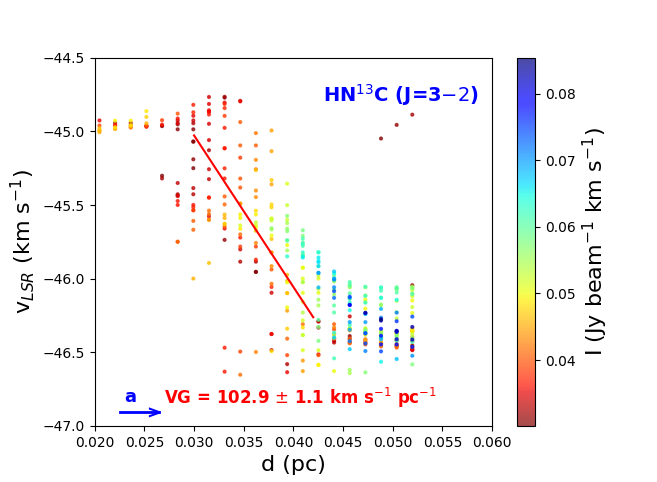}
\includegraphics[width=3.4in,height=2.8in,angle=0]{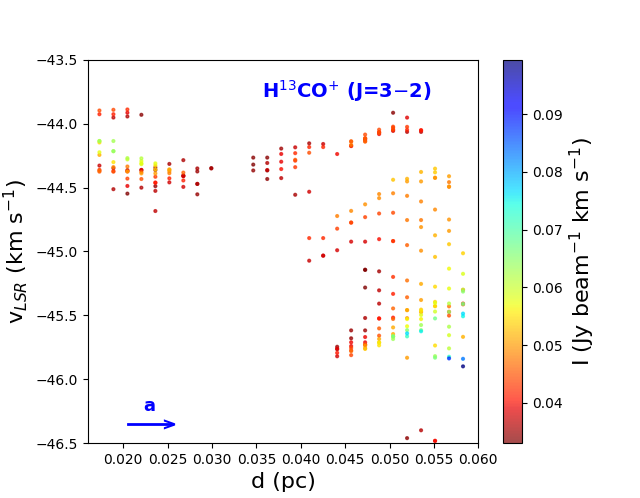}\\
	\caption{Upper: Integrated intensity map of the 1.2 mm dust continuum emission at $\sim$0.80$''$ resolution (same as DCN). Four rectangular regions \textbf{a} (\texttt{Region 1}), \textbf{b} (\texttt{Region 2}), \textbf{c} (\texttt{Region 3})  and \textbf{d} (\texttt{Region 4}) shown in blue, maroon, red, and green, exhibit smooth velocity gradients, indicative of gravitational infall. Middle and Lower: Velocity gradient of four spectral lines observed towards the blue arrow direction denoted by symbol \textbf{a} in the upper panel. The measured velocity gradients for the C$^{18}$O and HN$^{13}$C spectral lines, where prominent velocity gradients are observed, are $34.9 \pm 3.4~\mathrm{km~s^{-1}~pc^{-1}}$ and $102.9 \pm 1.1~\mathrm{km~s^{-1}~pc^{-1}}$, respectively. The fitted lines are indicated by the red solid lines.}
    \label{fig:fig6}
\end{figure*}

\begin{figure*}[!ht]
\includegraphics[width=3.4in,height=2.8in,angle=0]{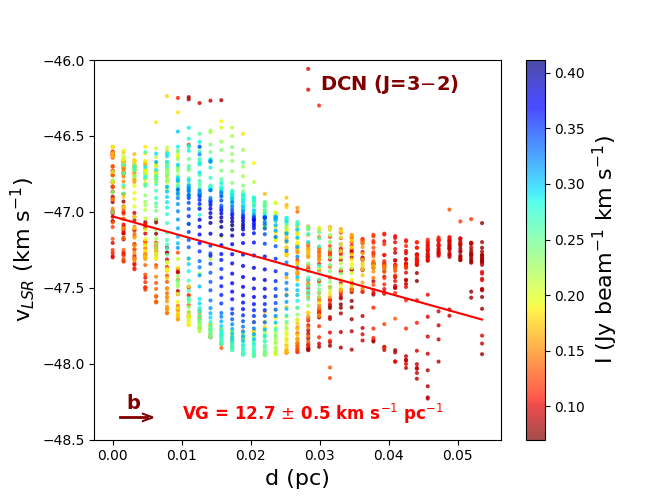}\includegraphics[width=3.4in,height=2.8in,angle=0]{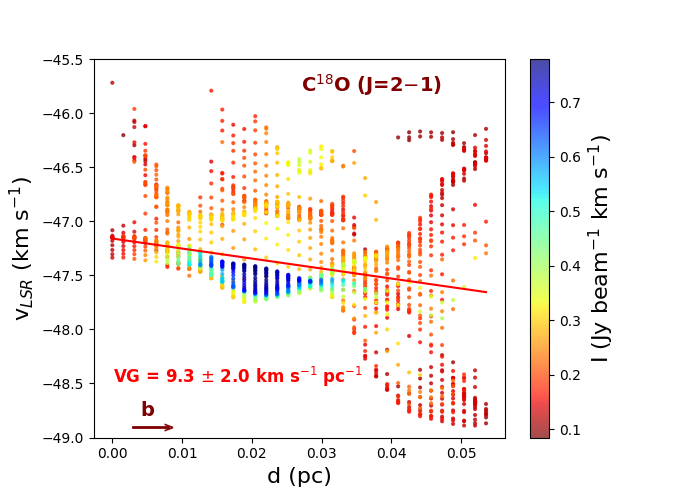}\\
\includegraphics[width=3.4in,height=2.8in,angle=0]{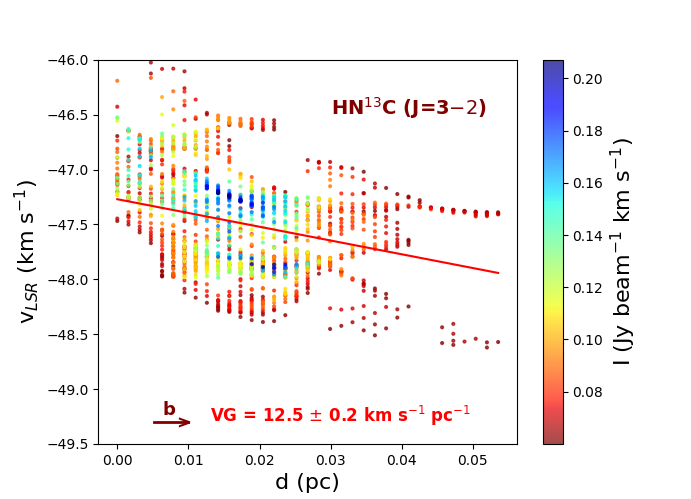}
\includegraphics[width=3.4in,height=2.8in,angle=0]{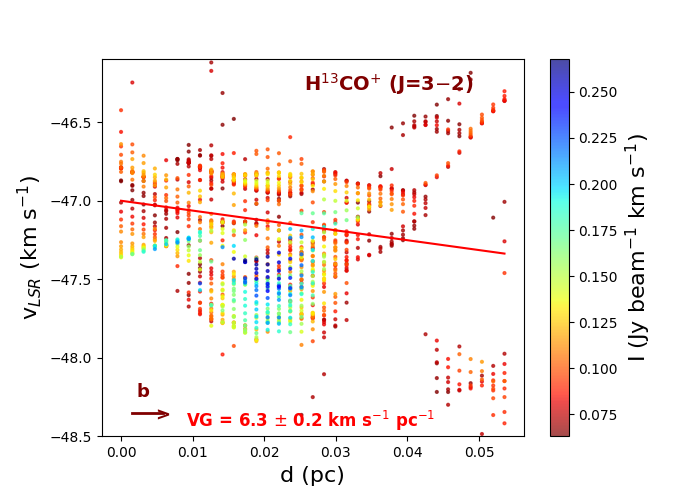}\\
	\caption{Velocity gradient of four spectral lines observed towards the maroon arrow direction denoted by symbol \textbf{b} in Fig. \ref{fig:fig6}. The measured velocity gradients are $12.7 \pm 0.5$, $9.3 \pm 2.0$, $12.5 \pm 0.2$, and $6.3 \pm 0.2~\mathrm{km~s^{-1}~pc^{-1}}$, respectively, and the corresponding best-fit lines are shown as red solid lines.}
    \label{fig:fig7}
\end{figure*}

\begin{figure*}[!ht]
\includegraphics[width=3.4in,height=2.8in,angle=0]{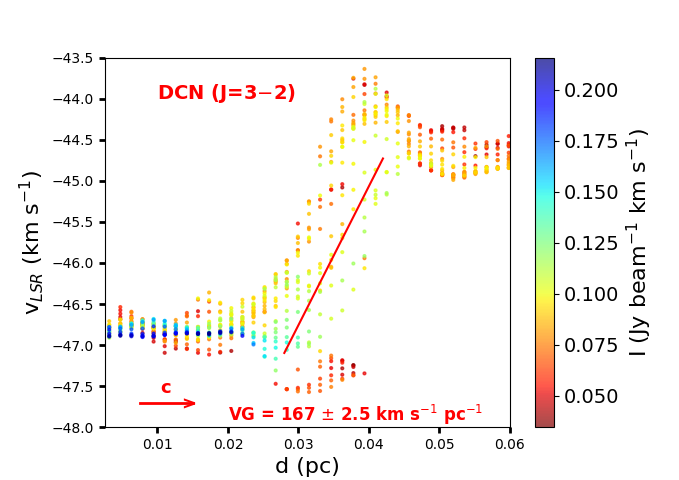}\includegraphics[width=3.4in,height=2.8in,angle=0]{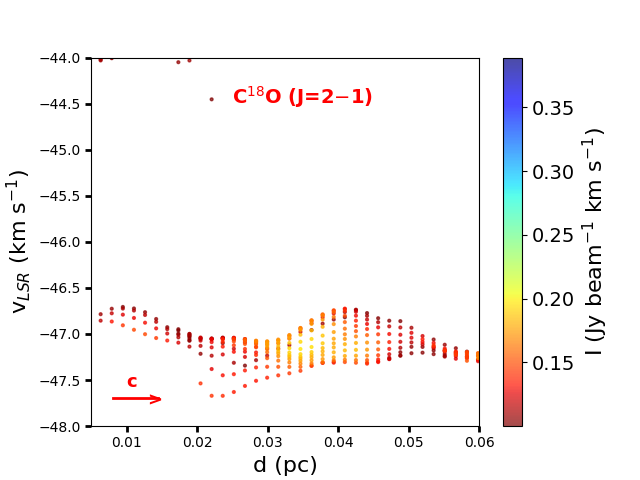}\\
\includegraphics[width=3.4in,height=2.8in,angle=0]{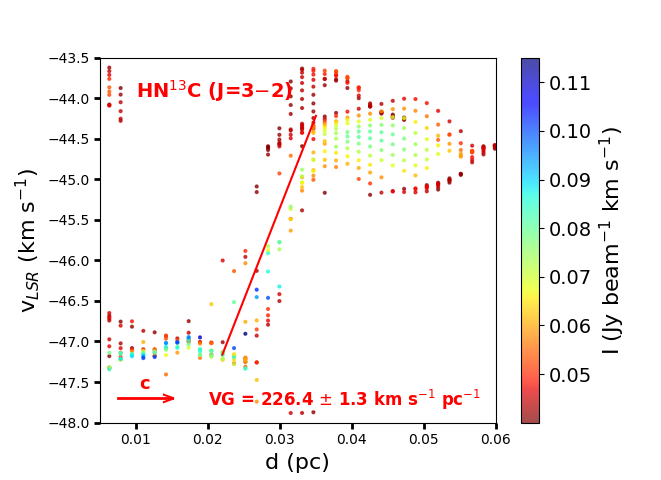}
\includegraphics[width=3.4in,height=2.8in,angle=0]{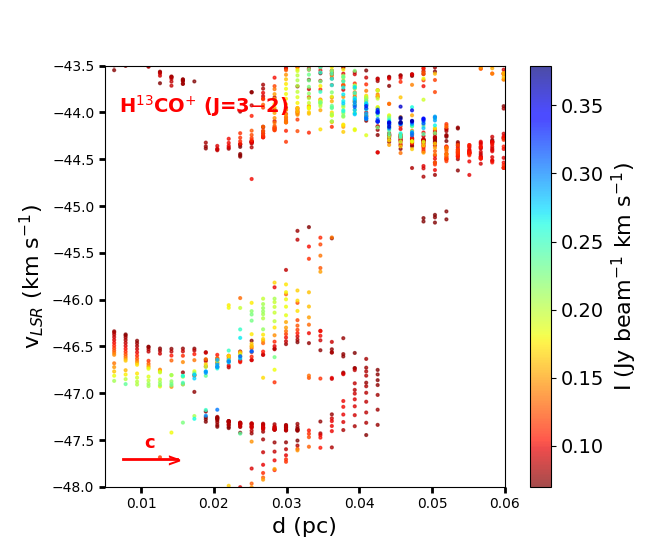}\\
	\caption{Velocity gradient of four spectral lines observed towards the red arrow direction denoted by symbol \textbf{c} in Fig. \ref{fig:fig6}. The measured velocity gradients for the DCN and HN$^{13}$C spectral lines, in regions exhibiting prominent velocity gradients, are $167.0 \pm 2.5~\mathrm{km~s^{-1}~pc^{-1}}$ and $226.4 \pm 1.3~\mathrm{km~s^{-1}~pc^{-1}}$, respectively. The corresponding best-fit lines are shown as red solid lines.}
    \label{fig:fig8}
\end{figure*}

\begin{figure*}[!ht]
\includegraphics[width=3.4in,height=2.8in,angle=0]{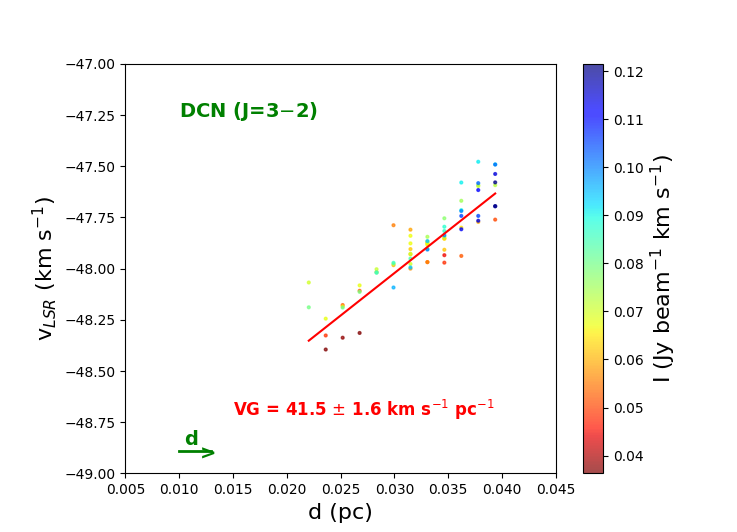}\includegraphics[width=3.4in,height=2.8in,angle=0]{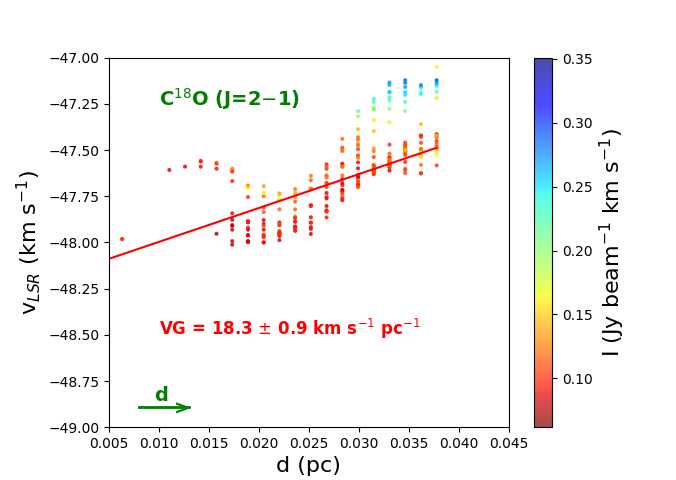}\\
\includegraphics[width=3.4in,height=2.8in,angle=0]{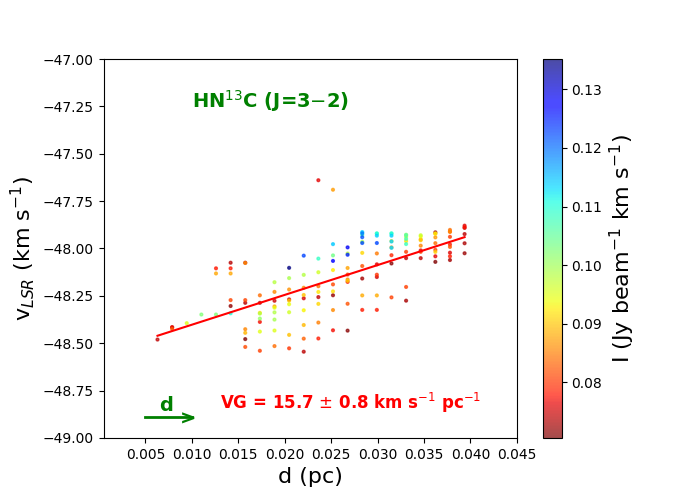}
\includegraphics[width=3.4in,height=2.8in,angle=0]{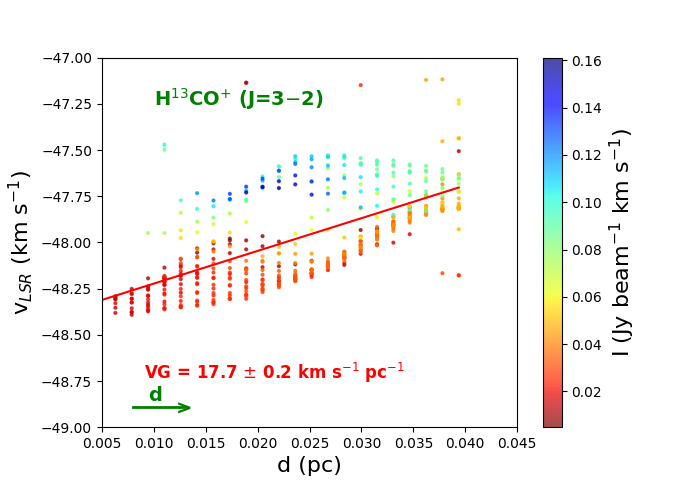}\\
	\caption{Same as Fig. \ref{fig:fig7} but towards the green arrow direction denoted by symbol \textbf{d} in Fig. \ref{fig:fig6}. The derived velocity gradients are $41.5 \pm 1.6$, $18.3 \pm 0.9$, $15.7 \pm 0.8$, and $17.7 \pm 0.2~\mathrm{km~s^{-1}~pc^{-1}}$, respectively. The corresponding best-fit relations are indicated by the red solid lines.}
    \label{fig:fig9}
\end{figure*}

\begin{figure*}[!ht]
\includegraphics[width=2.4in,height=2.0in,angle=0]{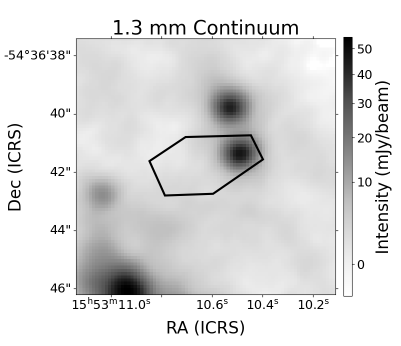}\includegraphics[width=2.4in,height=2.0in,angle=0]{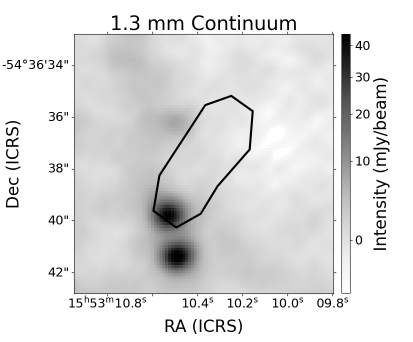}
\includegraphics[width=2.4in,height=2.0in,angle=0]{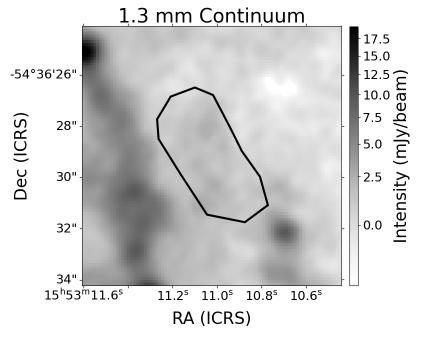}
\includegraphics[width=2.4in,height=2.0in,angle=0]{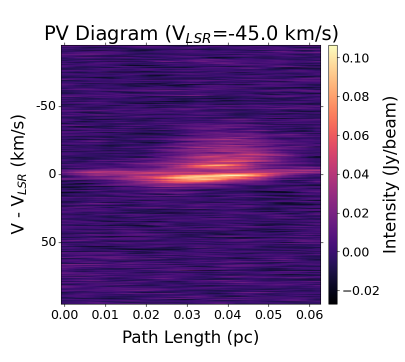}\includegraphics[width=2.4in,height=2.0in,angle=0]{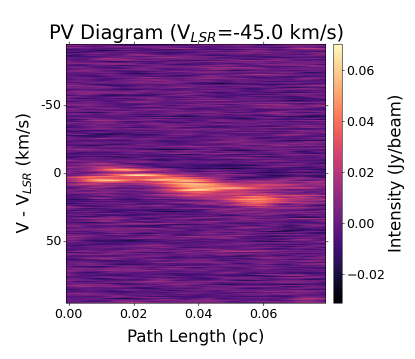}\includegraphics[width=2.4in,height=2.0in,angle=0]{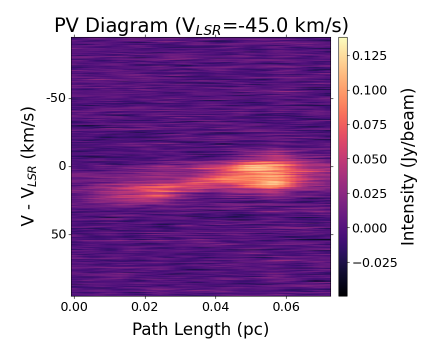}\\
	\caption{Upper panel: The gray-scale images of the 1.3 mm continuum emission toward the three outflow regions, which partially overlap with the areas where the large-scale velocity gradients were derived from the four spectral lines. The black polygons indicate the possible candidates for the SiO (5$-$4) outflows. Lower panel: The corresponding position-velocity (PV) diagrams of the SiO (5$-$4) emission using the inferno color scale. The average velocity gradients observed from the PV diagram are $\sim$ 500$-$600 km s$^{-1}$pc$^{-1}$, which are significantly higher than the gradients obtained from the four spectral lines analyzed in this work. Figures are adopted from the work of \cite{2024ApJ...960...48T}.}
    \label{fig:fig10}
\end{figure*}

\begin{figure*}[!ht]
\centering
\includegraphics[width=7.4in,height=2.4in,angle=0]{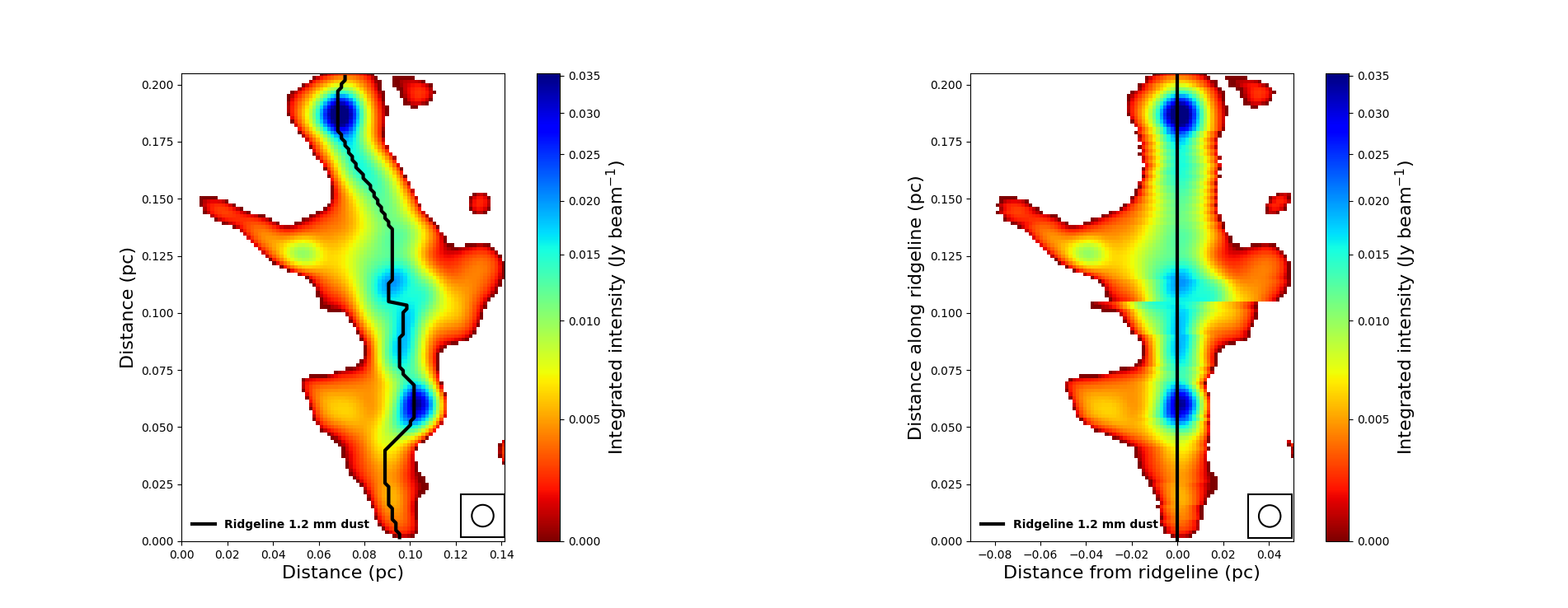}\\
\includegraphics[width=3.8in,height=3.2in,angle=0]{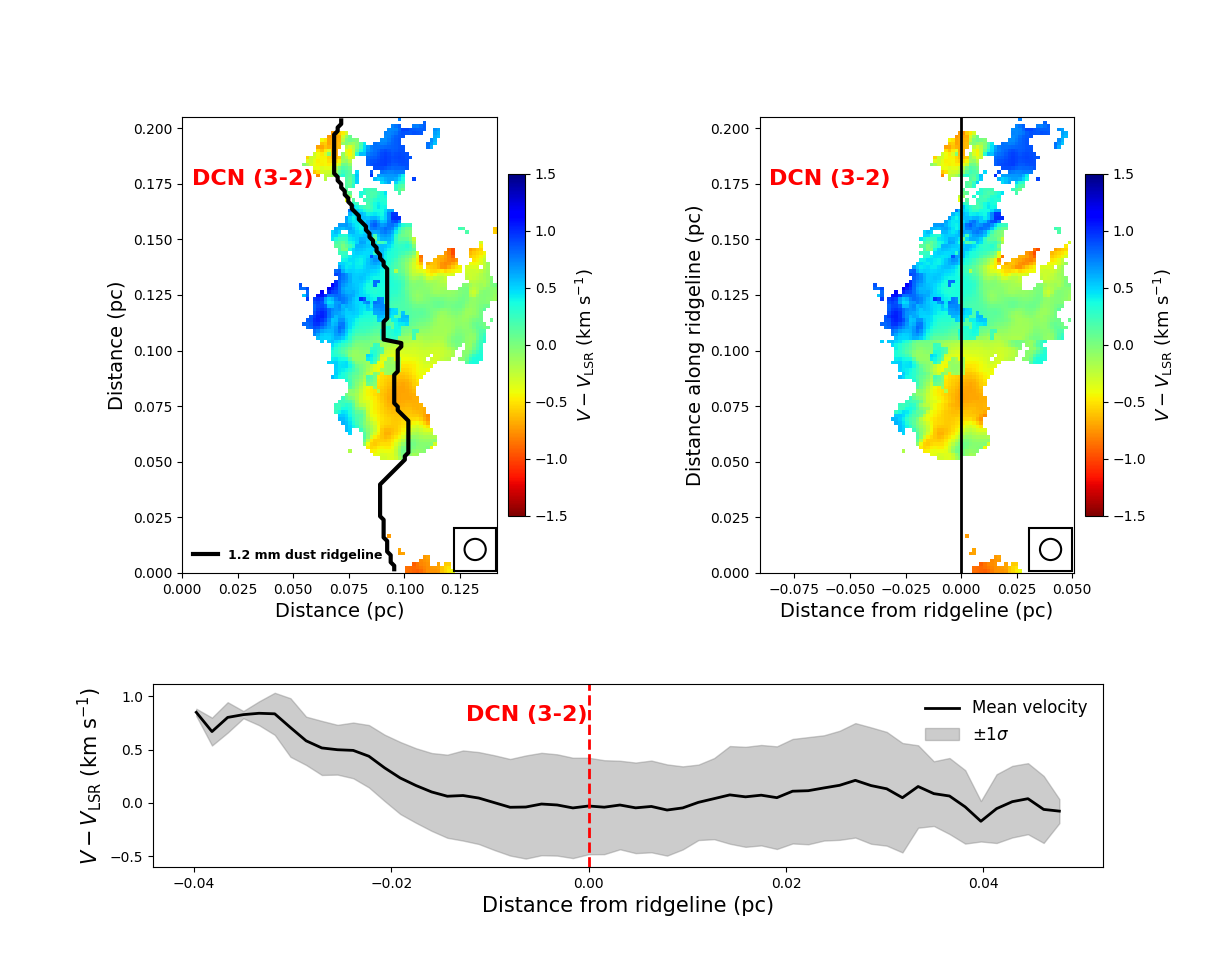}\includegraphics[width=3.8in,height=3.2in,angle=0]{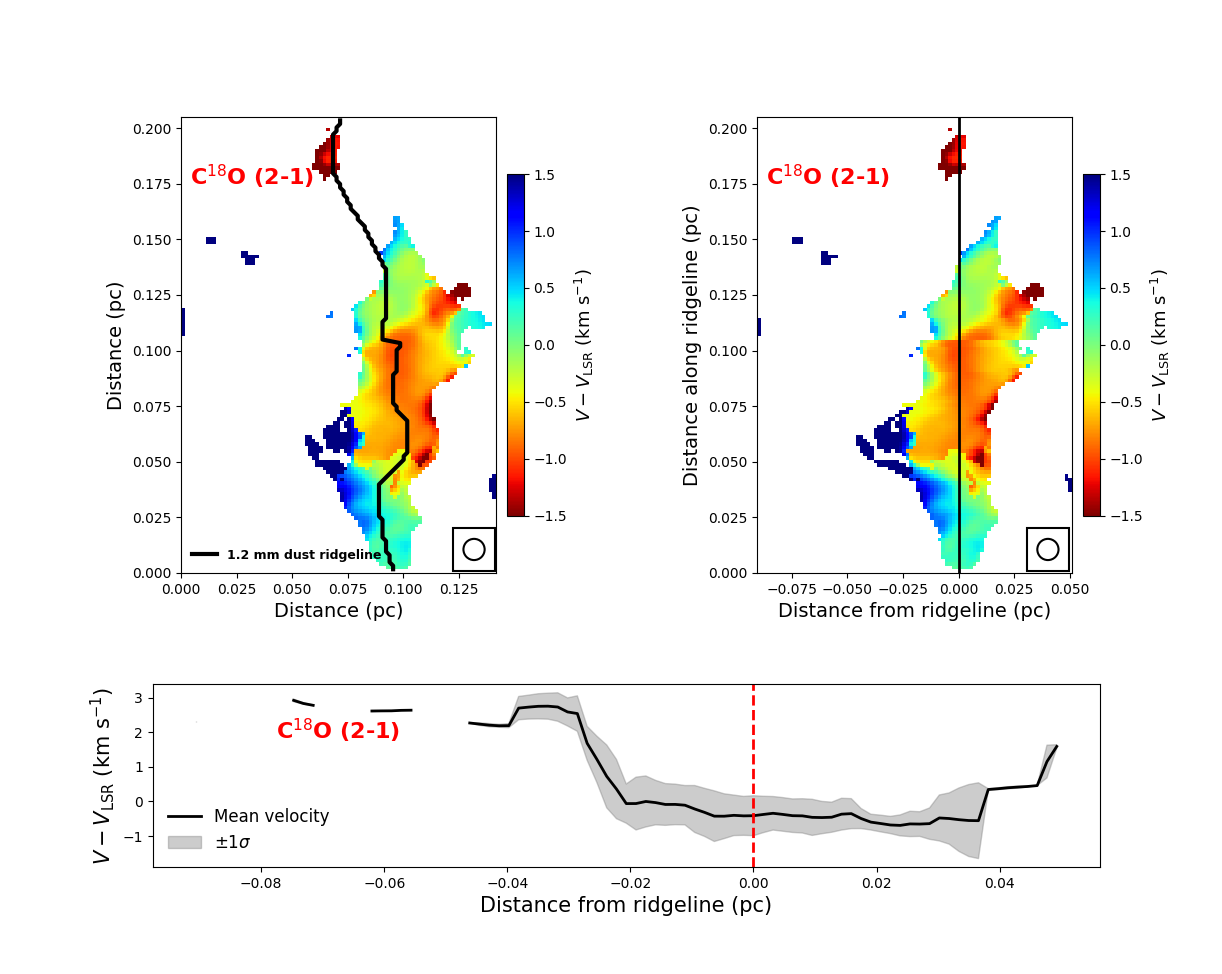}\\
\includegraphics[width=3.8in,height=3.2in,angle=0]{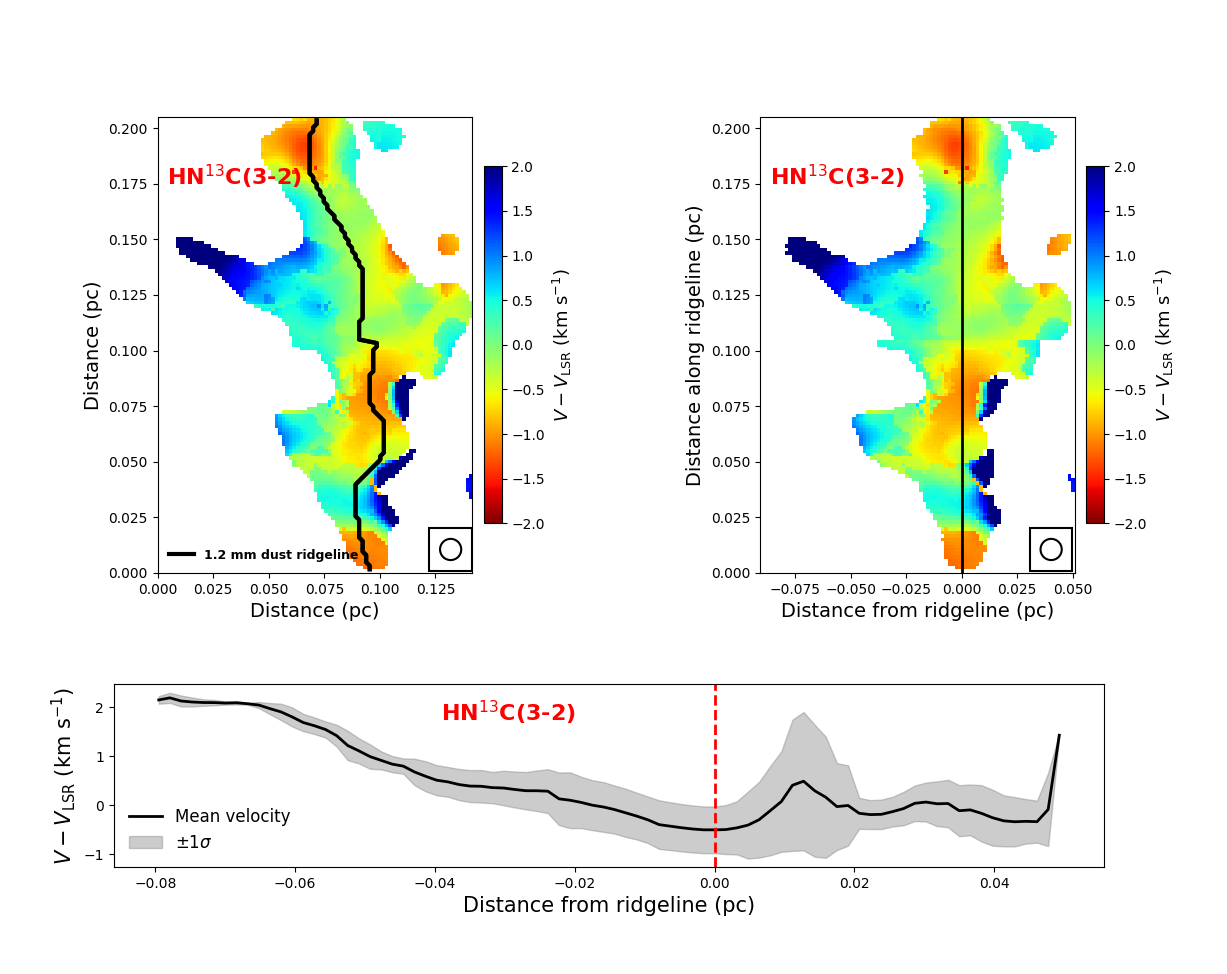}\includegraphics[width=3.8in,height=3.2in,angle=0]{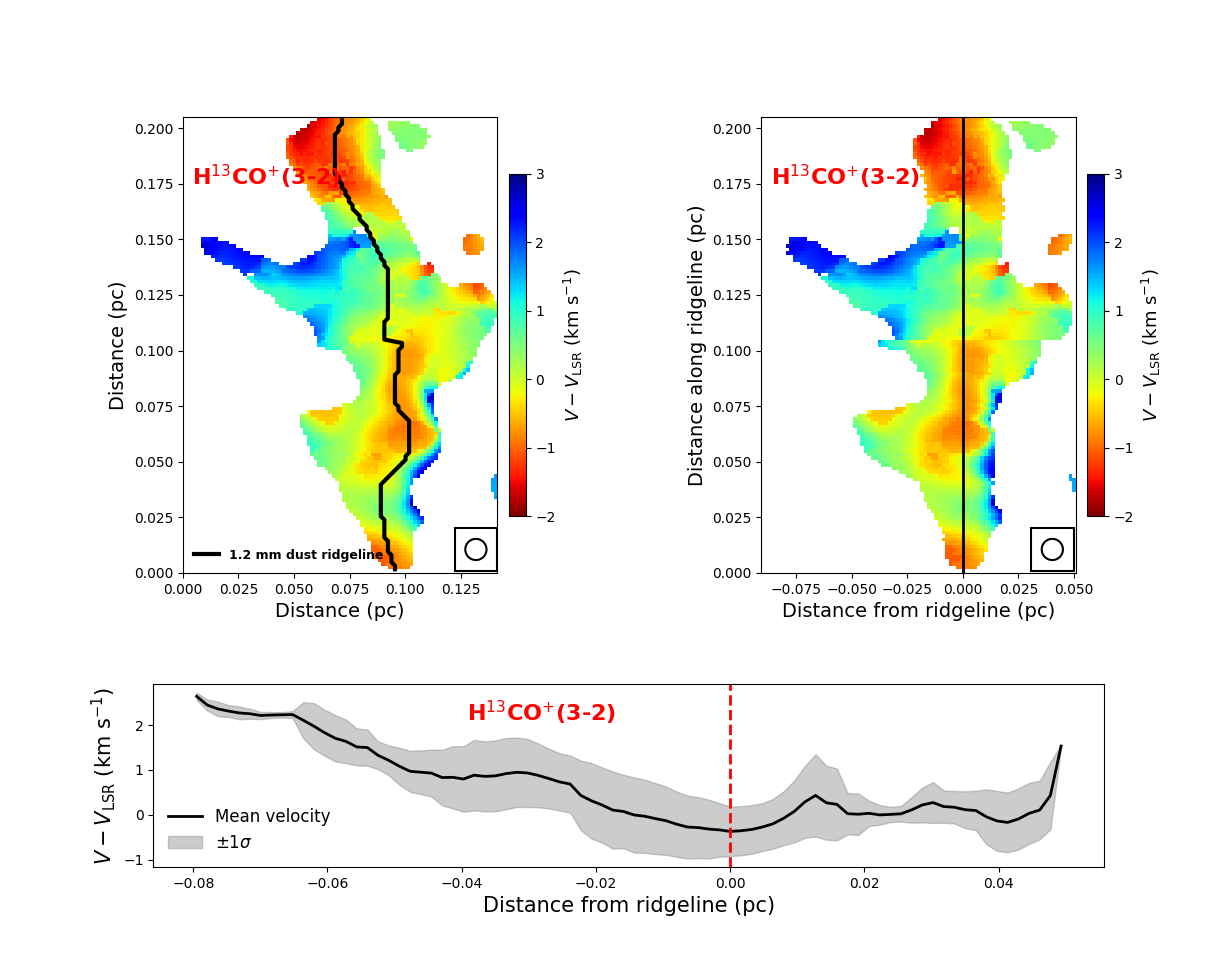}\\
	\caption{Upper panel: 1.2 mm dust continuum emission (before and after straightening along the ridgeline) of the filament located in the upper part of the observed region. Here, the black curve or straight line denotes the filament ridgeline. Middle and bottom panels: Intensity-weighted mean velocity (moment 1) maps of the four molecular transitions before and after straightening along the dust continuum ridgeline. In each panel the bottom figures show the corresponding radial velocity profiles obtained by averaging the moment 1 values perpendicular to the filament ridgeline. The profiles are shown to examine whether a systematic velocity gradient is present across the filament, which would be indicative of organized transverse gas motions caused by rotation.}
    \label{fig:fig10_0}
\end{figure*}

	 
    

Recently, a new method called the velocity gradient technique (VGT) has been developed by \cite{LY18} 
for determining the morphology of the $B_{\text{POS}}$. This technique employs the theory of MHD turbulence (\cite{1995ApJ...438..763G, 1999ApJ...517..700L, CV00}, see more in a book by \cite{BL19}), in particular the anisotropy in the velocity fluctuations in the magnetic field's local reference frame, as well as the theory of velocity caustics in thin-channel maps of the spectral line data cubes \citep{2000ApJ...537..720L}. Using this technique, the morphology of the $B$$_\text{pos}$ for each individual velocity component can be determined \citep{HY23}. Likewise, in the presence of strong gravity, rotation and strong outflow, the VGT rotates by 90$^\circ$ in respect to the magnetic field \citep{2020ApJ...897..123H, 2024MNRAS.530.1066L}. Due to the strong effect of gravity, moving gas particles that are frozen into the magnetic field can distort the field lines, causing the magnetic field to align with the gas motion. As a result, the velocity gradient direction traces the magnetic field direction. A similar situation may also arise in the presence of rotation or strong outflows.

Thus, by comparing the VGT results with dust polarization, it is also possible to characterize large-scale gravitational infall which is often observed in the massive star forming regions \citep{2024A&A...689A..74A,2025A&A...696A.202S,2025ApJ...979..233M,2025arXiv251003447S}. Details of this method are discussed in Appendix \ref{section:A0}.\\

VGT has been studied and tested numerically and observationally for different phases of the interstellar medium using different telescopes \citep{2019NatAs...3..776H,2022ApJ...934...45Z,2023MNRAS.523.1853S,2024MNRAS.528.3897S,2024MNRAS.530.1066L,2024MNRAS.528.3897S}. And it provides accurate correspondence with dust polarization. However, no prior studies have been conducted in the massive protocluster. In this work, we present the first analysis of this kind, utilizing high-resolution continuum and spectral line data towards the G327.29 massive star-forming region obtained with the ALMA 12-meter configuration. This target belongs to the large-scale ALMA-IMF program, a survey that investigates 15 massive protoclusters in detail \citep{2022A&A...662A...8M}.


The distance of the source $d$ is 2.5 $\pm$ 0.5 kpc \citep{2017A&A...601A..60C, 2022A&A...662A...8M}. The mass of the central clump (imaged area $\sim$ 1.3 $\times$ 1.3 pc$^{2}$) in G327.29 region denoted as protocluster is $\sim$ 5000 M\textsubscript{\(\odot\)}  \citep{2022A&A...662A...8M,2025A&A...694A..24M}. The systematic velocity ($V_{\text{LSR}}$) of the source G327.29 is $-$ 45.5 km s$^{-1}$. However, the region where we are comparing the VGT and the dust polarization is slightly blue-shifted and has a $V_{\text{LSR}}$ of $\sim$ $-$ 46.6 km s$^{-1}$ \citep{2023A&A...678A.194C,2026arXiv260323747K}. Based on the H41$\alpha$ emission and flux ratio of the 1.3 to 3.0 mm continuum emission, this protocluster is considered young \citep{2022A&A...662A...8M,2024ApJS..274...15G}. In Fig. \ref{fig:fig1}, we show the color image of the Spitzer (GLIMPSE) 8 $\mu$m emission  of the G327.29 star-forming region \citep{2003PASP..115..953B} and the contours represent the ATLASGAL 870 $\mu$m emission obtained from the work of \cite{2009A&A...504..415S}. In Fig. \ref{fig:fig1}, we also show the 1.3 mm continuum emission of the central 870 $\mu$m clump (protocluster) obtained from the ALMA-IMF large program and the 1.2 mm continuum emission area where we carried out the dust polarization observations using the ALMA 12m configuration. We selected this particular protocluster because high-quality polarization data are already available for the region. In addition, this protocluster is classified as young, implying that the feedback from newly formed H{\sc ii} regions is expected to be relatively weak. Furthermore, the area analyzed in this study is slightly offset from the central region of the protocluster, where the gas dynamics are likely to be more complex.\\

This paper is organized as follows. In \S~\ref{section_3}, we present data of 1.2 mm continuum and various spectral lines. We discuss spectral line decomposition using \texttt{Gausspy+} module in \S~\ref{section_4}. In \S~\ref{section_5}, we analyze the VGT for various spectral lines. In \S~\ref{section_6}, we discuss our main results. And finally we draw our main conclusions in \S~\ref{section_7}.



\section{Observational data}\label{section_3}

The spectral line and continuum data sets analyzed in this work were acquired from the ongoing study \citep{2026arXiv260323747K} and ALMA-IMF Large Program \citep{2022A&A...662A...8M,2022A&A...662A...9G,2023A&A...678A.194C}. A detailed description of each of these is provided in the following.\\

\subsection{DCN (J = 3$-$2) emission line}
We used the DCN (3–2) line cube from \citet{2023A&A...678A.194C}, which is part of the large ALMA-IMF program (project ID: 2017.1.01355.L; \citealp{2022A&A...662A...8M}). The rest frequency ($\nu_{0}$) of the DCN (3–2) transition is 217.23854 GHz \citep[and references therein]{2001A&A...370L..49M, 2023A&A...678A.194C}, and its critical density ($n_{\text{crit}}$) is approximately 1.5 × 10$^{7}$ cm$^{-3}$ at 20 K \citep{2019A&A...629A..10B}. This spectral line was observed using the 12 m configuration of the ALMA telescope. The corresponding data cube has a spectral resolution of $\sim$0.34 km s$^{-1}$ and a spatial resolution of $\sim$0.80$''$ (2500 AU), which is the lowest among the continuum and other line cubes analyzed in this study. The observed rms noise level is $\sim$ 8 mJy beam$^{-1}$. For the Velocity Gradient Technique (VGT) analysis, it is not necessary for all line cubes to have the same spectral resolution; rather, they must have identical spatial resolution. Therefore, we applied spatial smoothing and regridding to all other cubes to match the DCN (3–2) cube to make them all identical. This is because thin velocity-channel maps are expected to be dominated by velocity caustics. Therefore, the VGT can effectively trace the magnetic-field orientation when the channel width is smaller than the turbulent velocity dispersion. However, since we compare the VGT-derived magnetic-field orientations with those inferred from dust polarization, it is necessary to first smooth all datasets to a common angular resolution and regrid them onto the same spatial grid. In this way, the resulting data cubes are identical in their spatial properties, differing only in their spectral resolution.\\

\subsection{C$^{18}$O (J = 2$-$1) emission line}
We obtained the C$^{18}$O (2$-$1) data from the ALMA-IMF Large Program \citep{2022A&A...662A...8M}. The $\nu_{0}$ and $n_{\text{crit}}$ (at 20 K) of the line are 219.56035800 GHz and 9.93 $\times$ 10$^{3}$ cm$^{-3}$ respectively \citep{2016A&A...585A..44M,2025A&A...702A.133K}. The dataset corresponds exclusively to the ALMA 12 m array configuration. We first subtracted the continuum using CASA task \texttt{UVSUB}. Consequently, the final image cube was produced using the CASA task \texttt{TCLEAN}. We adopted a primary beam limited mask by setting the parameters \texttt{usemask} and \texttt{pbmask}, and employed the multiscale deconvolver (\texttt{deconvolver = 'multiscale'}) to recover extended structures. Four multiscale components were used in geometric progression, with the first scale set to 0, the second approximately equal to the beam size (in pixel units), and the subsequent scales following the geometric sequence (4, 16, 64). Increasing the number of scales beyond four did not result in any noticeable flux recovery. A cleaning threshold of 3$\sigma_{\text{rms}}$ and a \texttt{cyclefactor} of 4 were adopted to minimize cleaning artifacts. The pixel size was chosen one-fourth of the minor axis of the synthesized beam, which yielded the most effective cleaning performance. The resulting spectral and spatial resolutions of the cube are $\sim$0.33 km s$^{-1}$ and $\sim$0.80$^{''}$, respectively. The achieved rms noise level of this spectral line is $\sim$ 16 mJy beam$^{-1}$.\\ 


\subsection{HN$^{13}$C (J = 3$-$2) and H$^{13}$CO$^{+}$ (J = 3$-$2)  emission lines}
We adopted the HN$^{13}$C (J = 3$-$2)  and H$^{13}$CO$^{+}$ (J = 3$-$2) spectral line cubes from \cite{2026arXiv260323747K}. The $\nu_{0}$ and $n_{\text{crit}}$ of these transitions are 261.26351 GHz, $\sim$10$^{7}$ cm$^{-3}$ and 260.255617\,GHz and $\sim$ 9.5 $\times$ 10$^{5}$\,cm$^{-3}$, respectively \citep{2000ApJ...538..260G,2015ApJ...803...70S, 2022A&A...658A.128L}. These lines were observed simultaneously with the 1.2 mm continuum using the same ALMA 12 m array configuration. The spectral and spatial resolutions of the cubes are $\sim$0.07 km s$^{-1}$ and $\sim$0.30$^{''}$, respectively. To ensure consistency in analysis, the cube was smoothed and regridded to match the DCN (3$-$2) line cube. The observed rms noise level for these lines are $\sim$ 4 mJy beam$^{-1}$.\\

\subsection{Dust 1.2 mm continuum}
The continuum observation was conducted between March 2019 and April 2021 over three sessions (Project ID: 2019.1.01714.S; PI: Benjamin Wu) using the ALMA 12 m array configuration. Details of data reduction and analysis are presented in \cite{2026arXiv260323747K}. We adopted the final 1.2 mm continuum and polarization maps from \cite{2026arXiv260323747K}. The resulting angular resolution is approximately 0.35$''$, which corresponds to $\sim$1000 AU. To ensure consistency with the DCN (3$-$2) line data cube, the image was subsequently smoothed and regridded to a lower resolution. The observed rms noise level for the 1.2 mm continuum map is 0.4 mJy beam$^{-1}$ and for the polarization maps 0.04 mJy beam$^{-1}$ respectively.\\

\subsection{Dust temperature ($T_{\text{d}}$) map}
In addition to the spectral line and dust continuum emission, we make use of the dust temperature ($T_{\mathrm{d}}$) maps presented by \cite{2024A&A...687A.217D}. These maps were derived with the point process mapping (PPMAP) technique, a Bayesian approach that incorporates prior information on the opacity index ($\kappa_{\nu}$), the dust temperature ($T_{\mathrm{d}}$), and the resulting spectral energy distribution (SED). This method yields average physical properties along the line of sight. The analysis of \cite{2024A&A...687A.217D} combines the 1.3\,mm continuum data with SOFIA/HAWC+ (53, 89, and 214\,$\mu$m), APEX/SABOCA (350\,$\mu$m), and APEX/LABOCA (870\,$\mu$m) observations. The resulting dust temperature ($T_{\mathrm{d}}$) cubes have an angular resolution of 2.5$''$. Further details can be found in \cite{2024A&A...687A.217D}.


\subsection{1.2 mm core catalogue}
We also utilized the core catalog from \cite{2026arXiv260323747K}. The cores were extracted using the \texttt{Astrodendro} \citep{2008ApJ...679.1338R} and \texttt{TGIF} \citep[][]{tgif} modules. Integrated fluxes ($S_{\rm int}$), sizes ($d$), peak fluxes ($S_{\rm peak}$), and masses ($M_{\odot}$) of the cores were determined after subtracting the background diffuse emission. For the background subtraction, we define an annular region from which the background emission is estimated. The radius of this annulus depends on whether adjacent cores are present. The spatial distribution of the 1.2\,mm continuum cores in this region is shown in Fig.~\ref{fig:fig2}.

\subsection{SiO (5$-$4) outflow catalogue}
We also obtained the SiO (5$-$4) catalogue for this region from \cite{2024ApJ...960...48T}. This work was part of the ALMA-IMF large program, in which the SiO emission was observed at a spatial resolution of $\sim$ 2000 AU, a spectral resolution of $\sim$ 0.34 km s$^{-1}$, and a sensitivity of 2-12 mJy beam$^{-1}$. The polygonal regions of the outflow are overlaid on the 1.2 mm continuum in Fig.~\ref{fig:fig2}.

\section{Spectral-line decomposition and model-cubes construction}\label{section_4}
After we acquired all the line cubes, we performed a pixel-wise spectral line fitting to identify distinct gas components on the basis of their central velocities. This enables us to study the velocity gradient technique (VGT) for each individual gas cloud component, particularly when their magnetic field morphologies differ. The pixel-wise decomposition of all line cubes was performed using the \texttt{Gausspy+} module \citep{2019A&A...628A..78R}. The details of the spectral-line fitting using this module are presented in the Appendix \ref{section:A1}. After performing the pixel-wise fitting, we separated the gas clouds according to their systematic velocities. The detailed procedure for this separation is described in Appendix \ref{section:A2}. For the DCN and C$^{18}$O spectral lines, we generated three model cubes corresponding to the \texttt{first component}, \texttt{second component}, and the \texttt{total model} cube. For HN$^{13}$C and H$^{13}$CO$^{+}$ spectral lines, we produced seven model cubes, denoted as the \texttt{first}, \texttt{second}, \texttt{third}, and \texttt{fourth} components, as well as the combined \texttt{first+second}, \texttt{third+fourth}, and the \texttt{total model} cube. We analyze the VGT for each model cube individually; however, our main goal is to compare the VGT-derived angles with the dust polarization angles for the \texttt{total model} cube, since the integrated-intensity–weighted VGT-derived angle is analogous to the column-density–weighted dust polarization angle.\\

\begin{table*}[htb!]
\centering
\caption{Kolmogorov--Smirnov (KS) test results for the $\theta_\text{r}$ distributions.}
\label{tab:ks_test}
\begin{tabular}{lcccc}
\hline
Spectral Line & $N$ & KS Statistic ($D$) & $p$-value & Interpretation \\
\hline
DCN (3--2)                 & 153 &  0.14 & 0.003  & Strong evidence against uniformity \\
C$^{18}$O (2--1)           & 217 &  0.09 & 0.07   & Statistically inconsistent with uniformity \\
HN$^{13}$C (3--2)          & 295 &  0.11 & 0.0008 & Strong evidence against uniformity\\
H$^{13}$CO$^{+}$ (3--2)    & 299 &  0.09 & 0.02   & Statistically inconsistent with uniformity\\
\hline
\end{tabular}\\
\vspace{4mm}
\textbf{Notes.} Results of the Kolmogorov--Smirnov (KS) test comparing the observed $\theta_\text{r}$ distributions of the four spectral lines with a uniform distribution. Here, $N$ is the number of data points (per beam) observed from the analysis, $D$ is the KS statistic representing the maximum absolute difference between the empirical and theoretical cumulative distribution functions (CDF), and the $p$-value gives the probability of obtaining a KS statistic at least as large as the observed value under the null hypothesis that the data are\\ 
\hspace{-130mm} drawn from a uniform distribution.
\end{table*}

\section{Analysis of VGT }\label{section_5}
Before analyzing the velocity gradient technique (VGT), we first calculate the Alfven Mach number ($M_{\text{A}}$) and the ratio between the channel width ($\Delta_{\text{ch}}$) and the turbulent velocity dispersion ($\sigma_{\text{nth}}$) for the spectral lines, which we define by the symbol $a$. We find that the $M_{\text{A}}$ values range from 0.24 to 0.93, i.e., sub-Alfv\'enic system, and the parameter $a$ lies between 0.11 and 0.43 i.e., thin channel. Therefore, the anisotropy of the velocity fluctuations and the velocity-caustics effect are applicable in this region. However, the quoted $M_\text{A}$ range is conditional on the adopted critical density ($n_\text{crit}$) and magnetic-field strength ($B$) and represents region-averaged estimates. Thus, locally super-Alfv\'enic regions ($M_\text{A} > 1$) cannot be excluded without spatially resolved magnetic-field measurements, and the physical interpretation of the VGT--polarization relation remains conditional on the local applicability of these assumptions. A detailed description of how $M_{\text{A}}$ and the parameter $a$ are calculated is provided in Appendix \ref{section:A3}. We therefore analyze the velocity gradient technique (VGT) and compare it with dust polarization on both large scales (beyond the core) and core scales using four spectral lines. Figs. \ref{fig:fig3} and \ref{fig:fig4} show the comparison between the dust-polarization angles and the VGT-derived angles, as well as the spatial distribution of the angular measurement (AM) for the spectral lines. Fig. \ref{fig:fig5} presents the histograms of the AM for these lines. The alignment measure, $AM=\cos (2\theta_\text{r})$, is defined using the relative orientation, $\theta_\text{r}$ ($0^\circ \leq \theta_\text{r} \leq 90^\circ$), between the plane-of-sky magnetic field, $B_{\rm POS}$, inferred from the VGT and that derived from dust polarization observations. We note that the plots shown here are for the total model cubes; the component-wise analysis is provided in Appendix \ref{Appendix_C}. Due to the limited sensitivity of the dust continuum emission, the AM cannot be measured across the entire region, even though the spectral-line emission is detected throughout. From the spatial distribution of the AM measurements, the values appear to follow a bimodal pattern, which becomes even more evident in the AM histograms for the four spectral lines. In each histogram, we observe a clear bimodal distribution with peaks near +1 and −1.\\

We note that to provide a comprehensive assessment of the relative alignment, we examine both the alignment measure (AM), and the direct distribution of the relative angle ($\theta_\text{r}$). As demonstrated by the analysis presented in Appendix \ref{Appendix_C_0}, the nonlinear transformation from $\theta_\text{r}$ to AM, can enhance or may produce apparent structures in the AM distribution. Nevertheless, AM serves as a useful diagnostic because it directly quantifies the alignment tendency of each data point, distinguishing between preferential parallel, perpendicular, and intermediate orientations. In contrast, the $\theta_\text{r}$ distribution provides complementary information by showing how the data are distributed across different relative orientations.\\

Although the $\theta_\text{r}$ spans the range 0$^\circ$–90$^\circ$, its physical interpretation can be broadly grouped into three regimes. Angles between 0$^\circ$ and 30$^\circ$ indicate a preferential parallel alignment, angles between 30$^\circ$ and 60$^\circ$ represent a transitional or weakly aligned regime, and angles between 60$^\circ$ and 90$^\circ$ indicate a preferential perpendicular alignment. Therefore, while values such as 70$^\circ$, 80$^\circ$, and 90$^\circ$ differ quantitatively, they all correspond to the same qualitative tendency toward perpendicular alignment. Similarly, values such as 0$^\circ$, 10$^\circ$, and 20$^\circ$ all correspond to the same qualitative tendency toward parallel alignment, although they differ in the degree of alignment. Thus, the $\theta_\text{r}$ distribution preserves information about the exact relative orientation, whereas the AM distribution emphasizes the underlying alignment tendency of the data, making the two diagnostics complementary.\\

Figure~\ref{fig:fig5_0} presents the distributions of the relative angle, $\theta_\text{r}$, for the four molecular tracers. For DCN, the $\theta_\text{r}$ distribution exhibits two distinct peaks, one near $25^{\circ}$ and another near $70^{\circ}$, corresponding to preferential parallel and perpendicular alignments, respectively. In the case of C$^{18}$O, we identify a prominent peak near $75^{\circ}$, whereas a second peak is not prominently observed close to $0^{\circ}$. However, as discussed above, angles between $0^\circ$ and $20^\circ$ all represent preferential parallel alignment. Consequently, these data points are grouped together in the alignment measure, resulting in the secondary peak near $AM\approx +1$. For HN$^{13}$C, the $\theta_\text{r}$ distribution again exhibits two peaks, located near $15^{\circ}$ and $80^{\circ}$, indicating the coexistence of parallel and perpendicular alignments. Finally, the H$^{13}$CO$^{+}$ distribution is characterized by a single prominent peak around $85^{\circ}$, suggesting a strong preference for perpendicular alignment. Consequently, from the visual inspection of $\theta_\text{r}$ for four tracers, we argue that both DCN and HN$^{13}$C exhibit clear signatures of coexisting parallel and perpendicular alignments, implying that the influences of turbulence and gravity are both significant. In contrast, C$^{18}$O and H$^{13}$CO$^{+}$ show a stronger preference for perpendicular alignment, suggesting that gravity is the dominant process, while the signature of turbulence is comparatively less pronounced than in DCN and HN$^{13}$C.\\

Although the histogram of $\theta_\text{r}$  appears to indicate the presence of preferred orientation(s), we further performed a Kolmogorov--Smirnov (KS) test, adopting a uniform distribution as the null hypothesis. The KS test shows that all the four tracers reject the hypothesis of a uniform distribution. The detailed KS statistics are listed in Table~\ref{tab:ks_test}.\\

Now, we identify four regions—\texttt{Region 1}, \texttt{Region 2}, \texttt{Region 3}, and \texttt{Region 4}—in the AM distribution map where the AM values are close to −1. We selected these regions because they correspond to channels that are connected to the main filament and may serve as pathways through which material is transported toward the central clump via the arc-like structures. A visual inspection revealed that the AM values in these regions are predominantly close to $-$1. We therefore selected these four regions for further analysis. In particular, \texttt{Region 1} and \texttt{Region 2} exhibit sub-filamentary structures that are connected to the main filament, and therefore may serve as channels through which gas from the surrounding medium accretes onto the filament and feeds the cores. Likewise, \texttt{Region 3} exhibits an arc-like structure that connects to the central hub, suggesting that it may also act as a channel through which gas accretes and feeds by the multiple cores located in the central region. For \texttt{Region 4}, no such clear structure is visible; however, the AM values remain close to −1, indicating that it may still serve as a potential accretion channel. The lack of an evident structure is likely due to the limited sensitivity of the dust continuum emission. We have marked these regions on the 1.2 mm continuum map, where \textbf{a} and \textbf{b} represent the \texttt{Region 1} and \texttt{Region 2}, and \textbf{c} and \textbf{d} correspond to \texttt{Region 3} and \texttt{Region 4} respectively. This is depicted in the upper panel of Fig. \ref{fig:fig6}. We note that, in addition to strong gravitational infall, the presence of strong outflows or even rotation can also produce AM values close to −1. We therefore overlay the SiO (5$-$4) outflow regions identified by \cite{2024ApJ...960...48T} on the 1.2 mm continuum emission in Fig. \ref{fig:fig2} and on the AM distributions shown in Figs. \ref{fig:fig3} and \ref{fig:fig4}. From these overplots, we find that \texttt{Region 2} and \texttt{Region 3} are partially overlapped by the outflow region.\\

For DCN \texttt{total model}, the AM values are close to −1 in \texttt{Region 1} and \texttt{Region 3}, while AM measurements are absent in \texttt{Region 2} and \texttt{Region 4}. Similarly, for C$^{18}$O \texttt{total model}, the AM values are close to −1 in \texttt{Region 1} and \texttt{Region 4}, and no prominent results are obtained in the remaining regions due to the lack of measurable AM values. For HN$^{13}$C and H$^{13}$CO$^{+}$ \texttt{total model}, \texttt{Region 1} shows a mixture of AM values near −1 and +1. However, in \texttt{Region 2} and \texttt{Region 4}, the AM values are consistently close to −1. In Region 3, the AM values for HN$^{13}$C are close to −1, while for H$^{13}$CO$^{+}$ a mixture of both −1 and +1 values is observed. In addition to the total model cubes we also examined the similar analysis for the individual components of the spectral lines; however, no significant variation in the AM values was observed compared to the total model. Since the spatial distributions of each component are not identical and are less extended than in the total model, the AM measurement for each component appears in certain spatial positions relative to the total model, and is sometimes not consistent with the other components.\\

We now examine the position–velocity (PV) diagrams toward these regions for each spectral line to determine whether any smooth velocity gradient is present, which would further support the gravitational infall. In the later section, we explain why the observed velocity gradients are unlikely to be caused by outflow motion or rotation. In the middle and lower panels of Fig.~\ref{fig:fig6} we present the PV diagram towards the \texttt{Region 1} indicated by the blue rectangular area. The diagram is shown along the direction denoted by the symbol \texttt{a} for the four spectral lines. From the figures, we notice a smooth velocity gradient which is more prominently observed in the C$^{18}$O and HN$^{13}$C lines than in the DCN and H$^{13}$CO$^{+}$ spectral lines. We fitted the velocity gradients traced by the C$^{18}$O and HN$^{13}$C lines and obtained values of $34.9 \pm 3.4~\mathrm{km\,s^{-1}\,pc^{-1}}$ and $102.9 \pm 1.1~\mathrm{km\,s^{-1}\,pc^{-1}}$, respectively. Using the average of these two gradients and the mass of the region derived from the 1.2 mm dust continuum emission, we estimated a mass accretion rate of $(3.1 \pm 2.2) \times 10^{-4}~M_{\odot}\,\mathrm{yr}^{-1}$. The derived values are listed in Table \ref{tab:table1}. We note that this value is affected by projection effects and should be multiplied by \texttt{tan}$(i)$, where $i$ is the inclination angle between the direction of gravitational infall and the line of sight. Since $i$ cannot be directly constrained from the present observations, the true value remains uncertain. The mass of the region and its associated error are obtained by the following formulae:

\begin{equation}\label{Eqn4}
M_{\text{region}} = \frac{S^{\text{1.2mm}}_{\text{int}}~d^{2}}{\kappa_{\text{1.2mm}}~B_{\text{1.2mm}} (T_{\text{d}},\nu)},
\end{equation}

\begin{equation}\label{Eqn5}
    \Delta M_{\text{region}} = M_{\text{region}}\sqrt{\biggl(\frac{\Delta S^{\text{1.2mm}}_{\text{int}}}{S^{\text{1.2mm}}_{\text{int}}}\biggl)^{2} ~+~ \biggl(\frac{2\Delta d}{d}\biggl)^{2} ~+~  \biggl(\frac{\Delta B_{\text{1.2mm}} \Huge(T_{\text{d}},\nu)}{B_{\text{1.2mm}} (T_{\text{d}},\nu\Huge)}\biggl)^{2} }~~.
\end{equation}

Here, $S^{\rm 1.2mm}_{\rm int}$ represents the integrated flux density of the 1.2 mm continuum emission, $d$ is the distance to the source, and $T_{\rm d}$ is the dust temperature of the region. The dust opacity at 1.2 mm is assumed to be $\kappa_{\rm 1.2\,mm} = 0.01\,(\nu/228.9~{\rm GHz})^{\beta}$ cm$^{2}$ g$^{-1}$. The continuum emission was observed in three ALMA Band 6 spectral windows centered at approximately 244, 246, and 258 GHz, corresponding to a mean observing frequency of $\nu \approx 250$ GHz used in our analysis. \\

In Fig. \ref{fig:fig7}, we present the PV diagram toward \texttt{Region 2}, which is observed in all four spectral lines. The velocity gradients measured from the DCN, C$^{18}$O, HN$^{13}$C, and H$^{13}$CO$^{+}$ lines are $12.7 \pm 0.5$, $9.3 \pm 2.0$, $12.5 \pm 0.2$, and $6.3 \pm 0.2~\mathrm{km\,s^{-1}\,pc^{-1}}$, respectively. The average velocity gradient derived from these tracers is $10.2 \pm 2.6~\mathrm{km\,s^{-1}\,pc^{-1}}$. Using this average velocity gradient together with the measured mass of $20.5 \pm 10.1~M_{\odot}$, we estimate a mass accretion rate of $(2.1 \pm 1.2) \times 10^{-4}~M_{\odot}\,\mathrm{yr}^{-1}$.\\


\begin{table*}[htb!]
\centering
\caption{Mass Accretion Through Four Regions.}
\begin{tabular}{cccccc}
\hline
Region & Spectral line & Velocity gradient &
Mean velocity gradient &
Mass of the region &
Mass infall rate $\times$ \texttt{tan}($i$) \\
&
&
(km\,s$^{-1}$\,pc$^{-1}$) &
(km\,s$^{-1}$\,pc$^{-1}$) &
($M_\odot$) &
($M_\odot$\,yr$^{-1}$) $\times$ 10$^{-4}$ \\
\hline

\multirow{4}{*}{Region 1}
& DCN (3$-$2)              & $--$ & \multirow{4}{*}{68.9 $\pm$ 34.0}
& \multirow{4}{*}{4.4 $\pm$ 2.2}
& \multirow{4}{*}{3.1 $\pm$ 2.2} \\
& C$^{18}$O (2$-$1)        & 34.9 $\pm$ 3.4 & & & \\
& HN$^{13}$C (3$-$2)       & 102.9 $\pm$ 1.1 & & & \\
& H$^{13}$CO$^{+}$ (3$-$2) & $--$ & & & \\
\hline

\multirow{4}{*}{Region 2}
& DCN (3$-$2)             & 12.7 $\pm$ 0.5 & \multirow{4}{*}{10.2 $\pm$ 2.6}
& \multirow{4}{*}{20.5 $\pm$ 10.1}
& \multirow{4}{*}{2.1 $\pm$ 1.2} \\
& C$^{18}$O (2$-$1)       & 9.3 $\pm$ 2.0 & & & \\
& HN$^{13}$C (3$-$2)      & 12.5 $\pm$ 0.2 & & & \\
& H$^{13}$CO$^{+}$ (3$-$2) & 6.3 $\pm$ 0.2 & & & \\
\hline

\multirow{4}{*}{Region 3}
& DCN (3$-$2)             & 167.0 $\pm$ 2.5  & \multirow{4}{*}{196.7 $\pm$ 29.7}
& \multirow{4}{*}{25.2 $\pm$ 12.4}
& \multirow{4}{*}{51 $\pm$ 26} \\
& C$^{18}$O (2$-$1)        & $--$ & & & \\
& HN$^{13}$C (3$-$2)      & 226.4 $\pm$ 1.3 & & & \\
& H$^{13}$CO$^{+}$ (3$-$2) & $--$ & & & \\
\hline

\multirow{4}{*}{Region 4}
& DCN (3$-$2)              & 41.5 $\pm$ 1.6 & \multirow{4}{*}{23.3 $\pm$ 10.6}
& \multirow{4}{*}{29.4 $\pm$ 14.7}
& \multirow{4}{*}{7.0 $\pm$ 4.7} \\
& C$^{18}$O (2$-$1)       & 18.3 $\pm$ 0.9  & & & \\
& HN$^{13}$C (3$-$2)       & 15.7 $\pm$ 0.8 & & & \\
& H$^{13}$CO$^{+}$ (3$-$2) & 17.7 $\pm$ 0.2 & & & \\
\hline
\label{tab:table1}
\end{tabular}\\
\textbf{Notes.} Column 1: Regions where velocity gradients are identified. Column 2: Spectral lines used in the analysis. Column 3: Velocity gradients derived from each spectral line. Column 4: Mean observed velocity gradients of the region. Column 5: Mass of \\
\hspace{-0mm} the regions estimated from the 1.2 mm dust continuum emission. Column 6: Observed mass infall rates of the region. Here $i$ is \\
\hspace{-71mm}the unknown inclination angle between the infall direction and the line of sight.
\end{table*}

Likewise, toward \texttt{Region 3} and \texttt{Region 4}, we also compute the position–velocity (PV) diagrams for the four spectral lines. These regions are marked by the red and green rectangular areas in Fig.~\ref{fig:fig6} and are indicated by the red and green arrows labeled \textbf{c} and \textbf{d}, respectively. Toward \texttt{Region 3}, we identify a smooth velocity gradient across a spatial extent of 0.02 pc. The ends of this gradient are connected to two distinct velocity structures. The PV diagrams are shown in Fig. \ref{fig:fig8}. These characteristics suggest that the gas is infalling toward the central hub through the arc-like structure. Toward \texttt{Region 4}, we also observe a smooth velocity gradient across a spatial extent of 0.04 pc. The PV diagrams are presented in Fig. \ref{fig:fig9}.\\

In \texttt{Region 3}, prominent velocity gradients are observed in the DCN and HN$^{13}$C spectral lines, with values of $167.0 \pm 2.5$ and $226.4 \pm 1.3~\mathrm{km\,s^{-1}\,pc^{-1}}$, respectively. The average velocity gradient derived from these measurements is $196.7 \pm 29.7~\mathrm{km\,s^{-1}\,pc^{-1}}$. Using the mass of the region, $25.2 \pm 12.4~M_{\odot}$, derived from the continuum emission, we estimate a mass accretion rate of $(51 \pm 26) \times 10^{-4}~M_{\odot}\,\mathrm{yr}^{-1}$. In \texttt{Region 4}, the velocity gradients measured from the DCN, C$^{18}$O, HN$^{13}$C, and H$^{13}$CO$^{+}$ lines are $41.5 \pm 1.6$, $18.3 \pm 0.9$, $15.7 \pm 0.8$, and $17.7 \pm 0.2~\mathrm{km\,s^{-1}\,pc^{-1}}$, respectively. These measurements yield an average velocity gradient of $23.3 \pm 10.6~\mathrm{km\,s^{-1}\,pc^{-1}}$. Combining this value with the derived mass of $29.4 \pm 14.7~M_{\odot}$ results in a mass accretion rate of $(7.0 \pm 4.7) \times 10^{-4}~M_{\odot}\,\mathrm{yr}^{-1}$. Among the four regions, \texttt{Region 3} exhibits the highest mass accretion rate. A summary of the derived parameters is provided in Table~\ref{tab:table1}.\\

We now discuss why outflow activity is unlikely to be responsible for the angular measurement (AM) values approaching $-1$ in these four regions. First, the four molecular tracers used in our analysis (DCN, C$^{18}$O, HN$^{13}$C, and H$^{13}$CO$^{+}$) are not well-established tracers of molecular outflows. More importantly, velocity gradients associated with outflows are expected to be significantly larger than those measured from these lines. We find that \texttt{Region 2} and \texttt{Region 3} are partially overlapped by SiO (5$-$4) outflows. To assess the potential impact of these outflows, we examined their position-velocity (PV) diagrams. Fig. \ref{fig:fig10} shows the outflow regions overlaid on the 1.3 mm continuum emission, together with the corresponding PV diagrams of the SiO (5$-$4) emission \citep{2024ApJ...960...48T}. From these PV diagrams, we observe velocity gradients of $\sim$$500$--$600~\mathrm{km~s^{-1}~pc^{-1}}$, substantially larger than the average velocity gradients measured from the four spectral lines in \texttt{Region 2} and \texttt{Region 3}, namely $10.2 \pm 2.6~\mathrm{km~s^{-1}~pc^{-1}}$ and $196.7 \pm 29.7~\mathrm{km~s^{-1}~pc^{-1}}$, respectively. The large discrepancy between the velocity gradients traced by the SiO outflows and those measured from the four molecular lines indicates that the observed gradients are unlikely to be dominated by outflow contamination. We therefore rule out outflow activity as the primary cause of the AM values being close to $-1$ in these regions.\\

Likewise, we have also examined whether the observed smooth velocity gradients could originate from filament rotation. We particularly focus on \texttt{Region 1} and \texttt{Region 2}, as these regions correspond to the curved (arc-like) portions of the filament. We investigate whether the entire filament is undergoing rotation around its ridgeline. For this purpose, in Fig. \ref{fig:fig10_0}, we present the dust continuum emission of the filament and straighten the filament with respect to its ridgeline. Using the same transformation, we also re-map in the same figure the moment-1 velocity maps of the four molecular transitions along the filament ridgeline and examine whether a systematic velocity gradient is present across the ridgeline. However, from the radial velocity profiles, we find that the velocity gradient on the left side of the ridgeline is mainly contributed by \texttt{Region 1} rather than representing a coherent gradient across the entire filament. This gradient is prominent only within a limited radial range of approximately −0.05 pc to −0.01 pc, which is mainly dominated by \texttt{Region 1}, while beyond this region no systematic or smooth velocity variation is observed. Therefore, the observed velocity gradients in \texttt{Region 1} and \texttt{Region 2} are unlikely to be produced by global rotation of the filament. Instead, they may be associated with gas accretion or inflow from the surrounding medium onto the filament. However, recently \cite{2026A&A...711A.239S} identified signatures of rotation in the filamentary structure associated with the G012.80 protocluster using the N$_2$H$^+$ (1–0) molecular transition. Therefore, although rotation can produce smooth velocity gradients in some filamentary systems, we have examined this possibility in our case by investigating the velocity structure with respect to the filament ridgeline. Our analysis shows that the observed velocity gradients do not exhibit the characteristics expected from a coherent rotating filament, making rotation an unlikely explanation for the observed kinematic pattern. Likewise, for \texttt{Region 3} and \texttt{Region 4}, we do not identify a prominent, coherent axis about which the gas could rotate. These regions appear to be arc-like structures connected to the main central filament rather than independent rotating structures. Therefore, the possibility that rotation is responsible for the observed smooth velocity gradients in these regions is less likely.\\

\begin{table*}[htb!]
\centering
\caption{Characteristic of the cores obtained from the four spectral lines.}
\centering
\begin{tabular}{c  c  c c  c c c c c  }
\hline
\multirow{2}{*}{Core}  & RA & DEC & \multicolumn{4}{c}{AM value}  & Outflow \\
\cline{4-7} 
 ID & [ICRS] & [ICRS] & DCN & C$^{18}$O & HN$^{13}$C & H$^{13}$CO$^{+}$ & enclosed & \\ 
\hline
0 & 15:53:10.52 & -54:36:39.83  &  $+$0.4 $\pm$ 0.1 & $-0.978^{+0.020}_{-0.014}$  & $+$0.1 $\pm$ 0.1 & $-$0.2 $\pm$ 0.1 & \ding{51}\\

1 & 15:53:10.49 & -54:36:41.41  & $--$ & $--$ & $--$ & $--$   & \ding{51} \\

2 & 15:53:10.90 & -54:36:46.16 &  $+$0.6 $\pm$ 0.4 & $+$0.5 $\pm$ 0.3 & $-$0.2 $\pm$ 0.1 & $+$0.3 $\pm$ 0.4 &  \ding{51} \\

3 & 15:53:10.94  & -54:36:46.05  &   $+$0.6 $\pm$ 0.1 & $+$0.2 $\pm$ 0.4 & $-$0.2 $\pm$ 0.1 & $+$0.3 $\pm$ 0.3 & $-$\\

4 & 15:53:10.98  & -54:36:46.55  &  $+$0.6 $\pm$ 0.2 & $-$0.5 $\pm$ 0.3  & $-$0.2 $\pm$ 0.1 & $+$0.0 $\pm$ 0.3 &  \ding{51} \\

5 & 15:53:11.08  & -54:36:47.99 & $-$0.1 $\pm$ 0.0 & $+$0.2 $\pm$ 0.1   & $-$0.1 $\pm$ 0.1  & $-$0.2 $\pm$ 0.1  & $-$ \\

6 & 15:53:10.71  & -54:36:46.75 & $+$0.8 $\pm$ 0.1 & $+$0.9 $\pm$ 0.1  & $-$0.2 $\pm$ 0.1   &   $+0.993^{+0.006}_{-0.021}$   &  \ding{51} \\

7 & 15:53:11.02   & -54:36:42.79 &  $-0.990^{+0.032}_{-0.009}$ & $-$0.8 $\pm$ 0.2  & $-$0.7 $\pm$ 0.2   &  $+$0.9 $\pm$ 0.2   & $-$  \\

8 & 15:53:11.19 & -54:36:48.69 & $+$0.6 $\pm$ 0.1 & $-$0.8 $\pm$ 0.1   & $-$0.5 $\pm$ 0.2  &  $-0.989^{+0.031}_{-0.010}$   & $-$  \\

9 & 15:53:11.59 & -54:36:25.08  & $+$0.4 $\pm$ 0.1  &$--$  & $-$0.2 $\pm$ 0.1  & $+$0.6 $\pm$ 0.2   & $-$   \\

10 & 15:53:11.29 & -54:36:34.60 & $+0.996^{+0.004}_{-0.010}$    & $+$0.9 $\pm$ 0.1   & $+0.954^{+0.038}_{-0.069}$ & $+$0.9 $\pm$ 0.1 & $-$  \\

11 & 15:53:11.73 & -54:36:29.78 & $--$ & $--$  & $-$0.6 $\pm$ 0.1  & $-$1.0 $\pm$ 0.1    & $-$  \\

12 & 15:53:09.40  & -54:37:00.52 & $-$0.1 $\pm$ 0.0 & $+$0.1 $\pm$ 0.2  & $-$0.2 $\pm$ 0.1  & $-$0.1 $\pm$ 0.1 & \ding{51}\\

13 & 15:53:09.49 & -54:37:01.12 & $+0.996^{+0.004}_{-0.014}$   & $+$0.6 $\pm$ 0.1  & $-$ & $+$0.3 $\pm$ 0.4     &  \ding{51}\\

14 & 15:53:09.47 & -54:37:00.39 & $+$0.1 $\pm$ 0.0  & $+$0.3  $\pm$ 0.1  & $+$0.1 $\pm$ 0.1  & $+$0.0 $\pm$ 0.2 & \ding{51}\\

15 & 15:53:09.46 & -54:36:58.83 & $--$ & $--$ & $--$ &  $--$            &  \ding{51}\\

16 & 15:53:09.23 & -54:36:59.13 &  $--$ & $--$ & $--$ & $--$             &  $-$  \\

17 & 15:53:09.21 & -54:36:57.86   &  $-$0.7 $\pm$ 0.5 & $-$0.5 $\pm$ 0.2  & $-0.988^{+0.038}_{-0.011}$   & $-$0.7 $\pm$ 0.2 & \ding{51}\\

18 & 15:53:10.69  & -54:36:32.19 & $--$ & $--$  & $--$ & $--$           & $-$ \\

19 & 15:53:10.45 & -54:36:50.29 & $--$ & $-$0.1 $\pm$ 0.2   & $-0.970^{+0.034}_{-0.022}$    & $-0.899^{+0.059}_{-0.046}$   &  $-$ \\

20 & 15:53:09.66 & -54:36:44.22 & $--$ & $+$1.0 $\pm$ 0.1  & $+$0.3 $\pm$ 0.2  & $+$0.1 $\pm$ 0.3      &   $-$ \\

21 & 15:53:09.33 & -54:36:51.48 &  $--$ &  $-0.995^{+0.016}_{-0.004}$   & $-$0.4 $\pm$ 0.2  &  $-$0.5 $\pm$ 0.4  &  \ding{51}\\

22 & 15:53:09.45 & -54:36:51.94 &  $--$ & $-$0.5 $\pm$ 0.1  & $-$0.9 $\pm$ 0.1   & $-0.987^{+0.042}_{-0.012}$   & $-$   \\

23 & 15:53:10.03 & -54:36:53.59  &  $--$ & $--$  & $--$ & $--$           & $-$  \\

24 & 15:53:10.48  & -54:36:36.29  & $--$ & $--$  & $--$ & $--$           & $-$ \\

25 & 15:53:10.93 & -54:36:36.30  & $--$ & $--$ & $--$ & $--$            & $-$  \\

26 & 15:53:08.87 & -54:36:59.65  & $+$0.8 $\pm$ 0.2 &  $+0.978^{+0.020}_{-0.070}$    & $-$1.0 $\pm$ 0.1   & $-$0.1 $\pm$ 0.3  & \ding{51}  \\

\hline
\label{tab:table2}
\end{tabular}\\
\vspace{-2mm}
\hspace{-2.2mm}\textbf{Notes.} Column 1 : Core IDs taken from \cite{2026arXiv260323747K}. Columns 2 and 3 : Right Ascension (RA) and Declination (Dec) of each core. Columns 4–7 : Angular Measurement (AM) values toward the cores derived from the four spectral lines: DCN,\\
\hspace{-0mm}C$^{18}$O, HN$^{13}$C, and H$^{13}$CO$^{+}$ and the dust polarization. Column 8 :  Outflow association of the cores.
\end{table*}

In addition to the large-scale AM analysis, we also examine the AM toward the dense cores. We note that, while calculating the VGT-derived angles, we consider the most probable value of the VGT angle distribution obtained within each 20 $\times$ 20 pixel sub-block. This corresponds to a physical scale of $\sim$ 0.03 pc, which is approximately three times the average compact continuum core size derived from the ALMA-IMF observations ($\sim$ 0.01 pc). We also tested a smaller sub-block size of 10 $\times$ 10 pixels; however, the method was unable to reliably determine the most probable value of the VGT-derived angle distribution in several sub-blocks. In this sense, the derived AM values correspond to the regions that include the cores and their immediate surrounding environments, rather than representing only the compact continuum core component. However, we emphasize that the derived core size is not a unique quantity and depends on the adopted definition and methodology. For example, some studies use the FWHM obtained from a two-dimensional Gaussian fit as a proxy for the core size \citep[and references therein]{2025ApJ...980...87S}, while others adopt a larger characteristic size (e.g., twice the FWHM) to include more extended emission \citep{2026arXiv260323747K,2026arXiv260710029V}. Therefore, the physical boundary of a core remains ambiguous, as it is not uniquely established whether the surrounding lower-density material should be included as part of the core structure or how far this contribution should extend. This ambiguity can lead to variations in the reported core sizes. In comparison, nearby prestellar cores in the Taurus molecular cloud typically have characteristic sizes of $\sim$0.05 to 0.1 pc, which is larger than the compact continuum core sizes identified in the distant ALMA-IMF sources \citep{2022A&A...667A.119F, 2022MNRAS.516L..48K}.\\

The AM values toward the dense cores for the four spectral lines along with the uncertainties are listed in Table \ref{tab:table2}. The uncertainties of AM are obtained from the uncertainties of the polarization angles and the VGT-derived angles, and the angular measure (AM) which are discussed in Appendices \ref{Appendix_K} and \ref{Appendix_L}. We note that if a core is enclosed by regions with different AM values, we adopt the average of those values. In addition, we examined whether the cores are associated with SiO (5$-$4) outflows. While outflows can be a possible alternative explanation for AM values close to $-$1, we note that the molecular tracers used in this study are not ideal tracers of outflow activity. Nevertheless, we investigated the possibility that the observed signatures could be influenced by outflows. From the derived results, we find that the AM values measured toward cores 10, 13, and 20 are consistent with being close to $+$1 within the uncertainties for these spectral lines. Likewise, the AM values obtained from these spectral lines toward cores 11, 17, 19, 21, and 22 are consistent with values close to $-$1 within the associated uncertainties. Among these five cores, cores 17 and 21 are associated with SiO (5$-$4) outflows. Therefore, outflow activity could be a possible explanation for the AM values being close to $-$1 toward these two cores. For seven cores, no reliable AM measurements could be obtained. Therefore, determining their dynamical stage based on the AM analysis is beyond the scope of the present work.\\

We note that the VGT method determines the magnetic field orientation by identifying the direction of the maximum intensity gradient in individual thin velocity-channel maps. However, in regions containing multiple overlapping turbulent eddies with similar line-of-sight velocities, the determination of a unique gradient direction may sometimes become ambiguous. Consequently, the inferred magnetic field orientation can occasionally vary between tracers, and in some cases one spectral line may produce results that differ substantially from those derived from other spectral lines. For example, toward cores 2, 3, 6, and 9, the AM values derived from all spectral lines are consistent with values close to $+$1 within the uncertainties, except for HN$^{13}$C. However, when we examine the second velocity component of the HN$^{13}$C emission, we find AM values of approximately 0.9 for cores 2, 3, and 6 and $\sim$ 0.4 for core 9, which are consistent with those obtained from the other three spectral lines. This suggests that the apparent discrepancy may arise from the complexity of the velocity structure traced by HN$^{13}$C in these cores. Likewise, toward core 7, the AM values derived from all spectral lines except H$^{13}$CO$^+$ are consistent with values close to $-$1 within the associated uncertainties. In contrast, the AM value obtained from the \texttt{total\_model} cube of H$^{13}$CO$^+$ is close to $+$1. However, when we examine the \texttt{second component} of the H$^{13}$CO$^+$ emission, we obtain an AM value of approximately $-$0.7, which is broadly consistent with the values derived from the other spectral lines. For cores 0 and 8, particularly core 8, the AM values derived from all spectral lines except DCN are consistent with values close to $-$1 within the associated uncertainties. In contrast, for cores 5, 12, and 14, the AM values derived from all tracers lie between $-$1 and $+$1 and are not clearly consistent with either extreme. Therefore, we cannot draw any firm conclusions regarding the dynamical stage of these cores based on the AM analysis. For core 26, the core is only partially enclosed by the AM measurements obtained from the four spectral lines. Consequently, although we are able to derive AM values toward this core, it is not appropriate to infer its dynamical state from the present analysis.
Regarding the outflow activity, apart from the two cores (17 and 21) where AM values close to $-$1 coincide with the presence of SiO (5$-$4) outflows, we do not find any clear trend between AM values close to $-$1 and outflow activity. In some cases, the AM values remain close to $+$ 1 within the uncertainties, even in the presence of outflows. This suggests that outflows may be responsible for the observed AM signatures in few cores but not the primary reason for being AM values close to $-$1.\\

\section{Discussion}\label{section_6}
We apply VGT to a portion of the massive protocluster G327.29, offset from its central region, using four spectral lines: DCN (3–2), C$^{18}$O (2–1), HN$^{13}$C (3–2), and H$^{13}$CO$^{+}$ (3–2), and compare the results with 1.2 mm dust continuum emission. The angular measurement (AM) distributions for all spectral lines show peaks at +1 and -1, indicating that the region is influenced by both turbulence and gravity. However, when we examine the direct $\theta_\text{r}$ distributions, we find that both HN$^{13}$C and DCN exhibit two peaks, one towards $0^\circ$ and the other towards $90^\circ$, indicating the coexistence of preferential parallel and perpendicular alignments. In contrast, C$^{18}$O and H$^{13}$CO$^{+}$ do not exhibit a distinct peak near $0^\circ$, while its distribution is strongly concentrated toward $90^\circ$. This suggests that the signature of gravity is more dominant in C$^{18}$O and H$^{13}$CO$^{+}$, whereas the influence of turbulence is comparatively less pronounced than in HN$^{13}$C and DCN. This may occur because the DCN and HN$^{13}$C emission are more closely associated with the dense gas traced by the dust continuum emission compared to the C$^{18}$O and H$^{13}$CO$^{+}$ lines.\\

We examine the angular measurements both on large scales and on the core scale. On large scales, we identify evidence of gravitational infall (beyond core scale), which is further supported by the presence of a smooth velocity gradient. We derived velocity gradients of $\sim$ 9$-$226 km~s$^{-1}$~pc$^{-1}$. The corresponding mass accretion rates are of the order of a few $\times 10^{-4}$ to $10^{-3}$ M$_\odot$ yr$^{-1}$. These observed velocity gradients are significantly smaller than those derived from the SiO (5$-$4) outflow emission, where the measured velocity gradients are of the order of $\sim 500-600~\mathrm{km~s^{-1}~pc^{-1}}$. Thus, we rule out the possibility that the observed velocity gradients are strongly influenced by outflow motions. Likewise, we rule out the possibility that rotation is responsible for the observed velocity gradients in these regions. Although \texttt{Region 1} and \texttt{Region 2} correspond to arc-like portions of the filament, the absence of a coherent and continuous velocity gradient across the ridgeline (from one side of the filament to the other) argues against a global rotational motion. In the other two regions, we do not identify any well-defined rotational axis around which the gas could be rotating. Instead, the observed kinematic structures are more consistent with gravitational motions, such as gas inflow from the surrounding environment toward the main central regions where multiple cores are present. At the core scale, however, we find that the cores are influenced by both turbulence and gravity, rather than being purely gravity-dominated.\\

Several previous studies have demonstrated hierarchical mass accretion from large cloud scales down to small core scales in protoclusters \citep{2018ApJ...861...14C, 2024A&A...689A..74A,2025A&A...696A.202S,2025A&A...695A..51B,2025ApJ...979..233M}. For instance, \citet{2025A&A...696A.202S} analyzed the G351.77 protocluster using N$_{2}$H$^{+}$ (1–0), H$_{2}$CO (3–2), and DCN (3–2) lines and observed large-scale accretion from the surrounding environment into the central dense region, where multiple cores actively gather gas. Small-scale accretion manifests as “V-shaped” converging flows toward or around the cores. A similar phenomenon has been reported in the G353.41 protocluster, where large-scale accretion channels gas toward dense regions \citep{2024A&A...689A..74A}. From their analysis of the N$_{2}$H$^+$(1$-$0) data, they found that the converging gravitational inflow from the surrounding environment toward the cores exhibits velocity gradients of approximately 10$-$150 km~s$^{-1}$~pc$^{-1}$, with corresponding mass accretion rates of a few $\times 10^{-4}$ to $10^{-3}$ M$_{\odot}$ yr$^{-1}$. These values are broadly consistent with the mass accretion rates derived for the four regions analyzed in our study.\\


Despite these advances, the evolutionary stages of cores remain observationally uncertain. One major limitation arises from the uncertainties in magnetic field estimates derived from dust polarization, which provide indirect estimation rather than direct values. These uncertainties propagate into the virial parameter ($\alpha_{\text{vir}}$), making it difficult to reliably assess a core’s dynamical state \citep{2021ApJ...912..159P,2024ApJ...967..157L,2026arXiv260323747K}. Although Zeeman splitting offers a direct method to measure the line-of-sight magnetic field, observations in high-mass star-forming regions are still scarce \citep{2007ApJ...670.1159F,2010ApJ...725..466C,2008A&A...487..247F, 2008A&A...484..773V}. While statistical studies have established a correlation between magnetic field strength ($B$) and density ($\rho$), it is not generally applicable for determining magnetic fields at the scale of individual cores \citep{2010ApJ...725..466C,2025MNRAS.540.2762W} because the $B$–$\rho$ correlation depends on the gas accumulation process, with a variable power-law index $p$ \citep{2015MNRAS.451.4384T}. We are therefore unable to directly compare the AM values of the cores with the virial parameters ($\alpha_\text{vir}$) measured for those cores \citep{2026arXiv260323747K}. However, future high-resolution observations will allow us to probe core fragmentation in detail, helping to determine whether cores are Jeans-fragmented, magnetically regulated, or still unfragmented \citep{2018ApJ...853....5P, 2021ApJ...912..159P, 2025arXiv250906749Y}.\\



\section{Conclusions}\label{section_7}
We have applied the Velocity Gradient Technique (VGT) to study the G327.29 protocluster using the 1.2 mm dust continuum and four spectral lines: DCN (3$-$2), C$^{18}$O (2$-$1), HN$^{13}$C (3$-$2), and H$^{13}$CO$^{+}$ (3$-$2). The main findings of this study are summarized as follows:\\

(i) The region appears to be shaped by the combined influence of turbulence, magnetic fields, and gravity, rather than by any single dominant process. This interpretation is reflected in the $\theta_{\rm r}$ distributions of DCN and HN$^{13}$C, which exhibit distinct peaks toward both parallel and perpendicular orientations. In contrast, although the $AM$ distributions of C$^{18}$O and H$^{13}$CO$^{+}$ also exhibit bimodality, their $\theta_{\rm r}$ distributions show only a single prominent peak toward perpendicular alignment, suggesting that the signature of turbulence is less evident than in DCN and HN$^{13}$C.\\ 

(ii) The spatial distribution of the angular measurement (AM) reveals signatures of large-scale gravitational infall from the surrounding environment onto the filament and the dense central region. We calculated the velocity gradients from these spectral lines, which range from $\sim$ 9 to $\sim$ 226 $\mathrm{km~s^{-1}~pc^{-1}}$, and the corresponding mass infall rates, which vary from a few $\times$ 10$^{-4}$ to 10$^{-3}$\ M$_{\odot}\mathrm{yr^{-1}}$.\\

(iii) Taking the uncertainties into account, three cores exhibit AM values close to $+$1 consistently across all four spectral lines. Five cores show AM values close to $−$1 within the uncertainties, two of which are associated with molecular outflows. For seven cores, reliable AM values could not be determined because of the limited sensitivity of the polarization measurements. For the remaining cores, the AM values generally display consistent behavior among the different spectral lines, with most of the lines indicating the same trend and only one spectral line occasionally deviating from it. In a few cases (three to four cores), these deviations are likely attributable to measurement uncertainties. Overall, the AM values derived from the four spectral lines show a fairly consistent picture across the sample.\\

Overall, this study highlights the interplay of magnetic fields, turbulence, and gravity in the evolution of the G327.29 protocluster. Future high-resolution observations and similar analyses will further clarify the role of these processes in star formation within massive protoclusters.\\









 \begin{acknowledgements}
We thank the anonymous Reviewer for the valuable and constructive comments, which have helped us significantly improve the manuscript. A.K. would also like to acknowledge Fondecyt postdoctoral fellowship (project id: 3250070, 2025). A.S. gratefully acknowledges support by the Fondecyt Regular (project code 1220610), ANID BASAL project FB210003 and and the China-Chile Joint Research Fund (CCJRF No. 2312). A.L. acknowledges support from the National Science Foundation (NSF) under Grant No. AST-2307840. P.S. was partially supported by a Grant-in-Aid for Scientific Research (KAKENHI Number JP23H01221) of JSPS. P.S. was partially supported by a Grant-in-Aid for Scientific Research (KAKENHI No JP24K17100) of the Japan Society for the Promotion of Science (JSPS). R.A.G. acknowledges support from the STFC (grant ST/Y002229/1). 
\end{acknowledgements}




\bibliographystyle{aa} 
\bibliography{a.bib} 

\noindent\rule{\linewidth}{0.4pt}

\hspace{-5mm}
\hspace{-1mm}$^{1}$Departamento de Astronom\'{i}a, Universidad de Concepci\'{o}n, Casilla 160-C, Concepci\'{o}n, Chile\\
$^{2}$Department of Astronomy, University of Wisconsin-Madison, Madison, WI, USA\\
$^{3}$Institute for Advanced Study, 1 Einstein Drive, Princeton, NJ 08540, USA\\
$^{4}$Department of Astronomy, School of Science, The University of Tokyo, 7-3-1 Hongo, Bunkyo, Tokyo 113-0033, Japan\\
$^{5}$Academia Sinica Institute of Astronomy and Astrophysics, No.1, Sec. 4., Roosevelt Road, Taipei 10617, Taiwan\\
$^{6}$National Astronomical Observatory of Japan, National Institutes of Natural Sciences, 2-21-1 Osawa, Mitaka, Tokyo 181-8588, Japan\\
$^{7}$Scottish Universities Physics Alliance (SUPA), School of Physics and Astronomy, University of St. Andrews, North Haugh, St. Andrews KY16 9SS, UK\\

\hspace{-5mm}Corresponding author: A. Koley \\

\hspace{-5mm}\email{atanuphysics15@gmail.com}\\

\noindent\rule{\linewidth}{0.4pt}

\begin{appendix}

\appendix

\onecolumn


\section{Velocity gradient technique (VGT)}\label{section:A0}

VGT is a new method for tracing the morphology of the magnetic field in different phases of the interstellar medium \citep{LY18, 2019NatAs...3..776H,2022ApJ...934...45Z,2023MNRAS.523.1853S,2024MNRAS.530.1066L,2024MNRAS.528.3897S, LYP24}. The idea of VGT came from the anisotropy of the velocity fluctuation in a magnetized medium \citep{1995ApJ...438..763G,1999ApJ...517..700L}. 
Incompressible hydrodynamic (HD) turbulence follows the standard Kolmogorov law of turbulence, in which the turbulent velocity dispersion varies with length scale ($l$) with a power-law index ($\alpha$) of 0.33 \citep{1941DoSSR..30..301K}. In this case, cascading occurs isotropically. Thus, the magnitude of the velocity fluctuation on a particular length scale is the same in all directions. Likewise, in compressible HD turbulence, where shocks are frequently formed due to the supersonic nature of turbulence, the power law is steeper than the Kolmogorov law of turbulence and the value of $\alpha$ is $\sim$ 0.50 \citep{BURGERS1948171,1983ApJ...272L..45F}. In this case as well, the velocity cascade has no preferred direction.\\

In the case of incompressible magnetohydrodynamic (MHD) turbulence, due to the perturbation of the magnetic field by turbulence, two oppositely moving (parallel and antiparallel to the magnetic fields) wave packets (Elsasser fields) interact with each other. And, as a result, cascade occurs. Now, depending on the number of interactions required for a significant amount of energy cascade, two distinct theories exist. 
If the turbulence is driven at the injection scale with velocity $V_\text{L}$ less than Alfven velocity $V_\text{A}$, i.e., the Alfven Mach number $M_\text{A}=V_\text{L}/V_\text{A}<1$, the turbulence is weak at large scales (see \cite{1999ApJ...517..700L, Galtier00}. This is a strongly anisotropic turbulence with fluctuations happening in perpendicular directions. As the velocity decreases, the scale decreases at a scale $l_\text{tr}=LM_\text{A}^2$, the turbulence gets into the strong regime controlled by the critical balanced condition \citep{1995ApJ...438..763G, 2019smti.book.....X, BL19}. In that condition, the Alfven time-scale ($\tau_{\text{A}}$), the cascading time scale ($\tau_{\text{cas}}$) and the eddy turnover time scale ($\tau_{\text{s}}$) are equal. Here, $L$ is the injection or driving scale and $M_\text{A}$ is the Alfven Mach number.\\

If at the injection scale $L$ the velocity is greater than $V_\text{A}$, i.e. $M_\text{A}>1$, then at large scales the turbulence is mostly hydrodynamic and isotropic with the transition to MHD critically balanced regime taking place at $LM_\text{A}^{-3}$ \citep{2007ApJ...666L..69K}. For weak and strong turbulence, the velocity fluctuations are anisotropic, which results in anisotropic gradients of velocity amplitude. For instance, for sub-Alfv\'enic incompressible turbulence, the corresponding gradient of the velocity amplitude is as follows:







\begin{equation}
\nabla u_{l,\perp}  =  \left( \frac{u_{l,\perp}}{l_\perp}  \right) = \left( \frac{u_{L}}{L}\right) \left(\frac{l_\perp}{L}\right)^{-\frac{2}{3}}M_\text{A}^{\frac{1}{3}} ~,~~~~~~ M_{\text{A}} \leq 1.
\end{equation}

Here, $u_{l,\perp}$ is the velocity dispersion of the eddy at a length scale $l$ perpendicular to the magnetic field, $l_{\perp}$ is the length scale of the eddy perpendicular to the magnetic field, and $u_\text{L}$ is the velocity dispersion at the injection scale $L$. As, at lower scale, eddy is highly anisotropic means that the length scale along the magnetic field ($l_{\parallel}$) is $>>$  length scale of the eddy perpendicular to the magnetic field ($l_{\perp}$),
the maximum gradient of velocity fluctuation (locally) in the case of MHD turbulence is perpendicular to the direction of the magnetic field. The magnetic field in this case is the magnetic field affecting the eddies at hand, i.e., this is the local magnetic field \citep{1999ApJ...517..700L, CV00}. In other words, MHD turbulence reveals the directions of the local magnetic field in a way that is similar to how polarization samples the magnetic fields in a magnetized volume. This analogy is not exact, as gradients are not sensitive to the mean magnetic field, but it is useful \citep{LYP24}. For superAlfv\'enic turbulence, the magnetic field advection by eddies is essential. Nevertheless, the velocity gradient stays perpendicular to magnetic field \citep{HoL24}.\\


\hspace{-5mm}Now, the velocity gradient in each pixel in the image cube is calculated based on the theoretical concept of velocity caustic effect \citep{2000ApJ...537..720L}. According to this effect, in thin channels where the channel width ($\Delta_{\text{ch}}$) is less than the turbulent velocity dispersion ($\sigma_{\text{turb}}$), the gradient of the intensity fluctuation in each channel generates a gradient of velocity fluctuation. Consequently, the measurement of the intensity gradient in each thin channel in the image cube determines the gradient of velocity fluctuation. Each individual channel map in the image cube $I_\text{i}$ ($x$,$y$) ($\text{i}$=1,2,3....,$n$) is convolved with 3 $\times$ 3 Sobel convolutional kernels:

\begin{equation}
\nabla_\text{x} I_\text{i} (x,y) = s_\text{x} * I_\text{i} (x,y), ~~~~~\nabla_\text{y} I_\text{i} (x,y) = s_\text{y} * I_\text{i} (x,y), 
\end{equation}

\hspace{-5mm}$s_\text{x}$ = $\begin{pmatrix}
-1 & 0 & +1\\
-2 & 0 & +2\\
-1 & 0 & +1
\end{pmatrix}$~, ~~~ $s_\text{y}$ = $\begin{pmatrix}
-1 & -2 & -1\\
   ~~0 & ~~0 & ~~0\\
+1 & +2 & +1\\
\end{pmatrix}$.\\

\vspace{4mm}
\hspace{-5mm}Here the asterisk symbol denotes the convolution. $\nabla_\text{x} I_\text{i} (x,y)$ and $\nabla_\text{y} I_\text{i} (x,y)$ represent the pixel-wise gradient in the $i^{th}$ channel in the horizontal (X) and vertical (Y) directions, respectively. These are then used to calculate the pixel-wise orientation of the velocity gradient $\psi_\text{g}^\text{i} (x,y)$ using the formula:

\begin{equation}
    \psi_\text{g}^\text{i} (x,y) = tan^{-1} \left(\frac{\nabla_\text{y} I_\text{i} (x,y)}{\nabla_\text{x} I_\text{i} (x,y)}\right).
\end{equation}\\

After obtaining the pixel-wise velocity gradient angle, the following steps are necessary to obtain the orientation of POS magnetic fields:\\

\hspace{-5mm}(i) First, it is necessary to divide the pixel-wise angle map into 20 $\times$ 20 subblocks. We increase the sub-block size up to  60 $\times$ 60 subblocks and find that, due to over-smoothing, the complex structures may become smeared out. For further details, we refer the reader to \cite{2023MNRAS.523.1853S}. We also tested a smaller sub-block size of 10 $\times$ 10 pixels; however, the method was unable to reliably determine the most probable $\psi_\text{g,A}^{i}$ in each sub-block.\\

\hspace{-5mm}(ii) In the next step, it is required to plot the histogram of $\psi_\text{g}^\text{i}$ for each sub-block and compute the most probable value of $\psi_\text{g,A}^\text{i}$ after fitting a Gaussian profile to the histogram.\\

\hspace{-5mm} Thereafter, it is necessary to perform the same procedure for each channel to obtain the pseudo-Stokes parameters $Q_\text{g}$ and $U_\text{g}$ for each pixel ($x$,$y$) using the following formulae:

\begin{equation}
Q_\text{g} (x,y)=  \sum_{\text{i}=1}^{n} I_\text{i} (x,y)~~\text{cos}~(2 \psi_\text{g,A}^\text{i} (x,y)),
\end{equation}

\vspace{-3mm}
\begin{equation}
U_\text{g} (x,y)=  \sum_{\text{i}=1}^{n} I_\text{i} (x,y)~~\text{sin}~(2 \psi_\text{g,A}^\text{i} (x,y)).
\end{equation}

\hspace{-5mm}After that, the velocity gradient angle $\psi_\text{g} (x,y)$ in each pixel ($x$,$y$) is calculated as follows:

\begin{equation}
    \psi_\text{g} (x,y) =\frac{1}{2} \text{tan}^{-1} \left( \frac{U_\text{g}(x,y)}{Q_\text{g}(x,y)}  \right).
\end{equation}

\hspace{-5mm}Finally, the plane-of-sky (POS) magnetic field orientation, $\psi_\text{B}$ in each pixel ($x$,$y$) is obtained from the velocity gradient angle by $\psi_\text{B} (x,y)$ = $\psi_\text{g} (x,y)$ + $\pi$/2.\\

However, in a gravity-dominated system or in the presence of strong outflow, the VGT rotates further by 90$^{\circ}$ in respect to the magnetic field. In the presence of strong gravity, the magnetic field frozen into the gas adjusts itself so that its direction aligns with the velocity gradient. This interpretation assumes that the magnetic field is sufficiently coupled to the gas through flux freezing and that gravitational collapse proceeds on a timescale comparable to or shorter than competing dynamical processes. Under these conditions, gravitationally driven motions can drag the magnetic field lines, resulting in an increased alignment between the magnetic field orientation and the local velocity gradient. A similar situation occurs in the presence of strong outflows, where the poloidal magnetic field becomes aligned with the velocity gradient \citep{2020ApJ...897..123H,2024MNRAS.530.1066L}. As a result, the direction of the magnetic field obtained from the VGT and the dust polarization will be perpendicular to each other \citep{2020ApJ...897..123H}. The angular measurement (AM) parameter is used to determine whether the region is dominated by gravity or turbulence. This parameter is defined by \citep{2017ApJ...835...41G}:

\begin{equation}
\text{AM}~(x,y)  = 2~ \left(\text{cos}^{2}\theta_{\text{r}}(x,y)-\frac{1}{2}\right).
\end{equation}

Here, $\theta_{\text{r}}(x,y)$ = $|\psi_{\text{B}} (x,y) - \psi_{\text{D}} (x,y)|$, is the difference in the orientation of the $B_{\text{pos}}$ obtained from VGT ($\psi_{\text{B}} (x,y)$) and dust polarization ($\psi_{\text{D}} (x,y)$), where $\psi_{\text{D}} (x,y)$ = $\psi_{\text{p}} (x,y)$ +  $\pi$/2. $\psi_{\text{p}} (x,y)$ is the spatial distribution of the dust polarization angle. In the turbulence-dominated medium, $\theta_{\text{r}}$ = 0$^\circ$ and therefore AM = +1 and in the gravity-dominated medium, $\theta_{\text{r}}$ = 90$^\circ$ and thus AM = -1. Consequently, AM measurements provide information on the spatial characteristics of the region, such as whether it is dominated by turbulence or gravity.\\


\section{Spectral-line decomposition via \texttt{Gausspy+}}\label{section:A1}

\texttt{Gausspy+} module \citep{2019A&A...628A..78R} is based on the automated Gaussian decomposition (AGD) method. The module utilizes two spectral smoothing parameters, \texttt{decompose.alpha1} and \texttt{decompose.alpha2}, which are calculated automatically from the observed spectra. It is important to note that the smoothing is applied spectrally, not spatially. Spectral smoothing and derivatives (up to fourth order) are used to 
accurately locate the peak positions of the decomposed components. The original unsmoothed spectra were then decomposed using these estimated peak positions. Although using both parameters can improve decomposition for spectra with weak or narrow components, a single smoothing parameter is often 
sufficient. The $F_{1}$ score evaluates the accuracy of the decomposition on the training set. Additional parameters in the module include \texttt{significance}, \texttt{snr\_noise\_spike}, \texttt{refit\_rchi2}, and \texttt{min\_fwhm}. Here, \texttt{min\_fwhm} sets the minimum number of channels for a Gaussian component, which we fixed to 2 to ensure appropriate sampling of the spectral profile. The two smoothing parameters, \texttt{decompose.alpha1} and \texttt{decompose.alpha2}, were varied between 1.4 and 4.0 for the four spectral lines. For the remaining parameters, \texttt{significance} was fixed at 6 for all lines, while \texttt{snr\_noise\_spike} and \texttt{refit\_rchi2} were set to 6, 5, and \texttt{True}, respectively. The resulting \texttt{F$_1$} scores ranged from 80\% to 85\% for the four spectral lines. After pixel-wise fitting, we obtained the peak intensity, central velocity, and full width at half maximum (FWHM) for each component. Using these pixel-wise components for all line cubes, we analyzed the spatial distribution of the component numbers and constructed histograms of the center velocities ($V_{\rm LSR}$) to distinguish different gas components in the region.\\

\section{Spatial distribution and histogram of the number of components ($n$) of the spectral lines}\label{section:A2}
\begin{figure*}
	\includegraphics[width=2.6in,height=2.3in,angle=0]{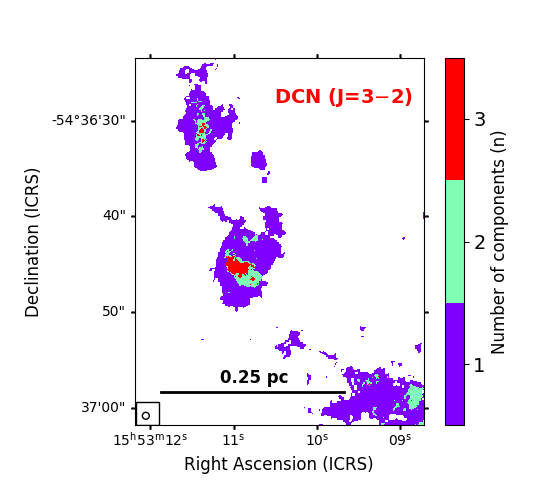}\includegraphics[width=2.6in,height=2.3in,angle=0]{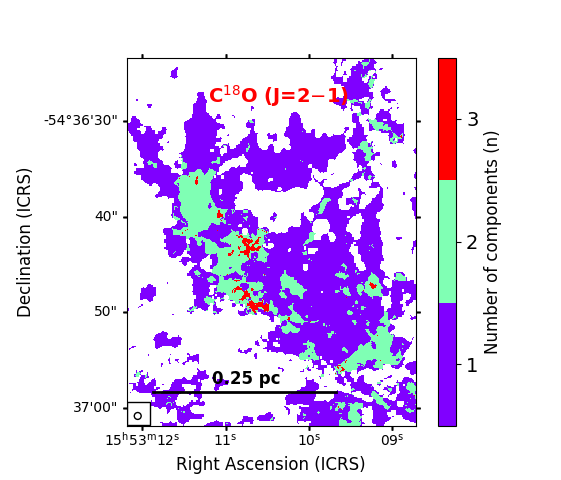}\includegraphics[width=2.6in,height=2.3in,angle=0]{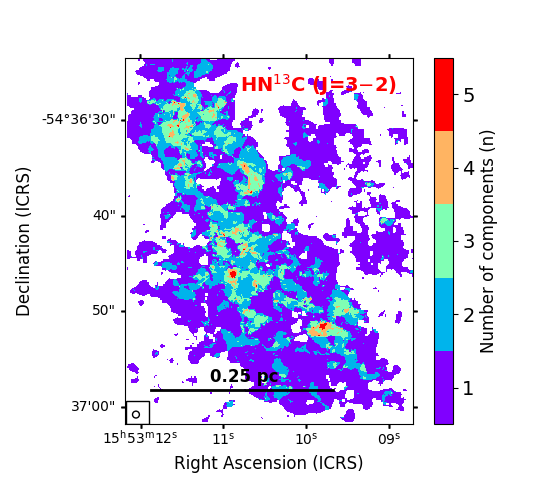}\\ 
    
    \includegraphics[width=3.2in,height=2.3in,angle=0]{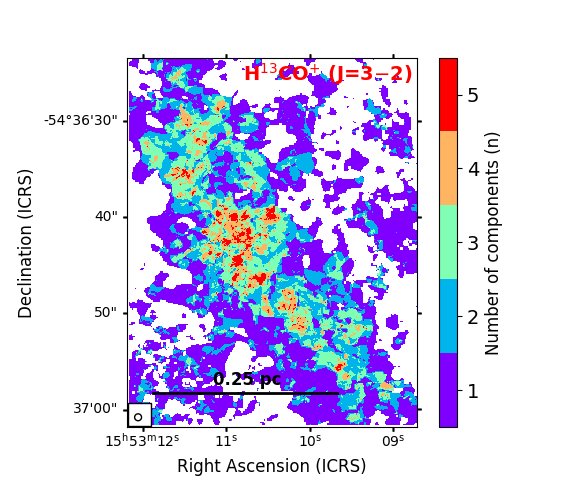} \includegraphics[width=2.2in,height=2.3in,angle=0]{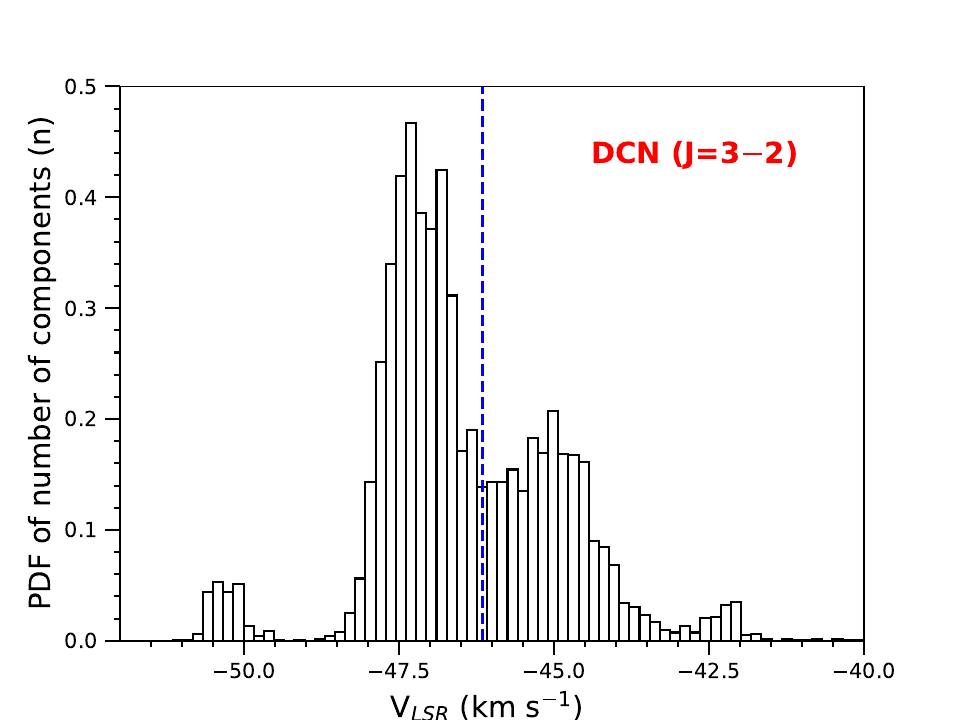}\includegraphics[width=2.2in,height=2.3in,angle=0]{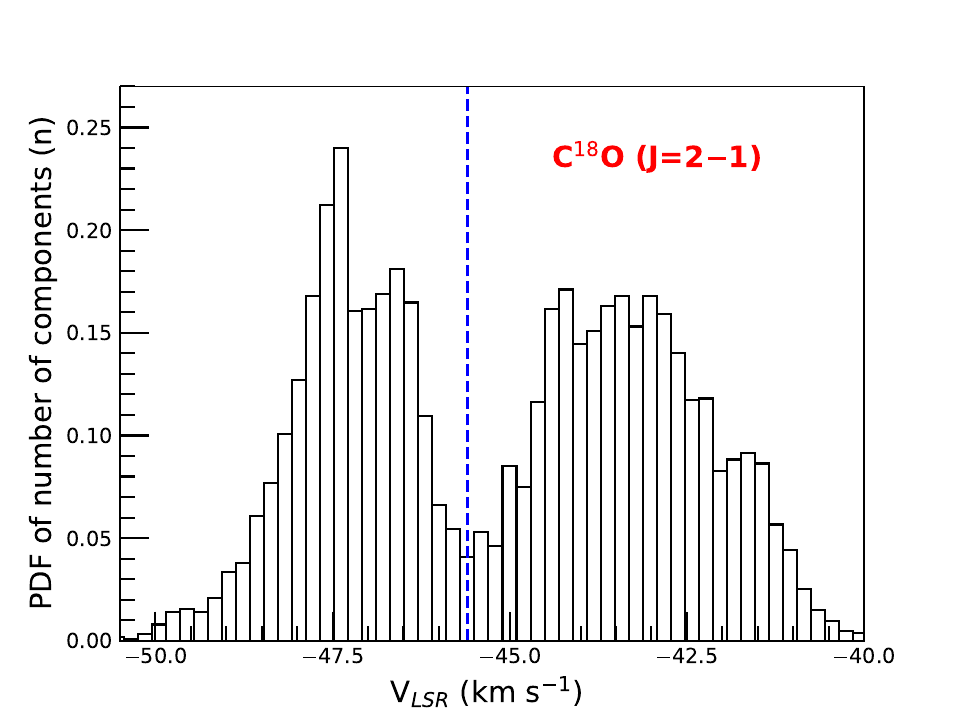}\\
~~~~~~~~~~~~~~~~~~~\includegraphics[width=2.2in,height=2.3in,angle=0]{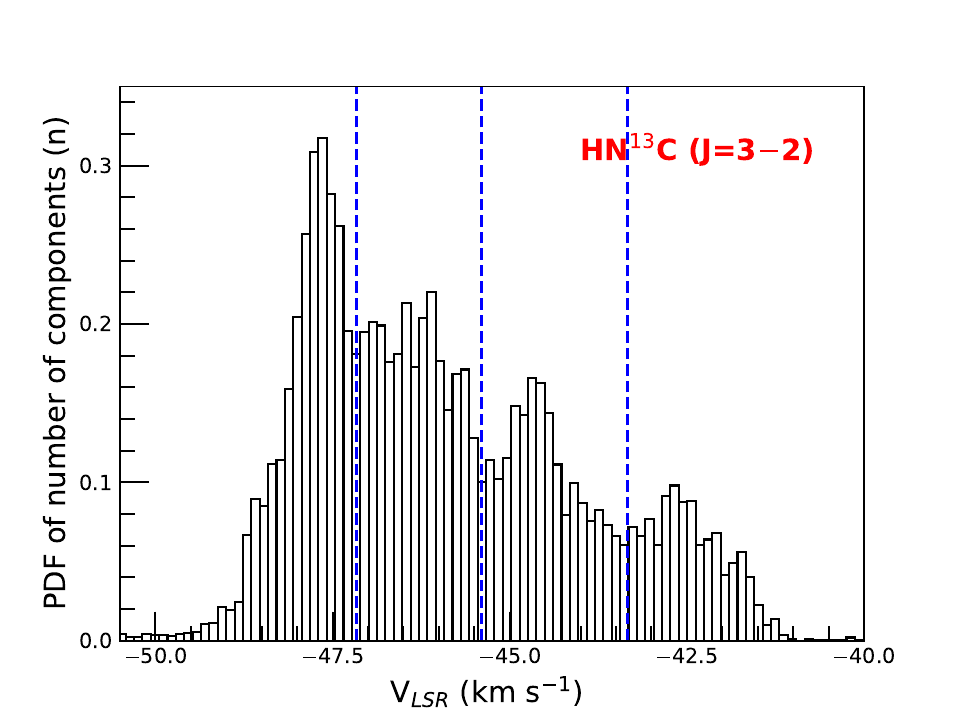}
     \includegraphics[width=2.4in,height=2.3in,angle=0]{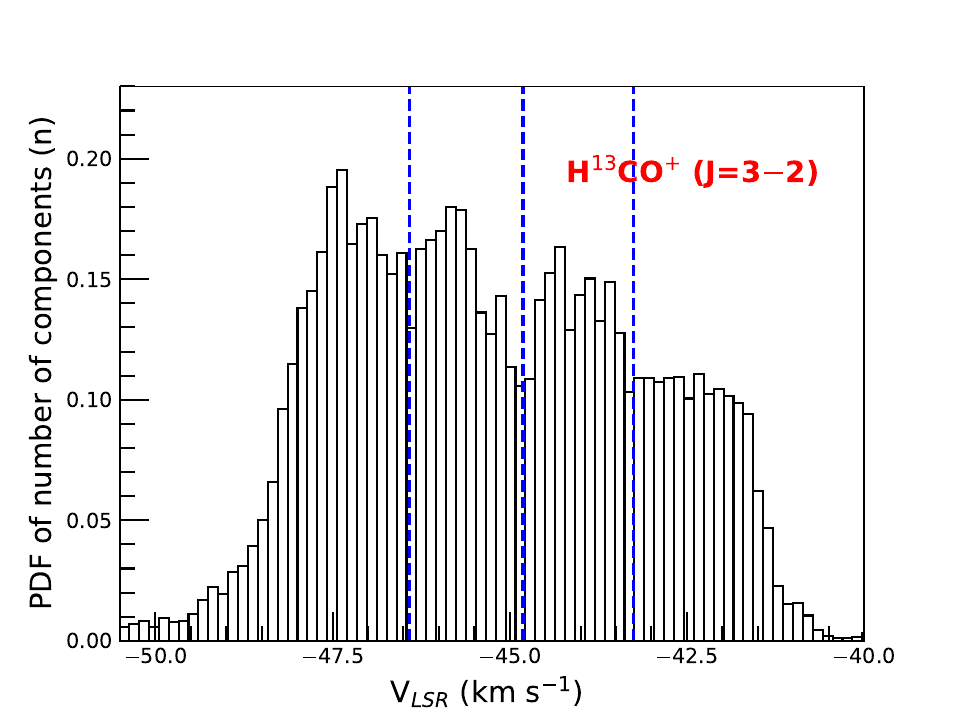}\\

	\caption{Spatial distributions of the number of components ($n$) and histogram plots of the center velocities ($V_{\text{LSR}}$) of  the pixel-wise fitted DCN (3-2), C$^{18}$O (2$-$1), HN$^{13}$C (3$-$2), and H$^{13}$CO$^{+}$ (3-2) line components towards the observed region. }
    \label{fig:figC}
\end{figure*}

Fig. \ref{fig:figC} presents the spatial distribution of the number of components along with the corresponding histograms for all four spectral lines. For the DCN and C$^{18}$O spectral lines, most regions show a single component, but the number of components ($n$) increases toward the dense areas, reaching up to three. The histogram of $n$ shows two prominent peaks separated by a valley at approximately $-$46.1 km s$^{-1}$ for the DCN line and at $-$45.1 km s$^{-1}$ for the C$^{18}$O line. Based on the valley velocity, we construct three model cubes: one containing components with velocities below the valley (\texttt{first component}), one with velocities above the valley (\texttt{second component}), and one including all components (\texttt{total model}). From the histogram plots of $n$ for the HN$^{13}$C and H$^{13}$CO$^{+}$ spectral lines, we identify three valleys, which are marked with blue dashed lines. Based on the velocities at the valley positions, we create four model cubes, denoted as \texttt{first}, \texttt{second}, \texttt{third}, and \texttt{fourth}. We also construct combined model cubes: \texttt{first+second} and \texttt{third+fourth}, along with a \texttt{total model} cube that includes all components.

\section{Dust temperature ($T_{\text{d}}$), Sonic Mach number ($M_{\text{s,3D}}$), Alfven Mach number ($M_{\text{A}}$), Ratio of turbulent velocity dispersion and channel width ($a$) in G327.29 region}\label{section:A3}

\begin{figure*}[hbt!]
	\includegraphics[width=3.6in,height=2.4in,angle=0]{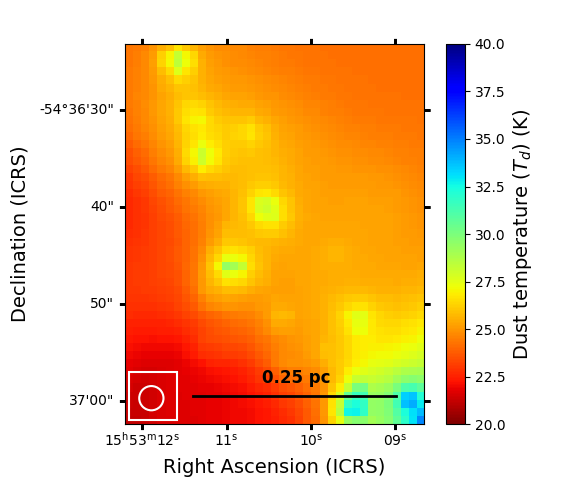}\includegraphics[width=3.2in,height=2.4in,angle=0]{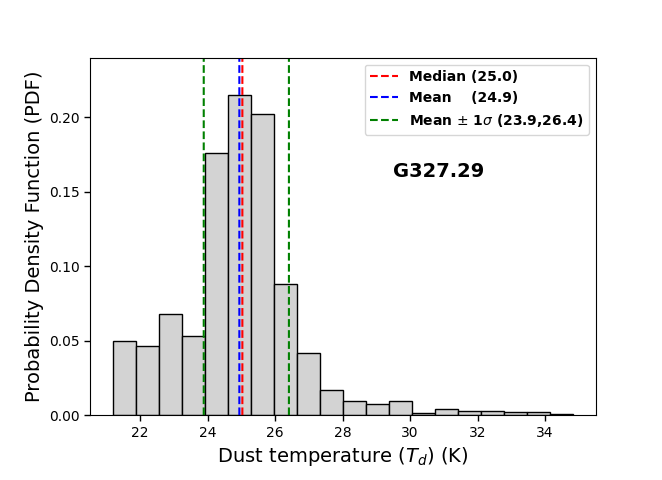}
	\caption{Spatial distribution and histogram plot of dust temperature ($T_{\text{d}}$) map in the G327.29 region.}
    \label{Fig:FigA2}
\end{figure*}

We show the spatial distribution and the histogram plot of dust temperature ($T_{\text{d}}$) in the G327.29 protocluster in Fig. \ref{Fig:FigA2}. The median and mean values of $T_{\text{d}}$ are 25.0 K and 24.9 K, respectively.  Furthermore, we show the spatial distributions and histogram plots of the effective velocity dispersion ($\sigma_{\text{eff}}$) for the DCN (3$-$2), C$^{18}$O (2$-$1), HN$^{13}$C (3$-$2) and H$^{13}$CO$^{+}$ (3$-$2) lines in Fig. \ref{Fig:FigB1}. The median and mean values of $\sigma_{\text{eff}}$ for the DCN line are 0.72 and 0.83 km s$^{-1}$, respectively. Likewise, the median and mean values of $\sigma_{\text{eff}}$ for the C$^{18}$O (2$-$1), HN$^{13}$C (3$-$2) and H$^{13}$CO$^{+}$ (3$-$2) lines are 0.56, 0.65, 0.44, 0.50, 0.52 and 0.60, respectively.\\


For calculating the $\sigma_{\rm eff}$ we use the following formula \citep{2025A&A...702A.133K}:\\

\begin{equation}
  \sigma_{\text{eff}} = \sqrt{\frac{w_{1}}{w} \sigma^{2}_{1} + \frac{w_{1}}{w} \sigma^{2}_{2}+ ..............}~~~~~~~~.
\end{equation}

Here, $w_{1}, w_{2}, \dots$ represent the integrated intensities of individual components, and $w$ is the total integrated intensity of the spectra. The mean values of $\sigma_{\rm eff}$ are 0.83, 0.65, 0.50, and 0.60 km s$^{-1}$ for DCN, C$^{18}$O, HN$^{13}$C, and H$^{13}$CO$^{+}$, respectively. After subtracting the thermal contribution ($\sigma_{\rm th}$), the resulting non-thermal velocity dispersions ($\sigma_{\rm nth}$) are 0.826, 0.645, 0.493, and 0.594 km s$^{-1}$ for the respective lines. 
We also calculate the three-dimensional sonic Mach number ($M_\text{s,3D}$) in the VGT–examined region, which ranges from 2.9 to 4.8 across the different spectral lines, indicating that the turbulence in this region is supersonic. To determine the Alfvén Mach number ($M_{\rm A}$), we require the non-thermal velocity dispersion ($\sigma_{\rm nth}$), the mass density ($\rho$), and the magnetic field strength ($B$). The values of $\rho$ and $B$ were estimated from the 1.2 mm dust continuum emission, yielding $\rho \sim 7.5 \times 10^{-18}$ g cm$^{-3}$ and $B \sim 2.7$ mG \citep{2026arXiv260323747K}. This mass density corresponds to a number density of $n \sim 1.9 \times 10^{6}$ cm$^{-3}$, assuming a mean molecular weight of $\mu=2.37$. Since direct measurements of the gas density associated with individual molecular tracers are challenging, we adopt the corresponding critical densities ($n_{\rm crit}$) of these spectral lines at 20 K as a proxy for the density traced by each transition. This temperature is appropriate for the region, where the mean dust temperature ($T_\text{d}$) derived from the continuum emission is $\sim 25$ K. In Section~\ref{section_3}, we mention the critical densities ($n_{\rm crit}$) of these spectral lines at 20 K. To estimate the magnetic field strength corresponding to these densities from the dust-derived density, we assume a power-law relation between the magnetic field strength and gas density: $B \propto \rho^\alpha$. We consider two different values of $\alpha$. The first value, $\alpha=0.66$, is adopted from the work of \citet{2010ApJ...725..466C}, which is derived for the high-density regime ($n \geq$ 300 cm$^{-3}$). This value is obtained under the assumption of a purely isotropic gravitational collapse \citep{2015MNRAS.451.4384T}. The other one is taken from \citet{2025MNRAS.540.2762W}, who obtained $\alpha=0.50$ from Zeeman splitting observation, respectively. This value is consistent with magnetically regulated gravitational contraction \citep{2015MNRAS.451.4384T}. For these two values of $\alpha$, we estimate the magnetic field strength corresponding to the critical densities of each molecular transition. Using these density-dependent magnetic field strengths, we then calculate the corresponding ranges of the Alfvén Mach number ($M_{\rm A}$), which are listed in Table~\ref{tab:table3}. The values of $M_\text{A}$ for these lines range from 0.24 to 0.93, indicating that the system is sub-Alfv\'enic.  It is important to note that the $M_\text{A}$ range quoted here is obtained under the adopted values of $n_\text{crit}$ and the magnetic-field strength. The resulting values represent region-averaged estimates and should not be interpreted as spatially resolved measurements of the local Alfv\'enic Mach number. Thus, the possibility of locally super-Alfv\'enic regions cannot be excluded with the present data, because such a conclusion would require spatially resolved measurements of the magnetic-field strength. Consequently, the physical interpretation of the observed VGT--polarization relation in terms of the underlying magnetohydrodynamic regime depends on the local applicability of the adopted assumptions for density and magnetic-field strength. \\


Next, we evaluate the ratio of the channel width ($\Delta_{\rm ch}$) to the turbulent velocity dispersion ($\sigma_{\rm nth}$), which we define as $a$. The channels are classified as \texttt{thin} if $a < 1$ and \texttt{thick} if $a > 1$. The velocity–caustic effect is relevant in thin channels, where intensity fluctuations directly reflect velocity fluctuations within each channel. For the four spectral line cubes, we find that $a$ ranges from 0.11 to 0.43, indicating that the velocity–caustic effect is present. All measured quantities—$T_{\rm d}$, $\sigma_{\rm eff}$, $\sigma_{\rm nth}$, $M_\text{s,3D}$, $M_{\rm A}$, and $a$—are summarized in Table~\ref{tab:table3}. Based on the values of $M_{\rm A}$ and $a$, it is evident that the observed region within the G327.29 protocluster exhibits both anisotropic velocity fluctuations and a significant velocity–caustic effect. Consequently, the velocity gradient can be calculated at each pixel of the spectral line cubes to infer the plane-of-sky magnetic field direction ($B_{\rm pos}$).\\

\begin{table*}[ht!]
	\caption{Physical properties of the region derived from the dust continuum emission and molecular spectral line observations.}
	\begin{tabular}{  c c c c c c c c c}
		\hline
        
		Spectral lines & $T_{\text{d}}$ (K)  & $\sigma_{\text{eff}}$ (km s$^{-1}$)  &$\sigma_{\text{nth}}$ (km s$^{-1}$)  & $M_{{\text{s,3D}}}$  & $B$ (mG) &  $n_\text{crit}$ (cm$^{-3}$)& $M_{\text{A}}$ &  $a$ \\ [0.5 ex]
        \hline
         DCN (3$-$2)   & 24.9               & 0.83                                     & 0.826                                   &  4.8              & [7.6--10.6] & 1.5 $\times$ 10$^{7}$  &   [0.37--0.52]            &  0.41       \\ [0.5 ex] 
        C$^{18}$O (2$-$1)  & 24.9         & 0.65                                    &  0.645                                  &    3.7             & [0.08--0.19]   &  9.93 $\times$ 10$^{3}$   &  [0.40--0.93]              &  0.43         \\ [0.5 ex] 
         HN$^{13}$C (3$-$2)  & 24.9           &  0.50                                    & 0.493                                    &2.9                &   [6.2--8.1]    & 1.0 $\times$ 10$^{7}$  &   [0.24--0.31]            &0.14         \\ [0.5 ex] 
         H$^{13}$CO$^{+}$ (3$-$2)  & 24.9    &  0.60                                   &     0.594                                &  3.5              & [0.9--1.7]   & 9.5 $\times$ 10$^{5}$    &  [0.37--0.41]             &  0.11        \\ [0.5 ex] 
         
		\hline

		\hline
        \label{tab:table3}
	\end{tabular}
	
	\vspace{2mm}
	
\textbf{Notes.} Column 1: Name of the molecular spectral line. Column 2: Mean dust temperature ($T_{\rm d}$) adopted from \cite{2024A&A...687A.217D}. Column 3: Effective velocity dispersion ($\sigma_{\rm eff}$) derived from the molecular spectral line. Column 4: Non-thermal velocity dispersion ($\sigma_{\rm nth}$). Column 5: Sonic Mach number ($M_\text{s,3D}$). Column 6: Magnetic field strength ($B$) estimated using the two $B$--$n$ scaling relations reported by \cite{2010ApJ...725..466C} and \cite{2025MNRAS.540.2762W}. Column 7: Critical density ($n_{\rm crit}$) of the molecular transition at $\sim$20 K. Column 8: Alfven Mach number ($M_{\rm A}$). Column 9: Ratio of the spectral channel width ($\Delta_{\rm ch}$) to the non-thermal velocity dispersion ($\sigma_{\rm nth}$). If this ratio is denoted by $a$, values of $a$ $<$ 1 are classified as \texttt{thin slices}, whereas those with $a$ $>$ 1 are classified as \texttt{thick slices}. 
\end{table*}

\begin{figure*}[hbt!]
	\includegraphics[width=3.4in,height=2.4in,angle=0]{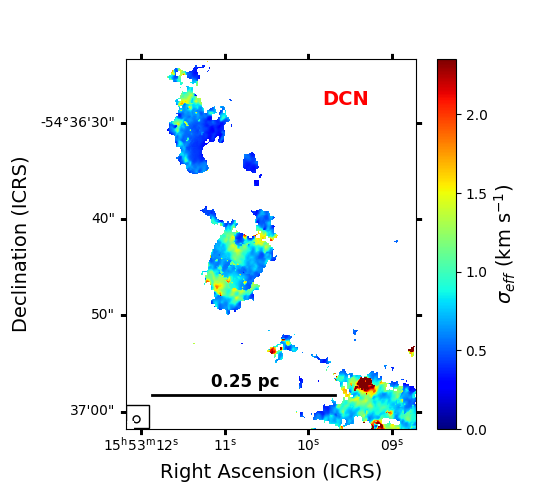}\includegraphics[width=3.2in,height=2.4in,angle=0]{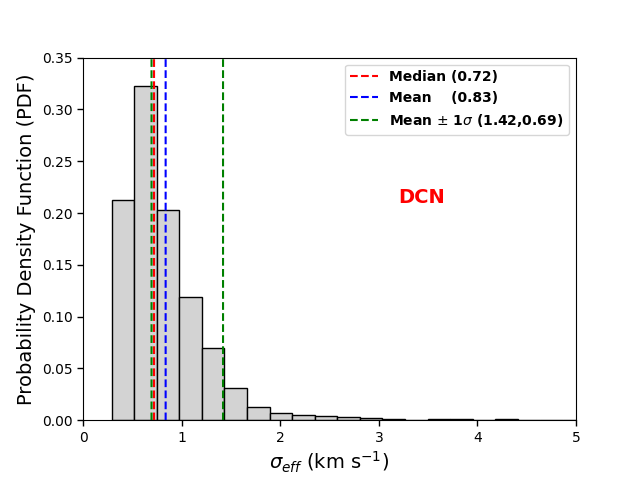}\\
    \includegraphics[width=3.4in,height=2.2in,angle=0]{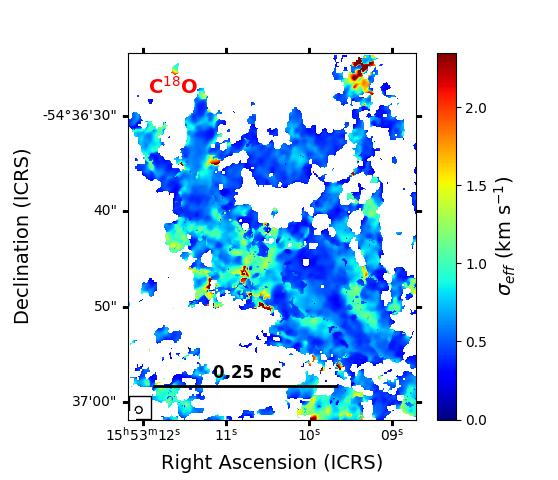}\includegraphics[width=3.2in,height=2.4in,angle=0]{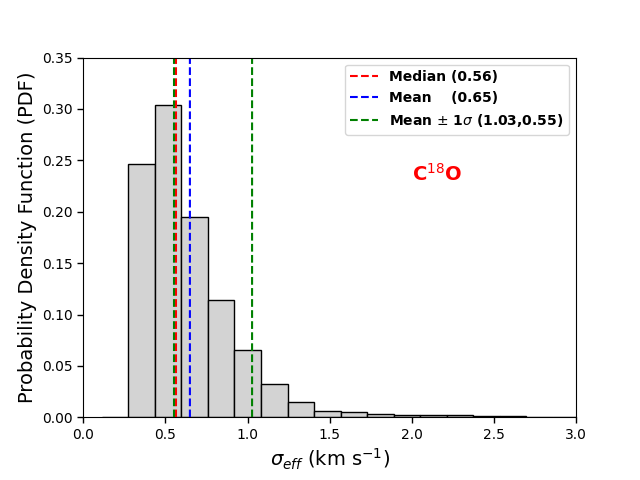}\\
    
   \includegraphics[width=3.4in,height=2.2in,angle=0]{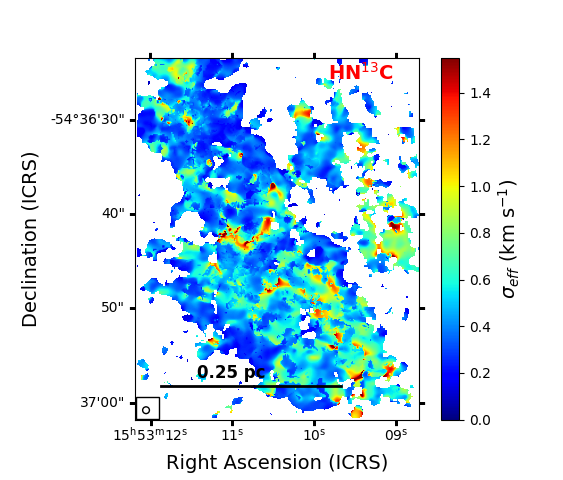}\includegraphics[width=3.2in,height=2.4in,angle=0]{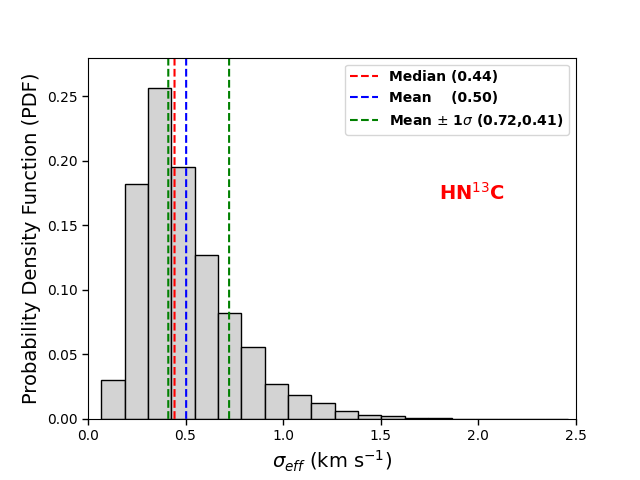}\\
    \includegraphics[width=3.4in,height=2.2in,angle=0]{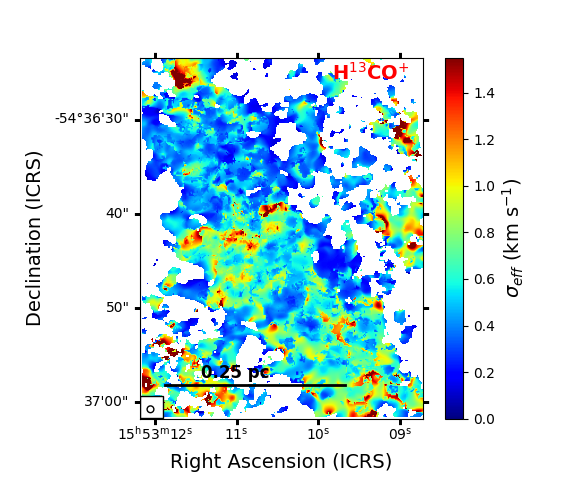}\includegraphics[width=3.2in,height=2.4in,angle=0]{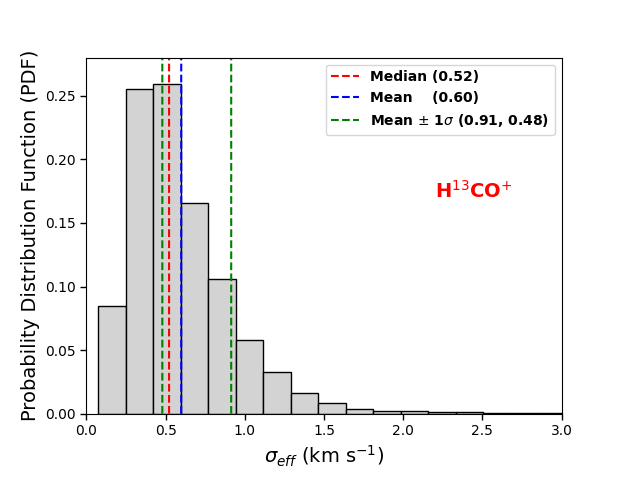}\\
    \caption{Spatial distribution and histogram plot of effective velocity dispersion ($\sigma_{\text{eff}}$) map of DCN, C$^{18}$O, HN$^{13}$C, and H$^{13}$CO$^{+}$ in the G327.29 region.}
    \label{Fig:FigB1}
\end{figure*}

\section{Component-wise VGT for the four spectral lines}\label{Appendix_C}

We show the component-wise VGT for four spectral lines components in Figs. \ref{fig:figC1}, \ref{fig:figE1}, \ref{fig:figE2}, and \ref{fig:figF1}.

\begin{figure*}
    \centering
    \includegraphics[width=2.4in,height=2.2in,angle=0]{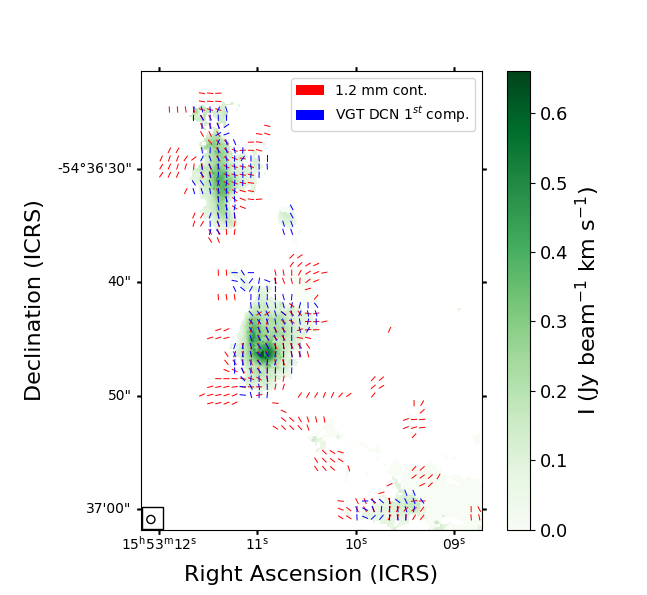}\includegraphics[width=2.4in,height=2.2in,angle=0]{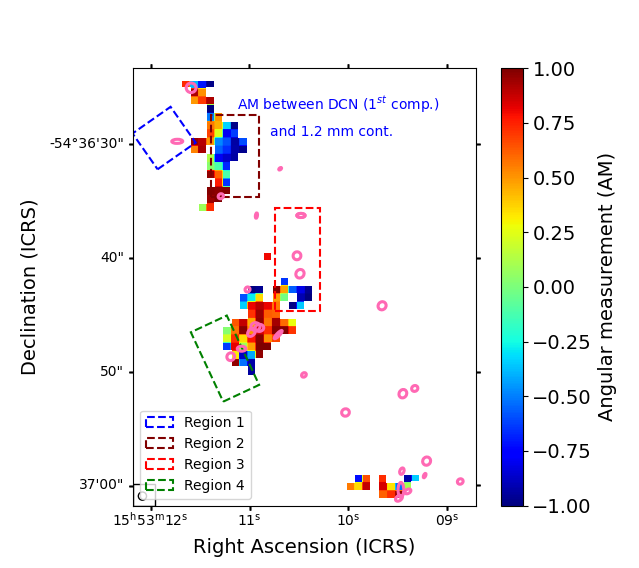}\includegraphics[width=2.3in,height=2.2in,angle=0]{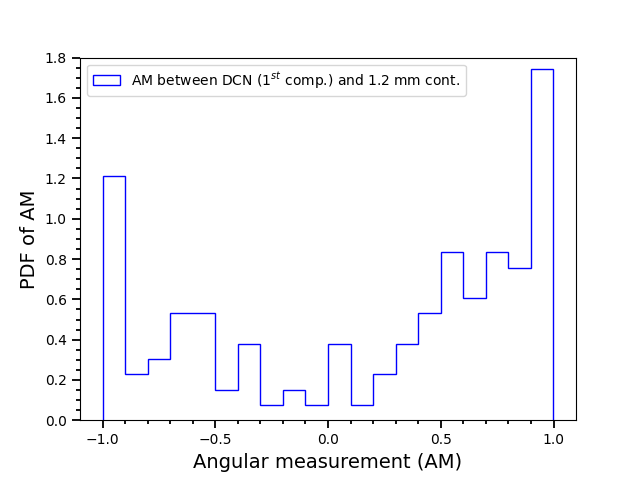}\\ 
    \includegraphics[width=2.4in,height=2.2in,angle=0]{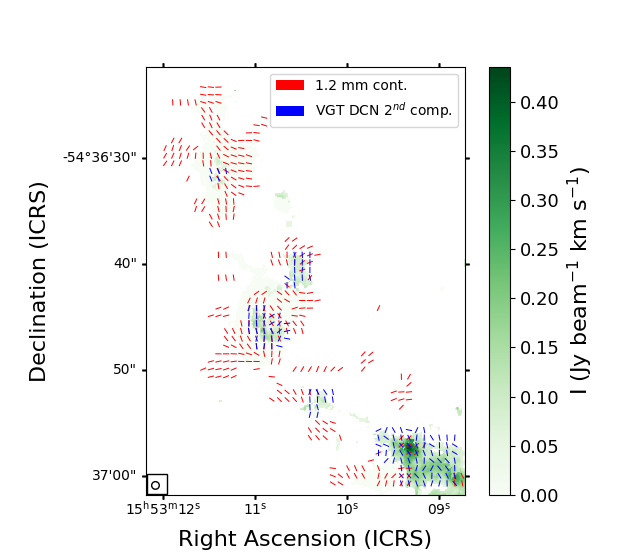} \includegraphics[width=2.4in,height=2.2in,angle=0]{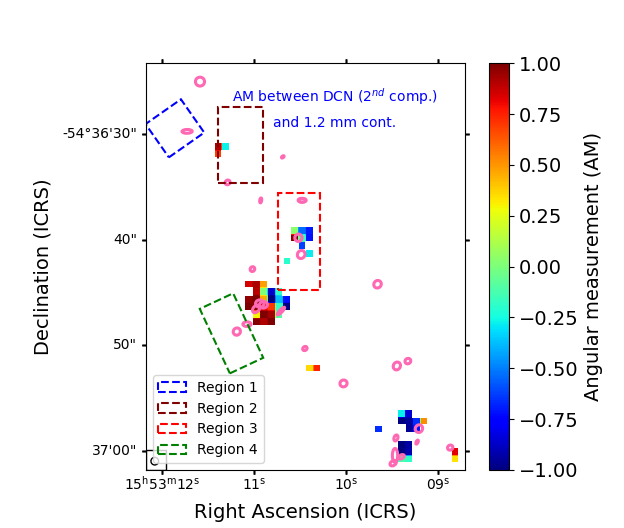}\includegraphics[width=2.4in,height=2.2in,angle=0]{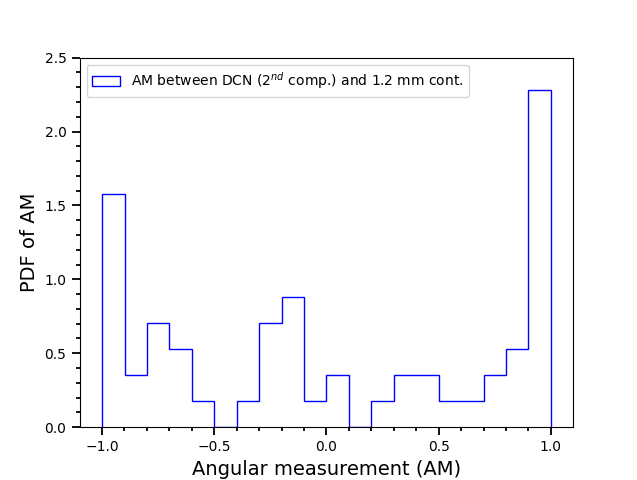}\\
    \includegraphics[width=2.4in,height=2.2in,angle=0]{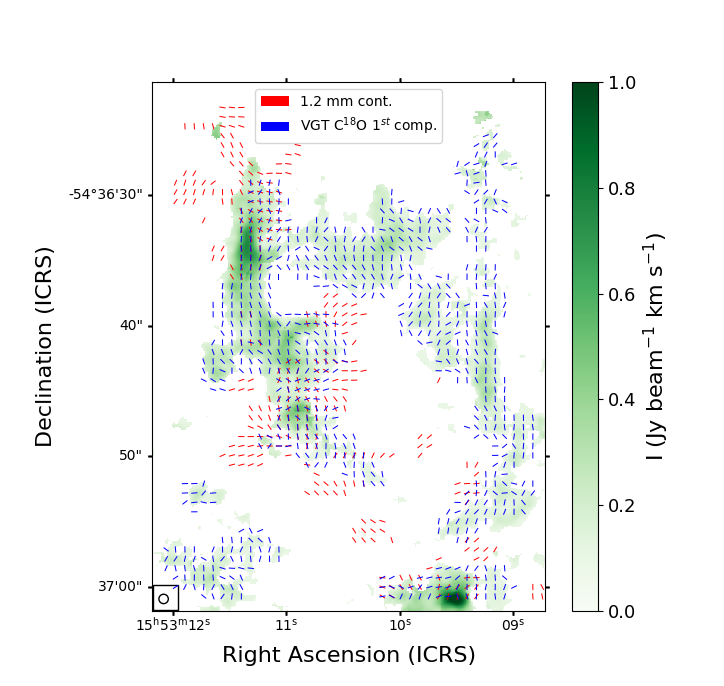}\includegraphics[width=2.4in,height=2.2in,angle=0]{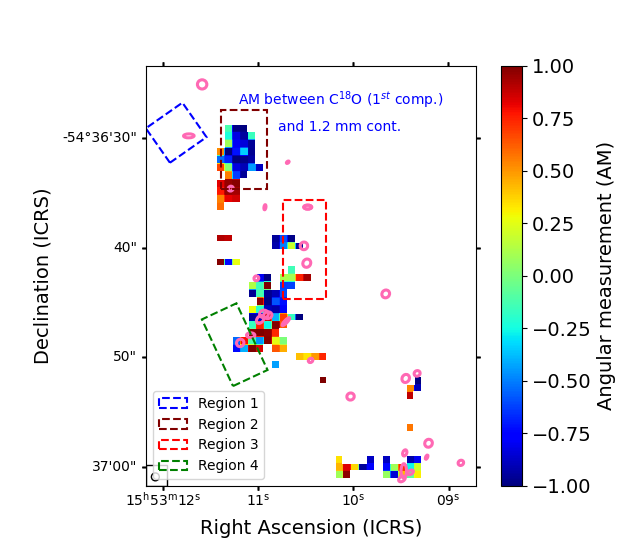}\includegraphics[width=2.4in,height=2.2in,angle=0]{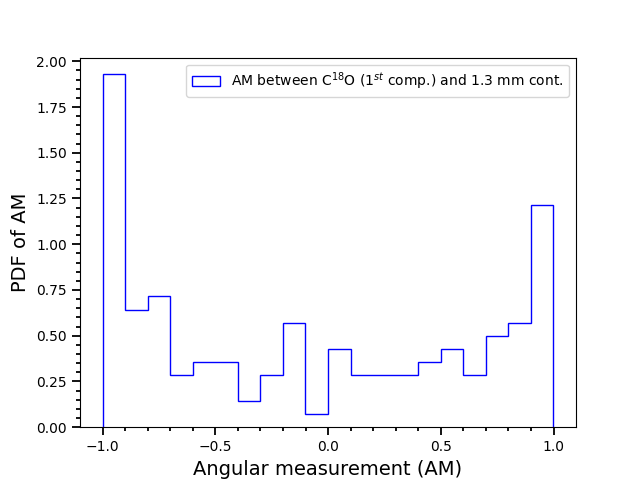}\\ 
    
    \includegraphics[width=2.4in,height=2.2in,angle=0]{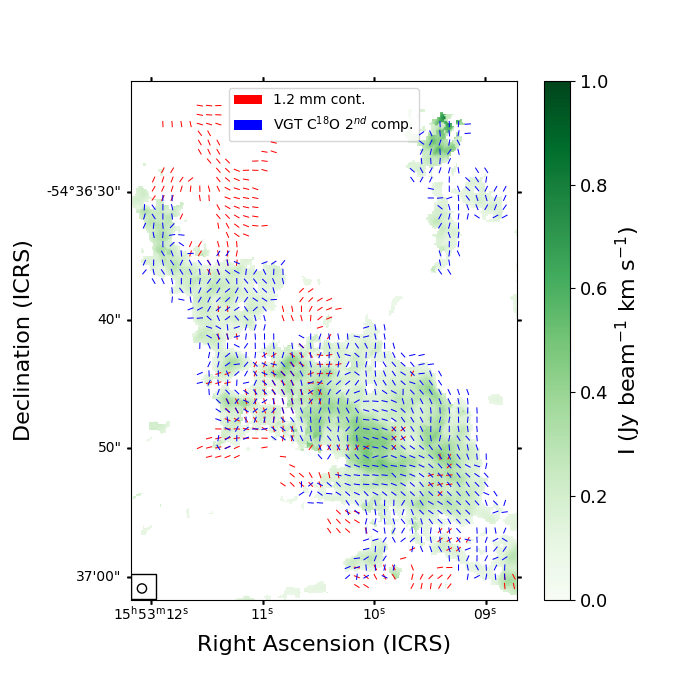} \includegraphics[width=2.4in,height=2.2in,angle=0]{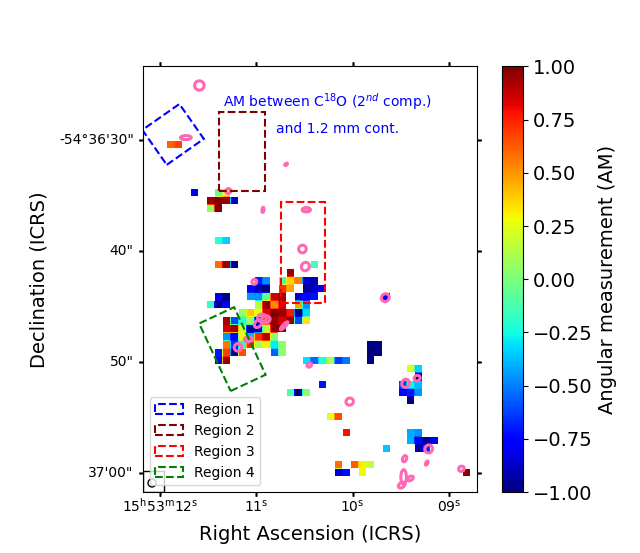}\includegraphics[width=2.4in,height=2.2in,angle=0]{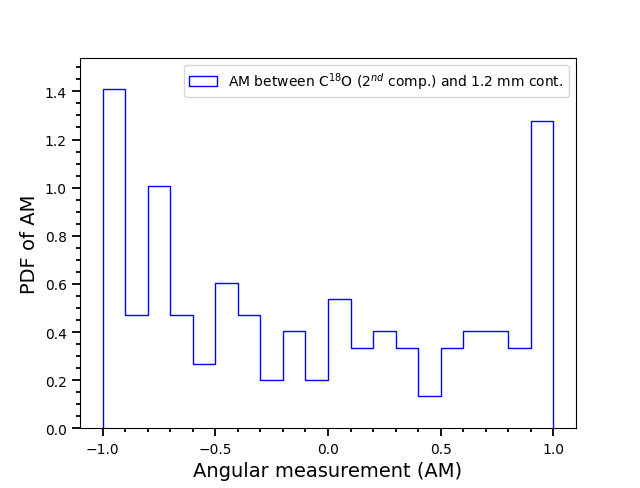}\\
	\caption{Continuation of Figs. \ref{fig:fig3} and  \ref{fig:fig5} but for \texttt{first} and \texttt{second} DCN and C$^{18}$O components. }
    \label{fig:figC1}
\end{figure*}

\begin{figure*}
    \centering
    \includegraphics[width=2.35in,height=2.2in,angle=0]{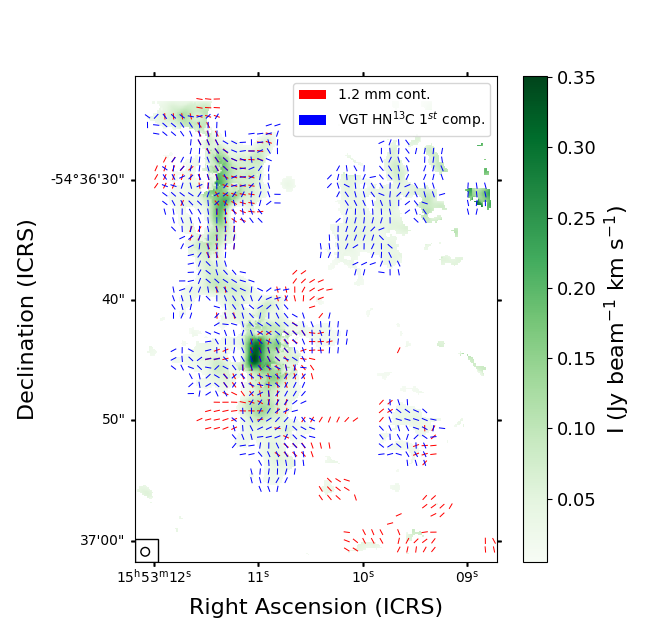} \includegraphics[width=2.4in,height=2.2in,angle=0]{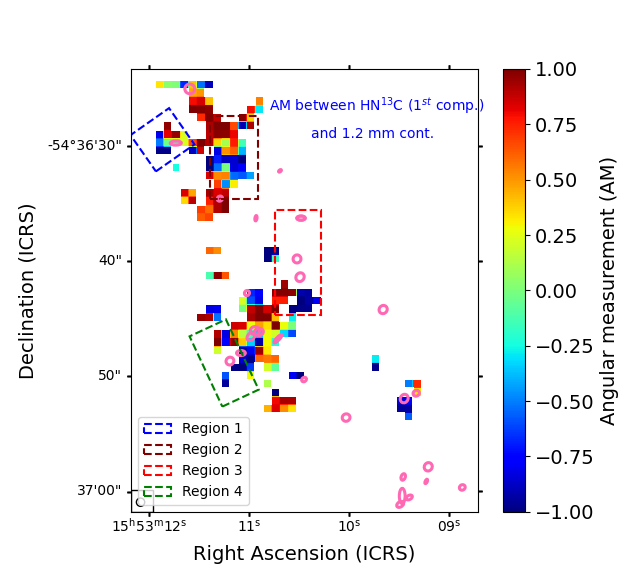}\includegraphics[width=2.4in,height=2.2in,angle=0]{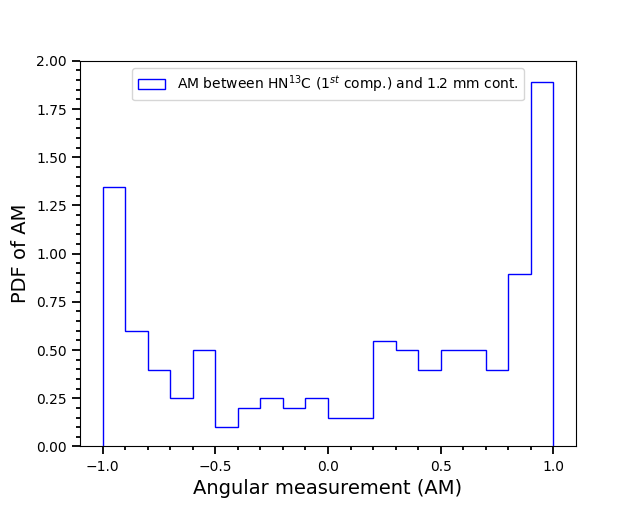}\\ 
    \includegraphics[width=2.4in,height=2.2in,angle=0]{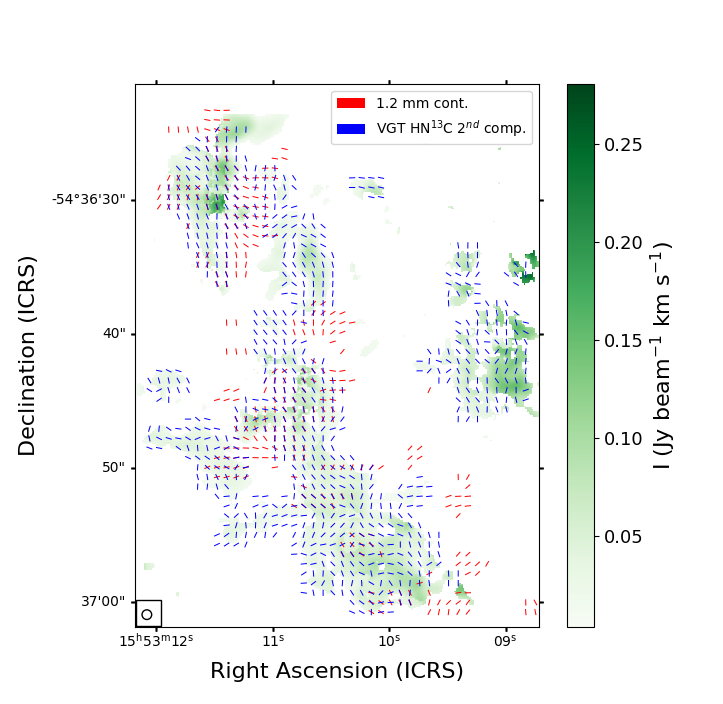} \includegraphics[width=2.4in,height=2.2in,angle=0]{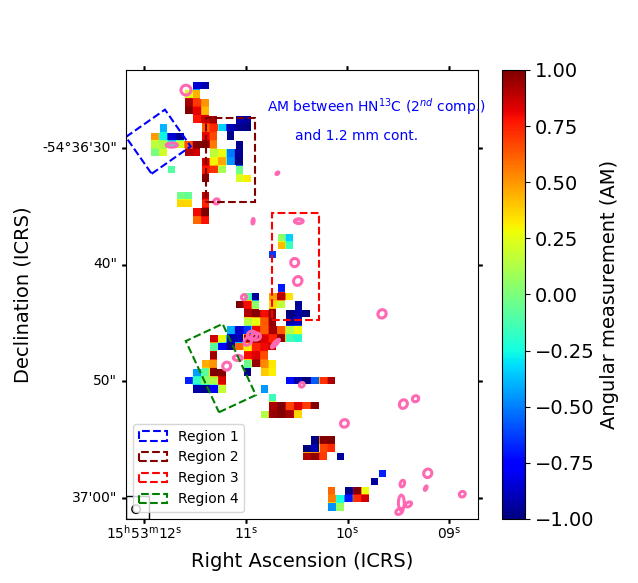}\includegraphics[width=2.4in,height=2.2in,angle=0]{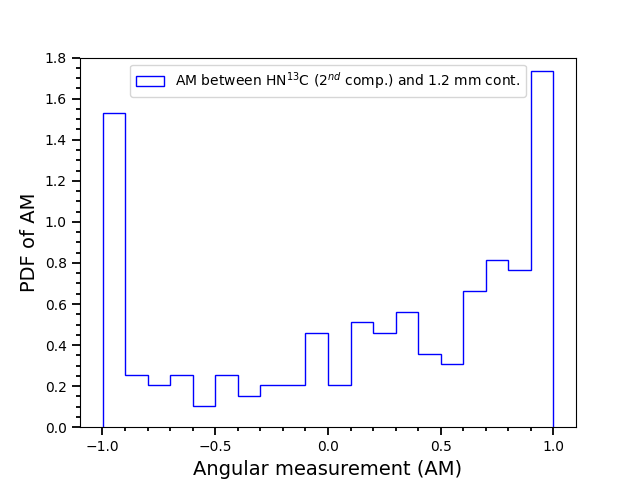}\\
    \includegraphics[width=2.4in,height=2.2in,angle=0]{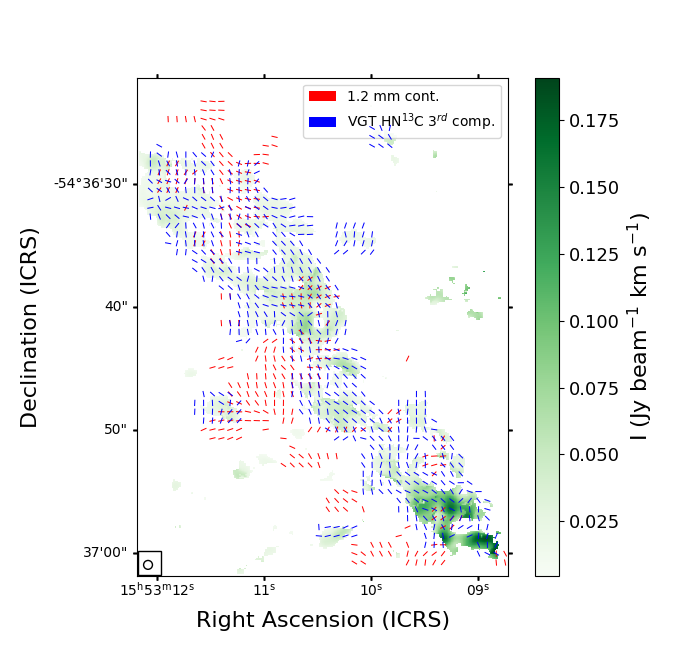} \includegraphics[width=2.4in,height=2.2in,angle=0]{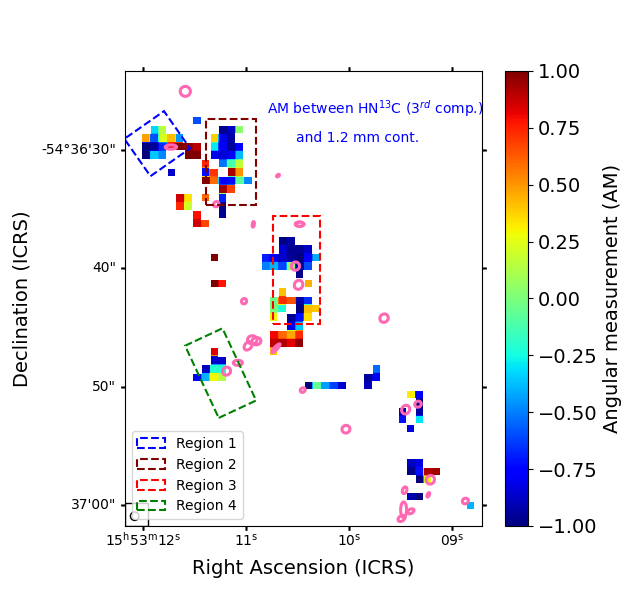}\includegraphics[width=2.4in,height=2.2in,angle=0]{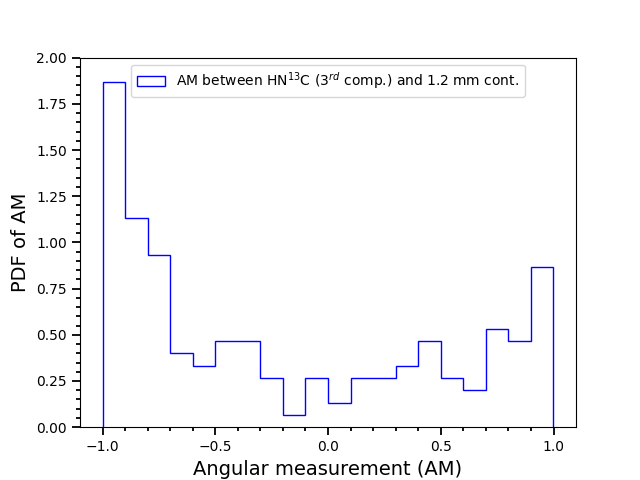}\\ 
    \includegraphics[width=2.4in,height=2.2in,angle=0]{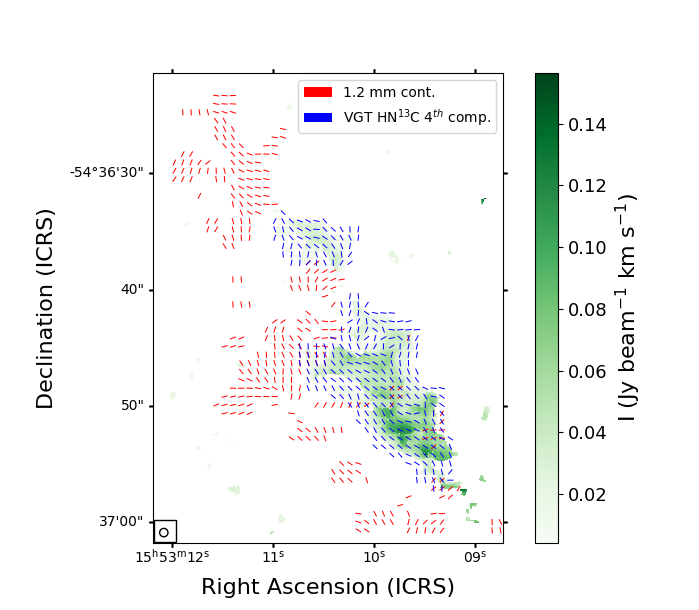} \includegraphics[width=2.4in,height=2.2in,angle=0]{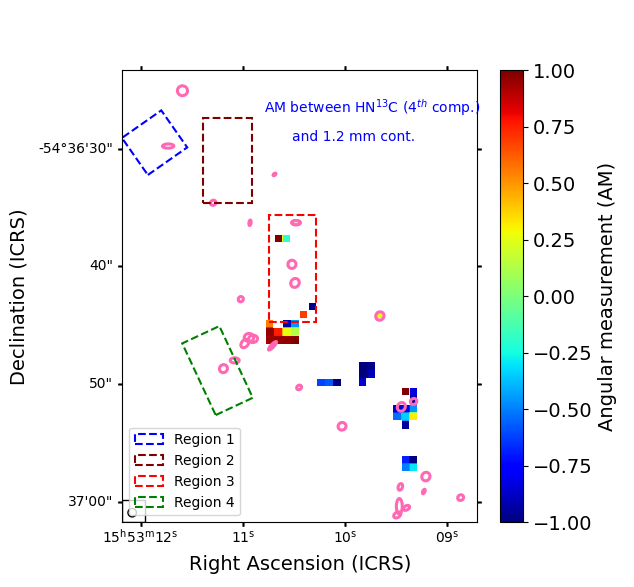}\includegraphics[width=2.4in,height=2.2in,angle=0]{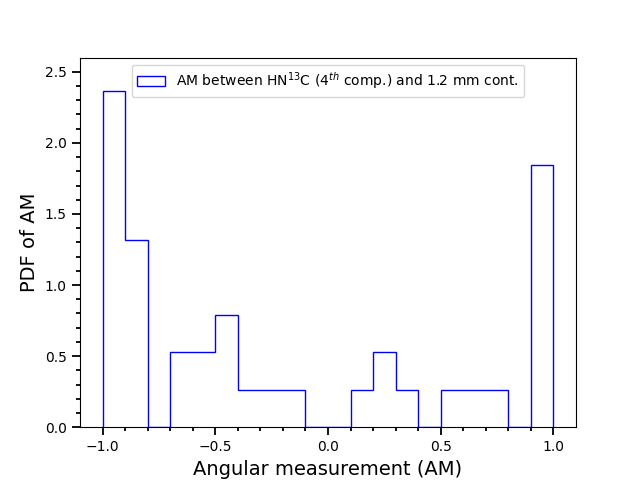}\\
	\caption{Continuation of Figs. \ref{fig:fig3} and  \ref{fig:fig5} but for \texttt{first}, \texttt{second}, \texttt{third}, and \texttt{fourth} HN$^{13}$C components.}
    \label{fig:figE1}
\end{figure*}

\begin{figure*}
    \centering
     \includegraphics[width=2.4in,height=2.2in,angle=0]{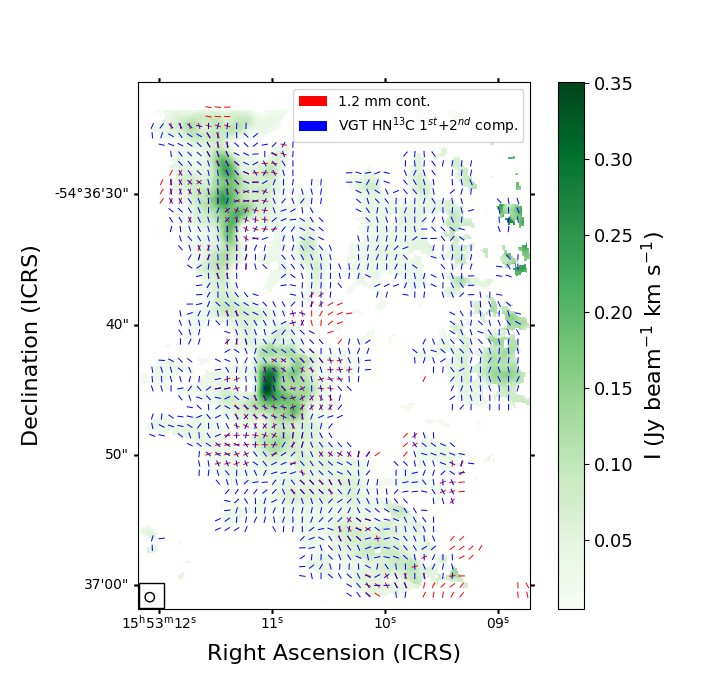} \includegraphics[width=2.4in,height=2.2in,angle=0]{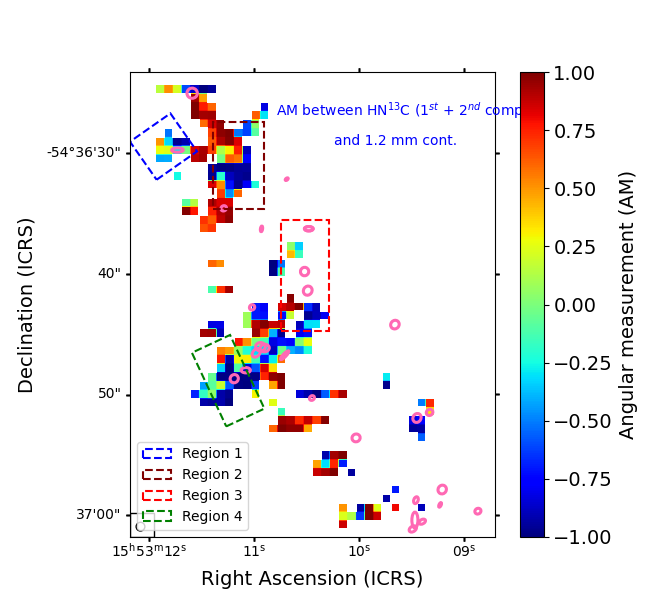}\includegraphics[width=2.4in,height=2.2in,angle=0]{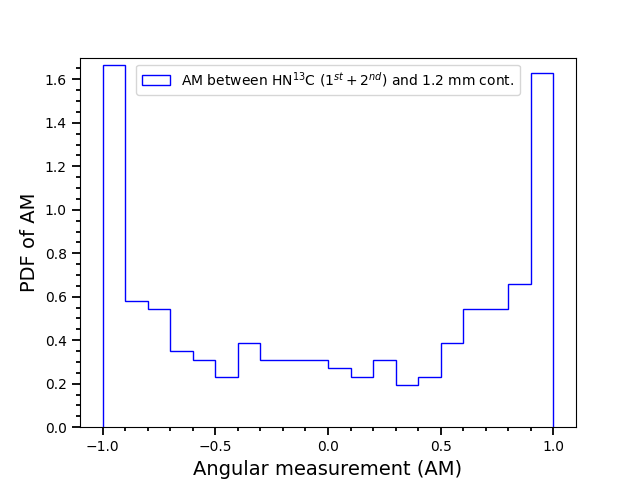}\\ 
     \includegraphics[width=2.4in,height=2.2in,angle=0]{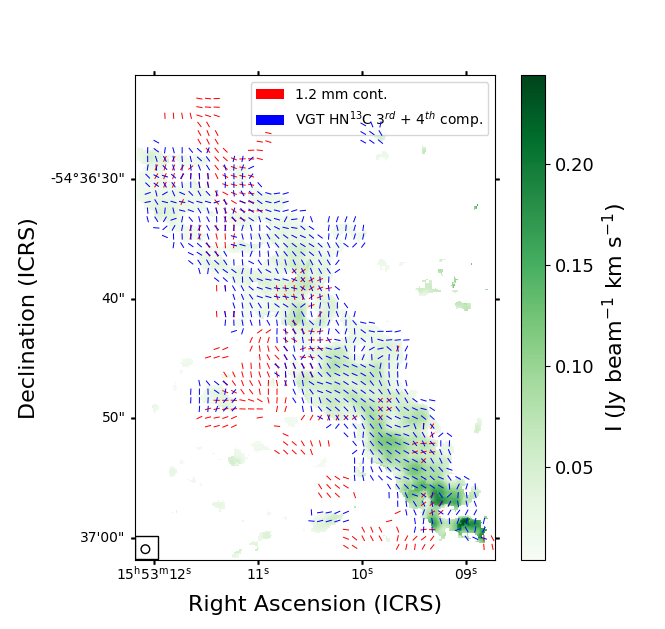} \includegraphics[width=2.4in,height=2.2in,angle=0]{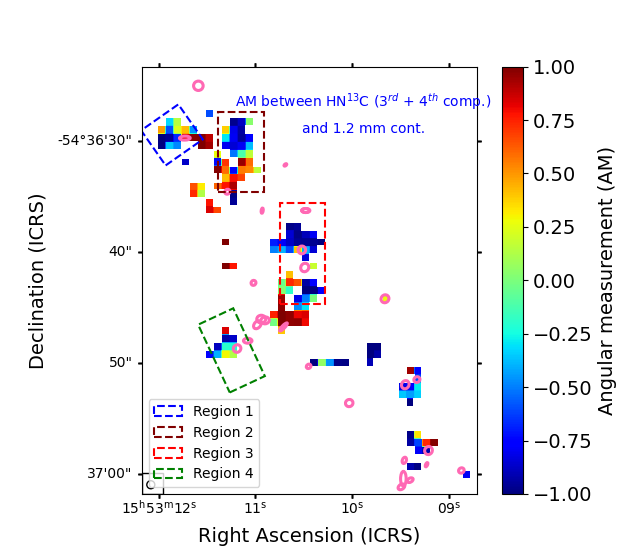}\includegraphics[width=2.4in,height=2.2in,angle=0]{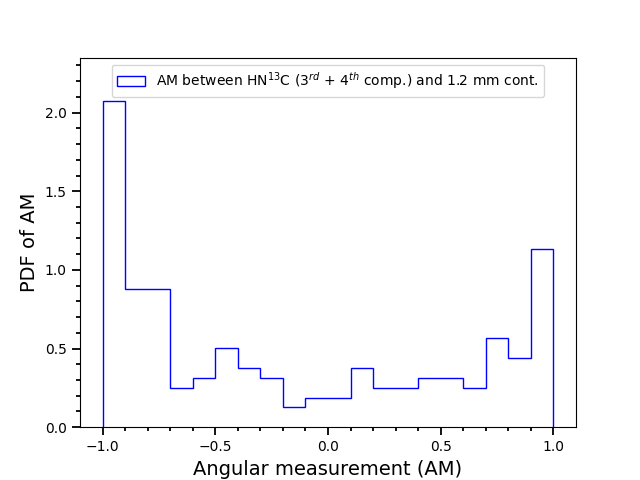}\\ 
       \includegraphics[width=2.4in,height=2.2in,angle=0]{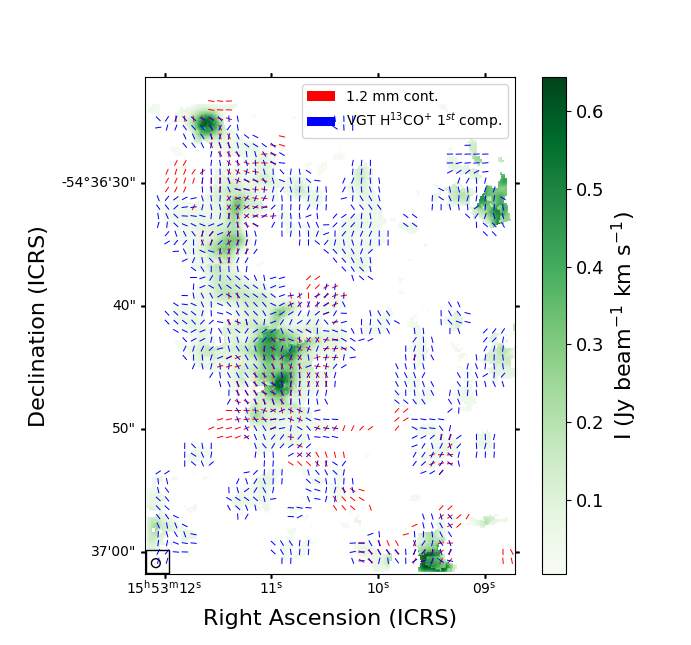} \includegraphics[width=2.4in,height=2.2in,angle=0]{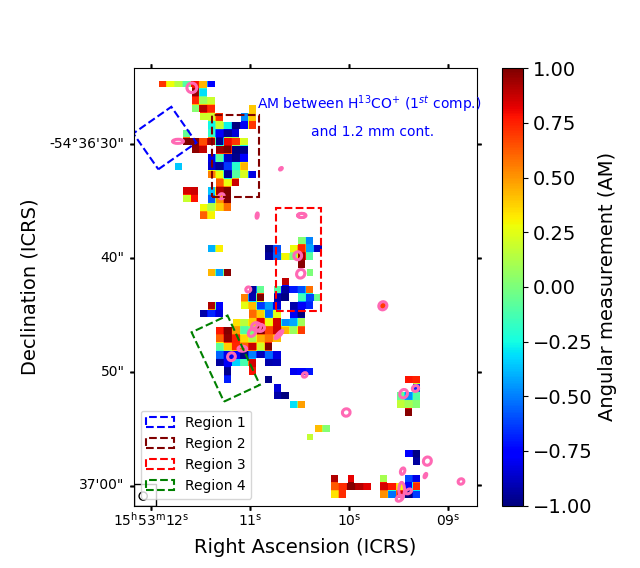}\includegraphics[width=2.4in,height=2.2in,angle=0]{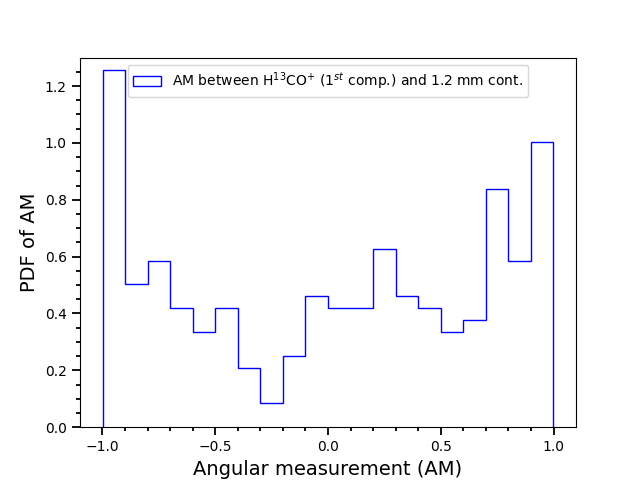}\\ 

    \includegraphics[width=2.4in,height=2.2in,angle=0]{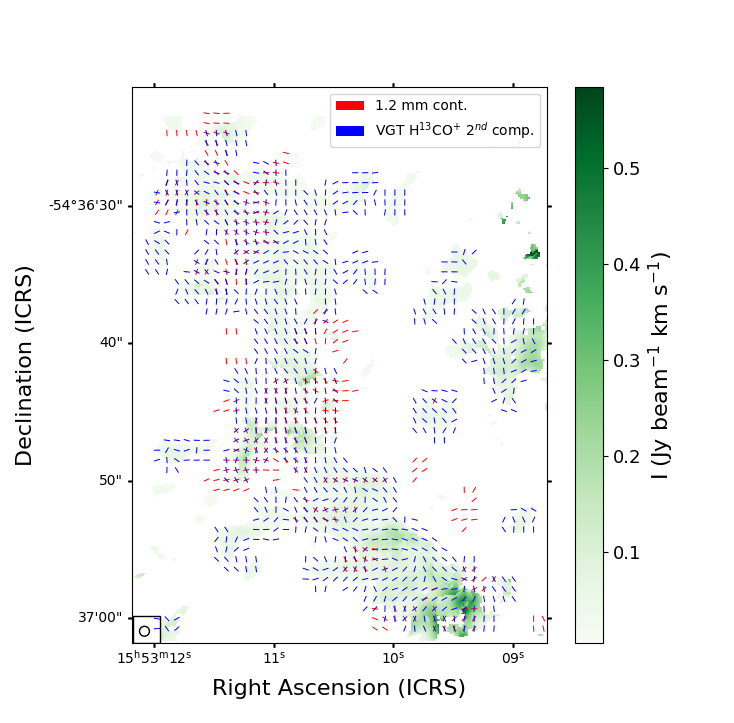} \includegraphics[width=2.4in,height=2.2in,angle=0]{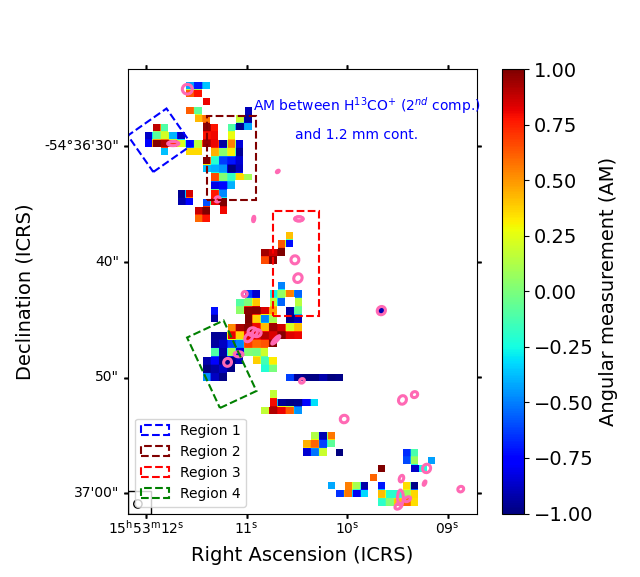}\includegraphics[width=2.4in,height=2.2in,angle=0]{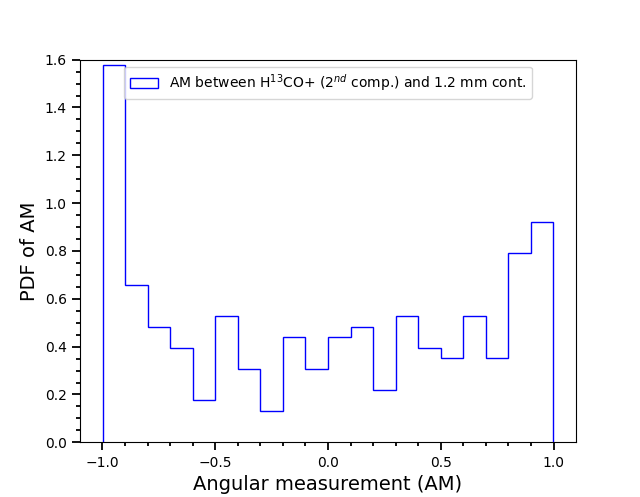}\\

	\caption{Continuation of Figs. \ref{fig:fig3} and  \ref{fig:fig5} but for \texttt{first+second}, and \texttt{third+fourth} HN$^{13}$C components and \texttt{first}, and \texttt{second} H$^{13}$CO$^{+}$ components.}
    \label{fig:figE2}
\end{figure*}



\begin{figure*}[ht!]
    \centering

     \includegraphics[width=2.4in,height=2.2in,angle=0]{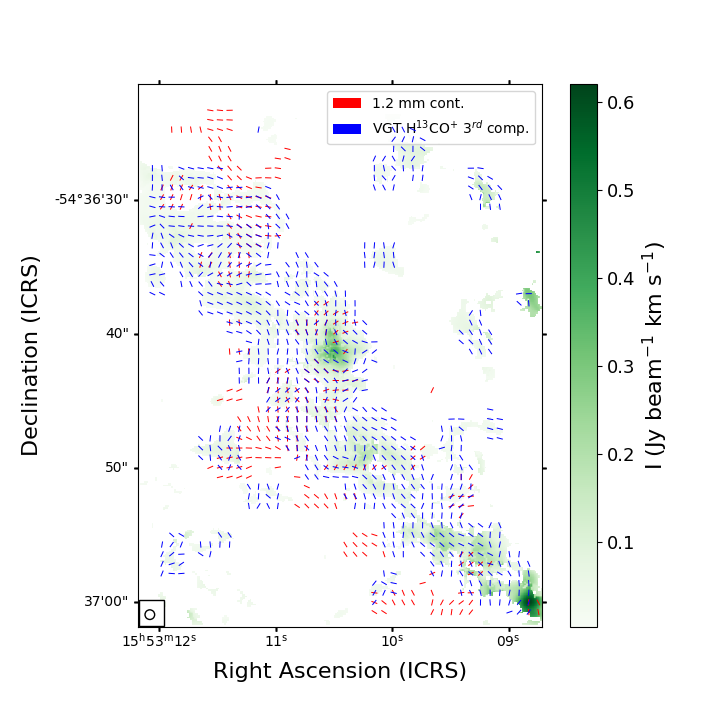} \includegraphics[width=2.4in,height=2.2in,angle=0]{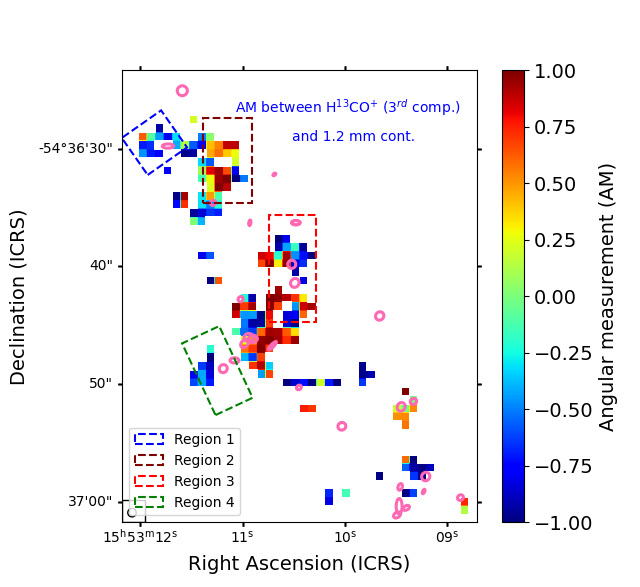}\includegraphics[width=2.4in,height=2.2in,angle=0]{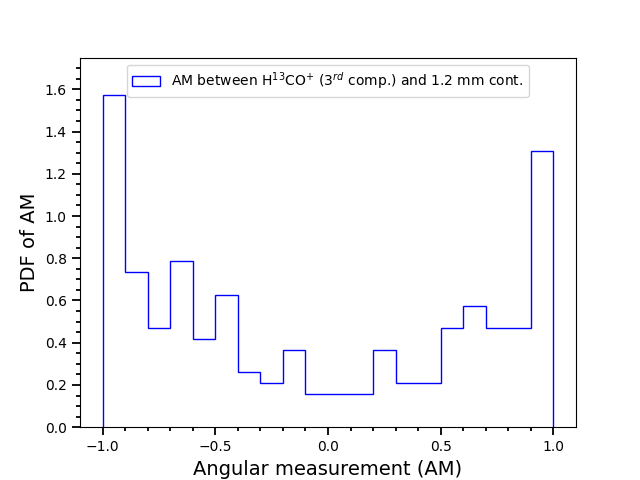}\\ 
     
      \includegraphics[width=2.4in,height=2.2in,angle=0]{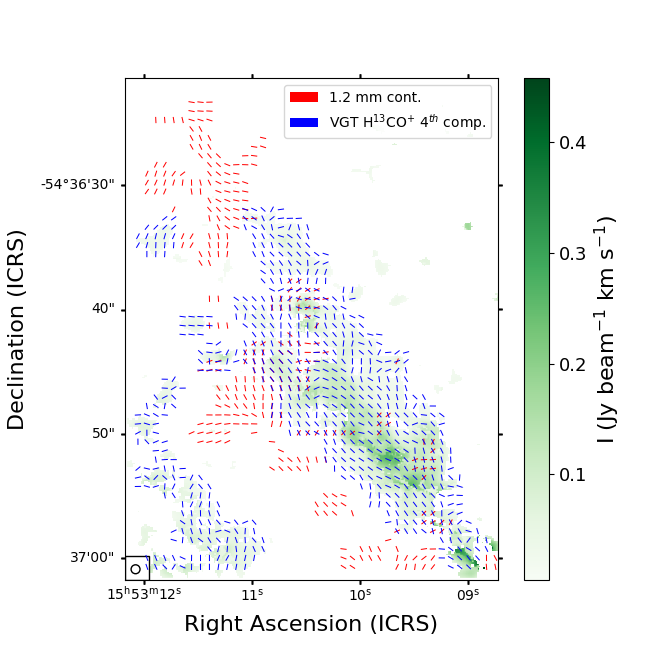} \includegraphics[width=2.4in,height=2.2in,angle=0]{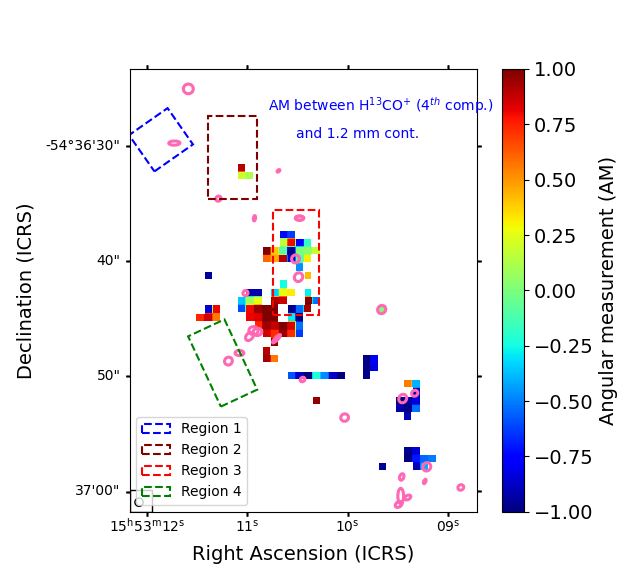}\includegraphics[width=2.4in,height=2.2in,angle=0]{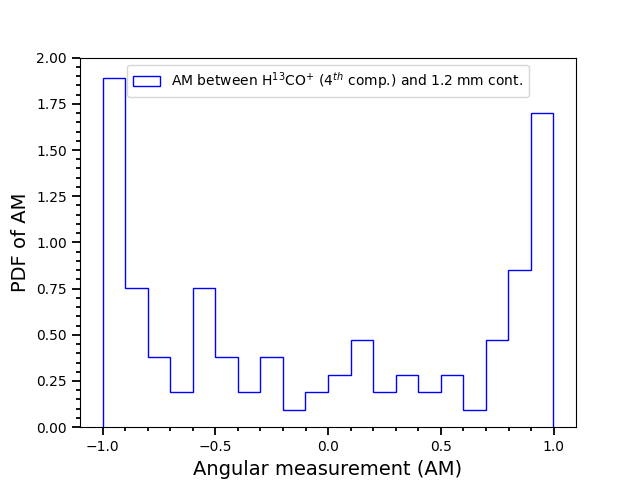}\\

     \includegraphics[width=2.4in,height=2.4in,angle=0]{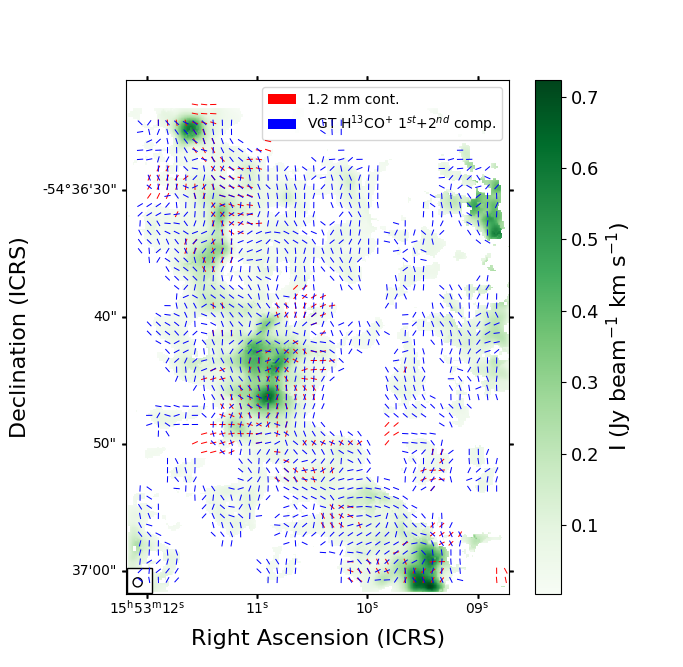} \includegraphics[width=2.4in,height=2.4in,angle=0]{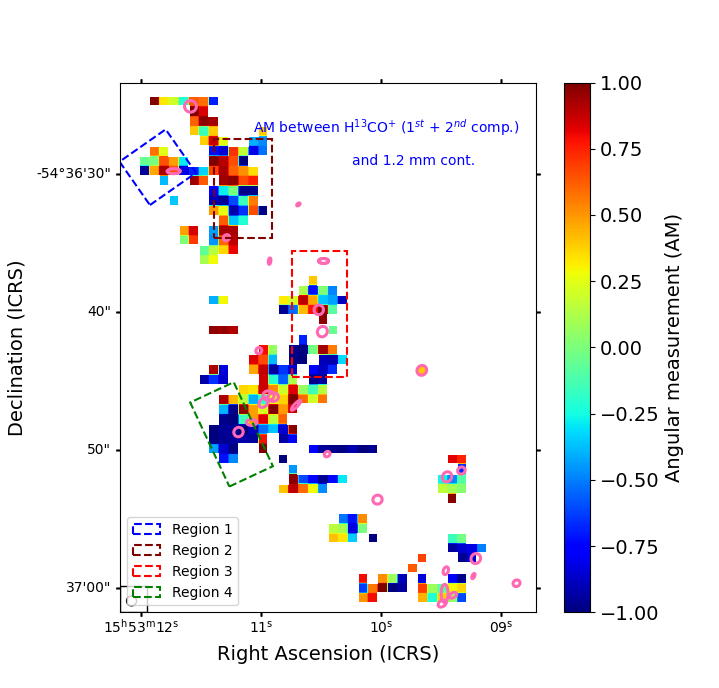}\includegraphics[width=2.4in,height=2.4in,angle=0]{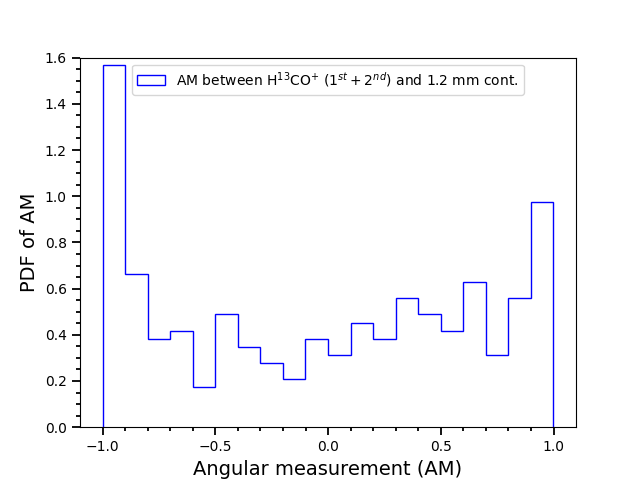}\\
     \includegraphics[width=2.4in,height=2.4in,angle=0]{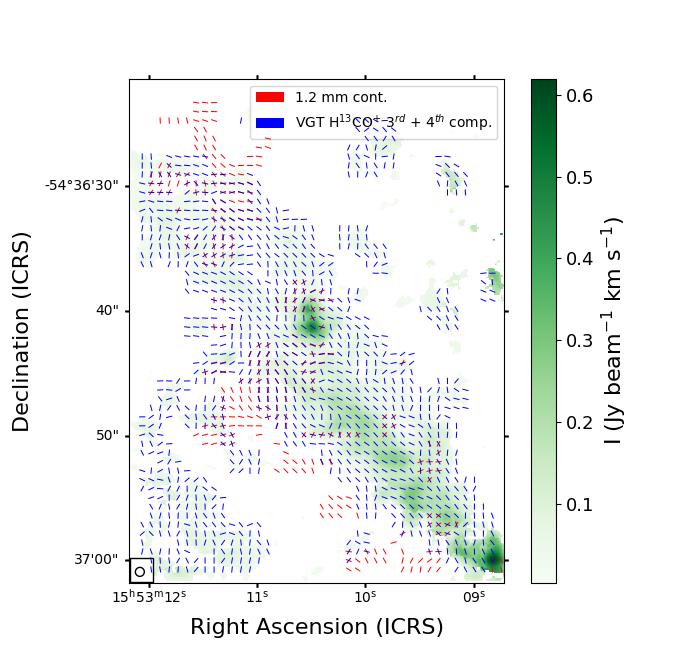} \includegraphics[width=2.4in,height=2.4in,angle=0]{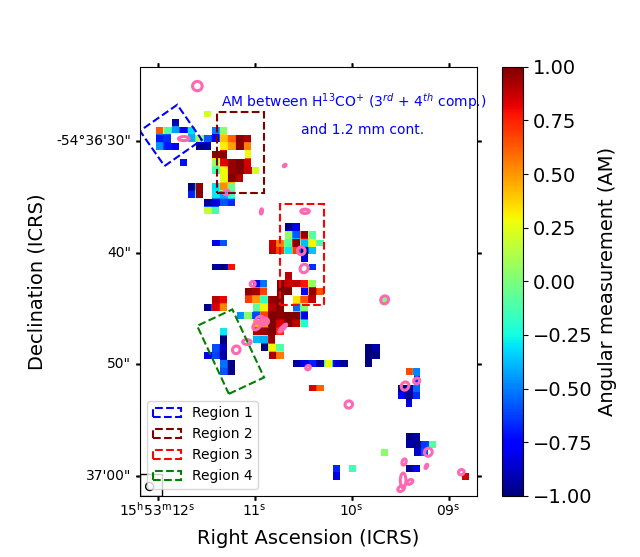}\includegraphics[width=2.4in,height=2.4in,angle=0]{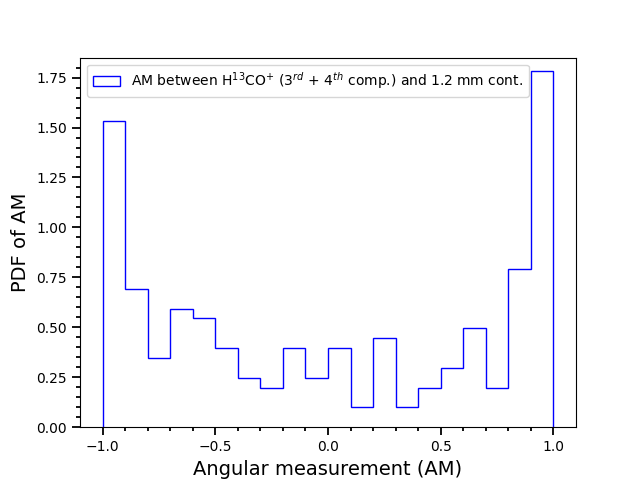}\\

	\caption{Continuation of Figs. \ref{fig:fig3} and  \ref{fig:fig5}  but for \texttt{third}, \texttt{fourth}, \texttt{first+second}, and \texttt{third+fourth}  H$^{13}$CO$^{+}$ components.}
    \label{fig:figF1}
\end{figure*}

\section{Investigation of the $\theta_\text{r} \rightarrow \text{AM}$  Transformation}\label{Appendix_C_0}

The alignment measure, $AM=\cos(2\theta_\text{r})$, is commonly used to quantify the parallel and perpendicular orientations of the VGT with respect to the magnetic field, where $\theta_\text{r}$ represents the relative orientation angle. However, this transformation is nonlinear and can significantly modify the underlying distribution of $\theta_\text{r}$. The derivative of the transformation, $|dAM/d\theta_\text{r}|=2\sin(2\theta_\text{r})$, describes how the relative orientation angle is mapped into the alignment measure. The magnitude of this derivative approaches zero at $\theta_\text{r}=0^\circ$ and $90^\circ$, while it reaches its maximum value at $\theta_\text{r}=45^\circ$. Therefore, a broad range of orientations close to $0^\circ$ and $90^\circ$ are compressed into narrow intervals near $AM=+1$ and $AM=-1$, respectively, causing data points with different $\theta_\mathrm{r}$ values to accumulate near the two extrema. In contrast, orientations around $45^\circ$ are stretched over a wider range of $AM$ values and do not produce similarly concentrated features.

\begin{figure*}
    \centering
    \includegraphics[width=7.4in,height=6.4in,angle=0]{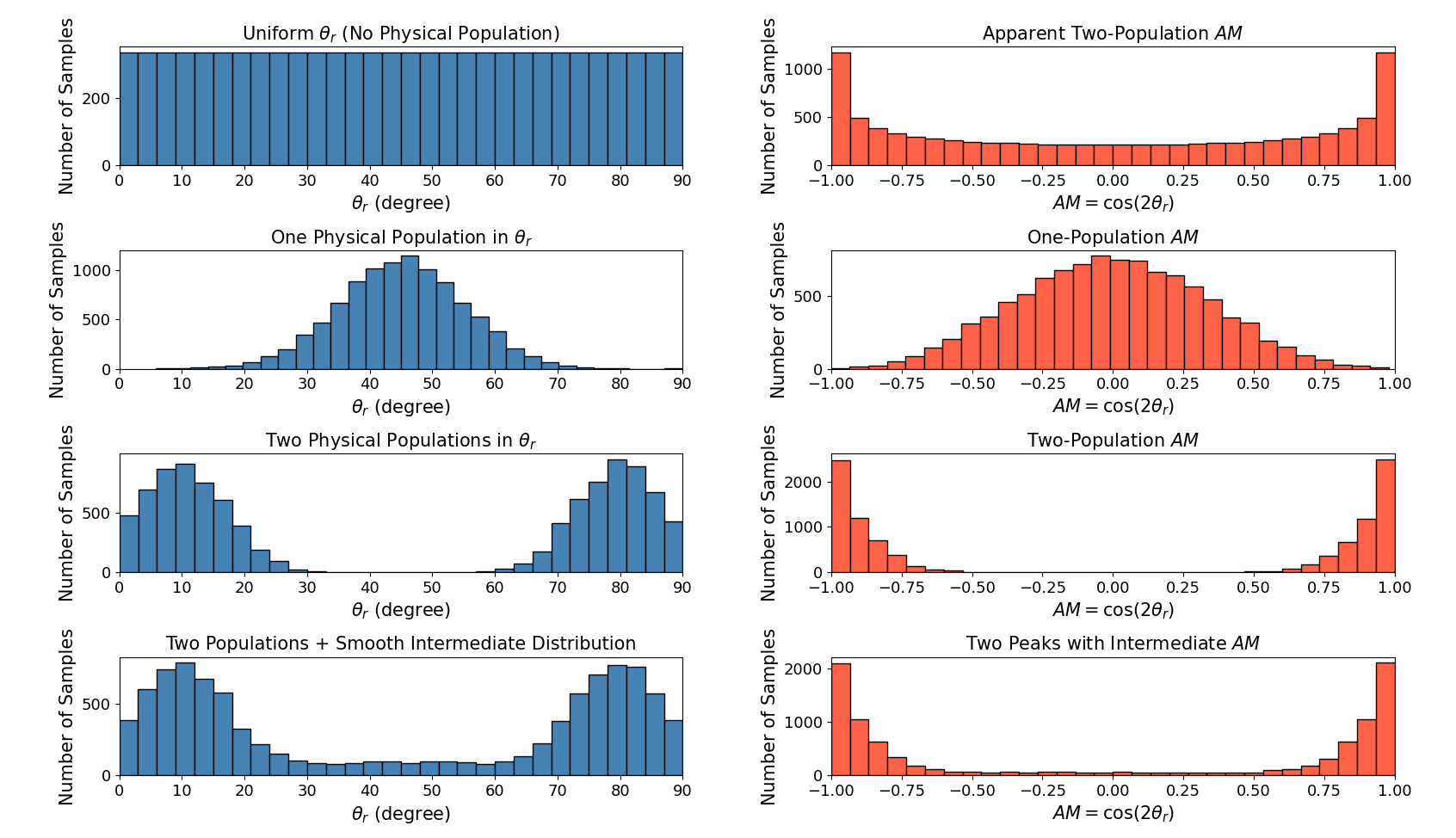}\\
	\caption{Illustrating the effect of the nonlinear transformation $AM=\cos(2\theta_\text{r})$ on the distribution of relative orientation angles ($\theta_\mathrm{r}$). The left panels show the input $\theta_\text{r}$ distributions, while the right panels show the corresponding $AM$ distributions for four representative cases: (i) uniform $\theta_\text{r}$, (ii) a single physical population, (iii) two physical populations, and (iv) two dominant populations with a smooth intermediate component. The top row illustrates that a uniform $\theta_\text{r}$ distribution can produce an apparently bimodal $AM$ distribution solely because of the nonlinear transformation.}
    \label{fig:figC0}
\end{figure*}

To investigate the impact of this nonlinear transformation, we perform the analysis with $10^4$ samples for different underlying $\theta_\mathrm{r}$ distributions and converting them into their corresponding $AM$ values. This is shown in Fig. \ref{fig:figC0}. Here, we consider four representative cases: (i) a uniform $\theta_\mathrm{r}$ distribution between $0^\circ$ and $90^\circ$, (ii) a single physical population with $\theta_\mathrm{r}$ centered around $45^\circ$, (iii) two distinct physical populations centered around $10^\circ$ and $80^\circ$, and (iv) two dominant populations with a smoothly distributed intermediate component. The result show that a uniform $\theta_\mathrm{r}$ distribution may produce an apparently bimodal $AM$ distribution due to the nonlinear mapping alone. A single population centered around $45^\circ$ remains concentrated near $AM\approx0$, while genuine bimodality in the underlying $\theta_\mathrm{r}$ distribution is preserved as two peaks near $AM=\pm1$. In the case with additional intermediate orientations, the two dominant $AM$ peaks remain visible but are accompanied by a population of intermediate $AM$ values, demonstrating that a broad range of physical orientation distributions can produce similar $AM$ morphologies. These results highlight that the presence of a bimodal $AM$ distribution alone does not necessarily indicate two distinct physical populations, and the underlying $\theta_\mathrm{r}$ distribution must be considered when interpreting alignment measurements. \\

Although the nonlinear transformation may introduce artificial features in the $AM$ distribution, the alignment measure also provides a useful representation of the relative orientation distribution. Unlike the direct $\theta_\text{r}$ distribution, where the distinction between preferred parallel ($\theta_\text{r}=0^\circ$) and perpendicular ($\theta_\text{r}=90^\circ$) orientations is spread over the angular space, $AM$ maps these two preferential orientations directly to the extrema, $AM=+1$ and $AM=-1$, respectively. Therefore, the $AM$ representation is particularly useful for identifying whether a population preferentially exhibits parallel or perpendicular alignment. In this sense, $AM$ provides a clearer visualization of the alignment tendency near the two extreme configurations, while the underlying $\theta_\mathrm{r}$ distribution remains essential for distinguishing genuine physical populations from features introduced by the nonlinear transformation.\\

\section{Uncertainty in polarization angle and VGT derived angle}\label{Appendix_K}







\begin{figure*}[ht!]

	\includegraphics[width=1.5in,height=2.1in,angle=0]{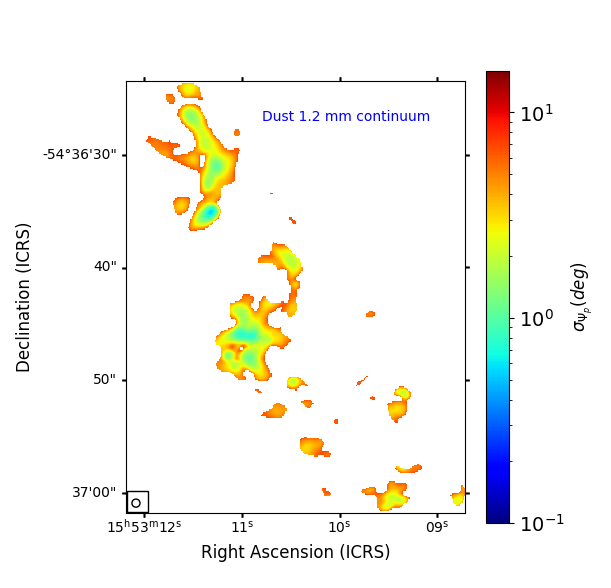}
    \includegraphics[width=1.5in,height=2.1in,angle=0]{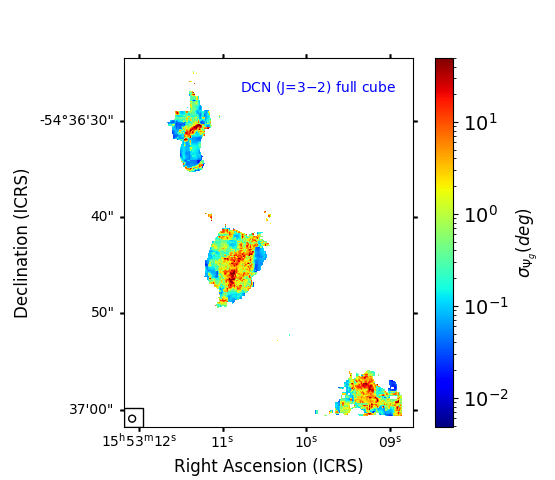}\includegraphics[width=1.5in,height=2.1in,angle=0]{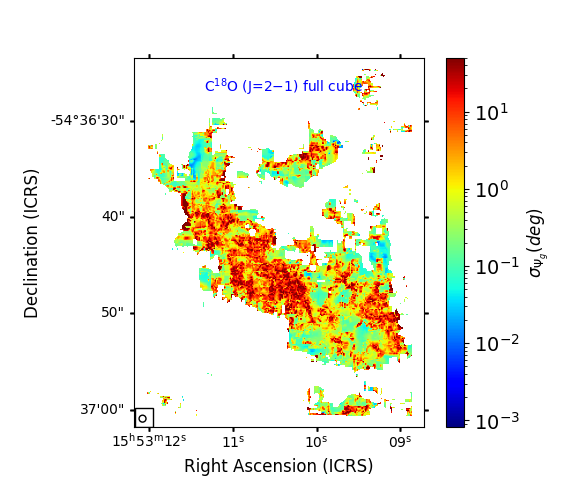}
    \includegraphics[width=1.5in,height=2.1in,angle=0]{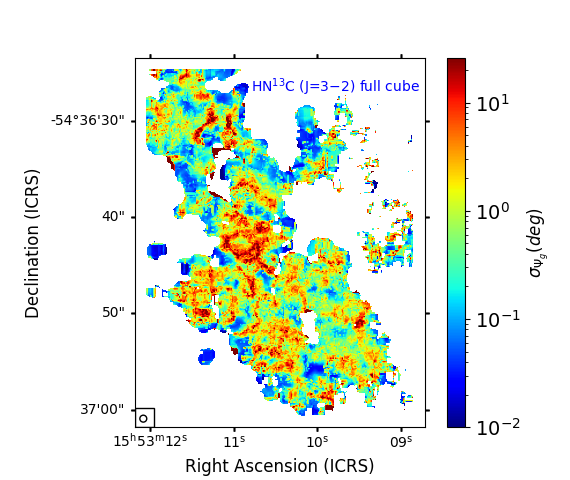}\includegraphics[width=1.5in,height=2.1in,angle=0]{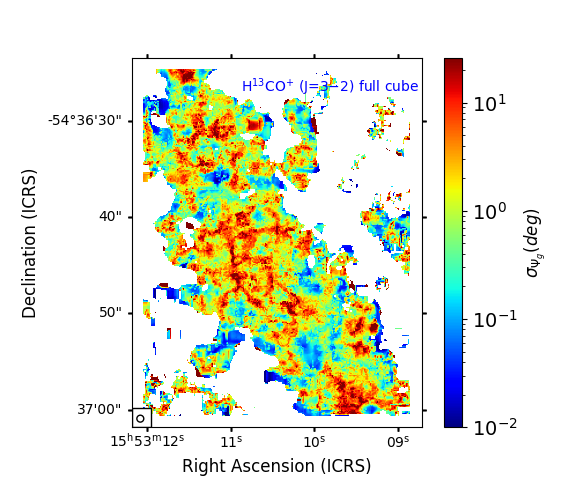}

	\caption{Pixel-wise error map of the 1.2 mm dust polarization angle and VGT derived angle.}.
    \label{fig:fig.K1}
\end{figure*}

We show the spatial distribution of the error maps of the polarization angle ($\psi_\text{p}$) and the VGT derived angles ($\psi_\text{g}$) in Fig. \ref{fig:fig.K1}. We calculated the errors using the formulae described in \cite{2024MNRAS.528.3897S}.\\

\section{Uncertainty in angular measurement (AM) for four spectral lines}\label{Appendix_L}
\begin{figure*}[ht!]

	\includegraphics[width=1.8in,height=2.1in,angle=0]{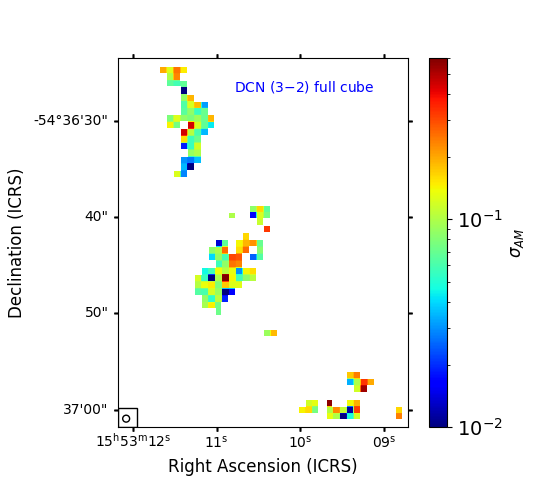}
    \includegraphics[width=1.8in,height=2.1in,angle=0]{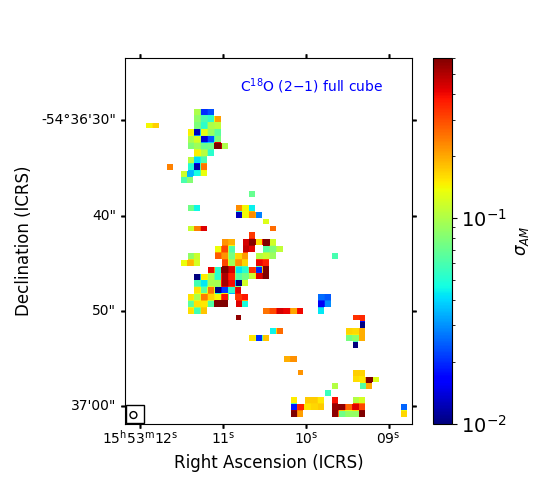}\includegraphics[width=1.8in,height=2.1in,angle=0]{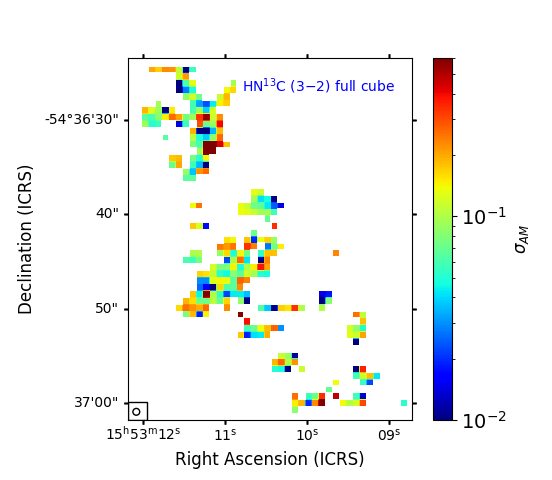}
    \includegraphics[width=1.8in,height=2.1in,angle=0]{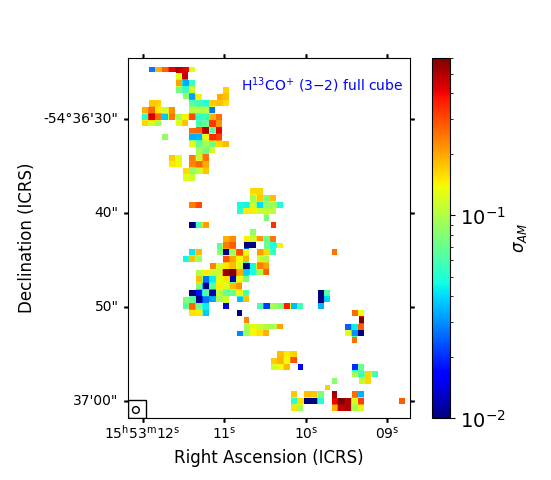}

	\caption{Spatial distribution of the error of the angular measurement ($\sigma_\text{AM}$) for four spectral lines.}.
    \label{fig:fig.L1}
\end{figure*}

We show the error of the angular measurements (AM) ($\sigma_\text{AM} (x,y)$) for four spectral lines in Fig. \ref{fig:fig.L1}. The errors were calculated using the following formulae:

\begin{eqnarray}
\sigma_\text{AM} (x,y) & = &  2 \Bigl| sin~2\bigl(\psi_\text{B} (x,y) - \psi_\text{D}(x,y)\bigl)\Bigl|~\sqrt{ \bigl\{\sigma^{2}_{\psi_\text{B}(x,y)} + \sigma^{2}_{\psi_\text{D}(x,y)}\bigl\} }, \nonumber \\ 
                       & = &  2 \sqrt{\bigl\{1-AM^{2} (x,y)\bigl\}}~\sqrt{ \bigl\{\sigma^{2}_{\psi_\text{B}(x,y)} + \sigma^{2}_{\psi_\text{D}(x,y)}\bigl\}. } 
\end{eqnarray}\\

The uncertainty in $AM(x,y)$ can become artificially small toward the limiting values, $AM=-1$ and $AM=+1$, because of the nonlinear transformation from $\theta_\mathrm{r}$ to $AM$. In particular, a direct first-order error propagation can underestimate the uncertainty near these endpoints, even when the VGT- and dust-derived angles, $\psi_\text{B}$ and $\psi_\text{D}$, have finite uncertainties. To properly estimate the uncertainty in $AM$ near these limiting values, we first calculate the mean values of $\psi_\text{B}$ and $\psi_\text{D}$ and their associated uncertainties. We then generate $10^{5}$ independent random realizations of $\psi_\text{B}$ and $\psi_\text{D}$ according to their respective uncertainties. For each realization, we calculate the corresponding relative angle $\theta_\mathrm{r}$ and $AM$, and construct the resulting $AM$ distribution. The median of this distribution is adopted as the representative $AM$ value, while the 16th and 84th percentiles are used to define the 68\% confidence interval and hence the asymmetric uncertainty on $AM$.\\

\twocolumn

\end{appendix}

\vspace{20 mm}


\end{document}